\documentclass[acmsmall]{acmart}

\usepackage{listings}

\usepackage{amsmath,amssymb,amsfonts}
\usepackage{graphicx}
\usepackage{xcolor}
\usepackage{subfigure}
\usepackage{booktabs}
\usepackage{multirow}
\usepackage{makecell}
\usepackage{threeparttable}
\usepackage{enumitem} 
\usepackage{url}
\usepackage{bbding}
\usepackage{diagbox}
\usepackage[ruled, linesnumbered]{algorithm2e}
\usepackage[breakable]{tcolorbox}
\usepackage{braket}
\usepackage{longtable}

\newtheorem{example}{Example}[section]
\newtheorem{criterion}{Criterion}

\AtBeginDocument{%
	\providecommand\BibTeX{{%
			\normalfont B\kern-0.5em{\scshape i\kern-0.25em b}\kern-0.8em\TeX}}}
		
\setcopyright{acmcopyright}
\copyrightyear{2025}
\acmYear{2025}
\acmDOI{10.1145/1122445.1122456}

\begin{document}
	
\title{A Black-box Testing Framework for Oracle Quantum Programs}

\author{Peixun Long}
\email{longpx@ihep.ac.cn}
\affiliation{
	\institution{Institute of High Energy Physics, Chinese Academy of Science}
	\country{China}
}

\author{Jianjun Zhao}
\email{zhao@ait.kyushu-u.ac.jp}
\affiliation
{
	\institution{Kyushu University}
	\country{Japan}
}

\begin{abstract}
Oracle quantum programs are a fundamental class of quantum programs that serve as a critical bridge between quantum computing and classical computing. Many important quantum algorithms are built upon oracle quantum programs, making it essential to ensure their correctness during development. Although software testing is a well-established approach for improving program reliability, no systematic method has been developed to test oracle quantum programs. This paper proposes a black-box testing framework designed for general oracle quantum programs. We formally define these programs, establish the foundational theory for their testing, and propose a detailed testing framework. We develop a prototype tool and conduct extensive experimental evaluations to evaluate the effectiveness of the framework. Our results demonstrate that the proposed framework significantly aids developers in testing oracle quantum programs, providing insights to enhance the reliability of quantum software.
\end{abstract}


\maketitle

\section{Introduction}
\label{sec:introduction}

Quantum computing is a rapidly evolving field that promises to revolutionize diverse domains by leveraging the principles of quantum mechanics to process information and solve computational problems~\cite{national2019quantum}. Unlike classical computing, quantum computing harnesses unique quantum phenomena such as superposition, entanglement, and quantum interference, enabling quantum algorithms to address specific problems with exponential or significant speedup~\cite{deutsch1985quantum,grover1996fast,shor1999polynomial}. These advances have opened new frontiers in optimization~\cite{farhi2014quantum}, encryption~\cite{mosca2018cybersecurity}, machine learning~\cite{biamonte2017quantum}, chemistry~\cite{mcardle2018quantum}, materials science~\cite{yang2017mixed}, and high energy physics~\cite{PRXQuantum.5.037001}.
As the field progresses, quantum computing is transitioning from theoretical exploration to practical applications, necessitating the development of high-quality quantum software. The complexity and non-intuitiveness of quantum systems present unique challenges in software design, verification, and testing~\cite{ying2012floyd,zhao2020quantum,PiattiniPPHSHGP2020,ali2022software}. To fully harness the potential of quantum computing, it is imperative to establish robust methodologies for developing reliable quantum software, particularly as quantum hardware advances toward scalability and practical usability.

Oracle quantum programs represent a fundamental class of quantum software that computes classical functions on quantum systems, thereby bridging the gap between quantum and classical computing. These programs are integral to many quantum algorithms, including Grover's algorithm~\cite{grover1996fast}, quantum arithmetic~\cite{Vedral_1996,Sahin2020QAonFourier}, and Hamiltonian simulation~\cite{lloyd1996universal}. Ensuring the correctness and reliability of such programs is vital for the broader adoption of quantum computing. However, the unique characteristics of quantum systems, such as superposition, entanglement, and measurement, present significant challenges to software testing, making it a complex yet essential task~\cite{miranskyy2020your,huang2019statistical}.
Although significant efforts have been made to test quantum programs~\cite{ali2021assessing,honarvar2020property,abreu2022metamorphic,wang2021application,fortunato2022mutation}, these methods have primarily focused on general testing of quantum programs. However, systematic methodologies specifically tailored for testing oracle quantum programs are lacking, which play a critical role in bridging the gap between quantum and classical computing.

To achieve this, we first formalize oracle quantum programs by providing a mathematical definition and discussing the foundational theories underlying their testing. According to these foundations, we develop a systematic black-box testing framework that consists of (1) designing equivalence classes for classical functions associated with the oracle quantum program, (2) deriving quantum equivalence classes from these classical equivalence classes, and (3) preparing input states based on these quantum equivalence classes and executing tests to complete the testing task. Furthermore, we implement a prototype testing tool and conduct comprehensive experimental evaluations using benchmark oracle quantum programs to demonstrate the effectiveness of our framework.

This paper makes the following contributions:

\begin{itemize}
\setlength{\itemsep}{2pt}
\item \textbf{Definition and Foundational Theory:} We present a formal mathematical definition of oracle quantum programs and effectively establish the theoretical basis for testing them.

\item \textbf{Testing Framework:} We propose a systematic black-box testing framework for oracle quantum programs that details the construction of equivalence classes, automated testing processes, and parameter optimization strategies.

\item \textbf{Prototype Tool and Evaluation:} We develop a prototype testing tool to implement our framework and evaluate its effectiveness through extensive experiments on benchmark oracle quantum programs.
\end{itemize}

The rest of this paper is organized as follows. Section~\ref{sec:background} introduces the basic concepts of quantum computing and oracle quantum programs. Section~\ref{sec:question} provides the formal definition of oracle quantum programs and outlines the core research questions. Section~\ref{sec:methods} presents our black-box testing framework and discusses its implementation details. Section~\ref{sec:evaluation} evaluates the effectiveness of the proposed framework. Section~\ref{sec:threat} addresses threats to the validity of our framework. Related work is discussed in Section~\ref{sec:relatedwork}, and Section~\ref{sec:conclusion} concludes the paper with directions for future research.

\section{Background}
\label{sec:background}
Quantum computing and oracle quantum programs serve as essential foundations to understand the theoretical and practical aspects of this study. This section provides an overview of the core concepts in quantum computing, including qubits, quantum gates, and circuits, which are essential to understand the functionality of quantum algorithms. Additionally, we explore the role and implementation of oracle quantum programs, which serve as a critical bridge between classical and quantum computing. Finally, we give a brief introduction to software testing and equivalence class partition, which will be involved in this paper.

\subsection{Quantum Computing}
\label{subsec:QC}

Quantum computing represents a transformative approach to information processing. It uses quantum mechanics to solve computational problems that are intractable for classical systems. This section provides an overview of its foundational concepts, including qubits, measurement, quantum gates, and circuits.

\vspace{2mm}
\noindent $\bullet$ \textbf{Qubit and Quantum State.}\hspace*{1mm}
The foundation of quantum computing is the quantum bit, or \textit{qubit}. Similar to a classical bit with two states (0 and 1), a qubit has two \textit{basis states} represented as $\ket{0}$ and $\ket{1}$.
Unlike classical bits, qubits can exist in a \textit{superposition} of the basis states, expressed as $a\ket{0}+b\ket{1}$, where $a$ and $b$ are complex numbers called \textit{amplitudes}, satisfying $|a|^2+|b|^2=1$.
This property enables an exponential number of amplitudes in a state, offering computational advantages over classical systems. For example, a 2-qubit system has four basis states: $\ket{00}$, $\ket{01}$, $\ket{10}$, and $\ket{11}$. The general state of a 2-qubit system can be written as $a_{00}\ket{00}+a_{01}\ket{01}+a_{10}\ket{10}+a_{11}\ket{11}$, where $|a_{00}|^2+|a_{01}|^2+|a_{10}|^2+|a_{11}|^2=1$. Generally, an $n$-qubit quantum system has $2^n$ basis states, corresponding to the binary strings of integers $0$ to $2^n-1$: $\ket{0}$, $\ket{1}$, $\ket{2}$, \dots, $\ket{2^n-1}$. They are called \textit{computational-basis states}.

A quantum state can be represented as a \textit{state vector} in the Hilbert space, where the elements of the vector correspond to the probability amplitudes. For example, the above single-qubit basis states can be represented as two-dimensional column vectors:
\begin{equation*}
\ket{0} = \left[
\begin{array}{c}
	1 \\
	0
\end{array}
\right], \qquad
\ket{1} = \left[
\begin{array}{c}
	0 \\
	1
\end{array}
\right], \qquad
a\ket{0}+b\ket{1} = \left[
\begin{array}{c}
	a \\
	b
\end{array}
\right]
\end{equation*}
\noindent Similarly, the general 2-qubit state can be expressed as a 4-dimensional vector $[a_{00}, a_{01}, a_{10}, a_{11}]^T$, where $^T$ is the transpose operation. Generally, an $n$-qubit state corresponds to a $2^n$-dimension column vector, i.e., there are $2^n$ amplitudes. In fact, the exponential increase in the amplitudes is the source of the power of quantum computing.

\vspace{2mm}
\noindent $\bullet$ \textbf{Measurement.}\hspace*{1mm}
Although qubits can contain exponential numbers of amplitudes, access to this information is strictly limited. In quantum devices, information in qubits can only be retrieved through \textit{measurement}. Measuring a quantum system produces a classical value with a probability determined by the corresponding amplitude. After measurement, the quantum system's state collapses to a basis state based on the observed value. For example, measuring a qubit $a\ket{0}+b\ket{1}$ yields the result 0 with probability $|a|^2$ and then collapses to $\ket{0}$, or the result 1 with probability $|b|^2$ and then collapses to $\ket{1}$. Note that the unique amplitude of computational-basis states is 1, so computational-basis states are similar to classical variables: measurement will always return a value and will not change the state. In contrast, measuring a superposition state will yield uncertain results. Owing to the fact that testing tasks involve observing running results, this property will impact testing tasks for quantum programs.

\vspace{2mm}
\noindent $\bullet$ \textbf{Quantum gates.}\hspace*{1mm}
To perform quantum computations, \textit{quantum gates} are applied to qubits. A quantum gate represents a \textit{unitary transform} acting on the corresponding state vector.
Unitary transform is a kind of reversible linear transform on Hilbert space which keeps the inner product of vectors, which can be represented as a \textit{unitary matrix}. Generally, the unitary matrix $U$ is a square matrix which satisfies $U^\dagger U=UU^\dagger = I$, where $U^\dagger$ is the \textit{conjugate transpose}\footnote{Take the complex conjugate of every element first, and then transpose the matrix.} of $U$. Specifically, a $n$-qubit gate can be represented by a $2^n \times 2^n$ unitary matrix. The output state is obtained by multiplying this matrix by the original state vector. For example, the $X$ gate, analogous to the classical NOT gate, flips the state of a qubit. The matrix representation of the $X$ gate is:
\begin{equation*}
X \triangleq \left[
\begin{array}{cc}
	0 & 1 \\
	1 & 0
\end{array}
\right]
\end{equation*}

\noindent Note that the basis states can be represented as $\ket{0} = [1,0]^T$ and $\ket{1} = [0,1]^T$. The application of the $X$ gate on $\ket{0}$ and $\ket{1}$ proceeds as follows:
\begin{equation*}
X\ket{0} = \left[
\begin{array}{cc}
	0 & 1 \\
	1 & 0
\end{array}
\right]\left[
\begin{array}{c}
	1\\
	0
\end{array}
\right] = \left[
\begin{array}{c}
	0\\
	1
\end{array}
\right] = \ket{1}, \qquad X\ket{1} = \left[
\begin{array}{cc}
	0 & 1 \\
	1 & 0
\end{array}
\right]\left[
\begin{array}{c}
	0\\
	1
\end{array}
\right] = \left[
\begin{array}{c}
	1\\
	0
\end{array}
\right] = \ket{0}
\end{equation*}

\noindent Another example is the $H$ gate, which generates superposition states from computational basis states:
\begin{align*}
&H \triangleq \frac{1}{\sqrt 2}\left[
\begin{array}{cc}
	1 & 1 \\
	1 & -1
\end{array}
\right] \notag
\\
&H\ket{0} = \frac{1}{\sqrt 2}\left[
\begin{array}{cc}
	1 & 1 \\
	1 & -1
\end{array}
\right]\left[
\begin{array}{c}
	1\\
	0
\end{array}
\right] = \frac{1}{\sqrt 2}\left[
\begin{array}{c}
	1\\
	1
\end{array}
\right] = \frac{1}{\sqrt 2}\left(\ket{0} + \ket{1}\right) \triangleq \ket{+} \\
&H\ket{1} = \frac{1}{\sqrt 2}\left[
\begin{array}{cc}
	1 & 1 \\
	1 & -1
\end{array}
\right]\left[
\begin{array}{c}
	0\\
	1
\end{array}
\right] = \frac{1}{\sqrt 2}\left[
\begin{array}{c}
	1\\
	-1
\end{array}
\right] = \frac{1}{\sqrt 2}\left(\ket{0} - \ket{1}\right) \triangleq \ket{-}
\end{align*}
\noindent Besides $H$ and $X$ gates, there are several other standard quantum gates, and their matrices are shown as follows:
\begin{align*}
	Z \triangleq \left[\begin{array}{cc}
		1 & 0\\
		0 & -1
	\end{array}
	\right], \quad
	S \triangleq \left[\begin{array}{cc}
		1 & 0\\
		0 & i
	\end{array}
	\right], \quad
	T \triangleq \left[\begin{array}{cc}
		1 & 0\\
		0 & e^{i\pi / 4}
	\end{array}
	\right], \quad CNOT \triangleq \left[\begin{array}{cccc}
			1 & 0 & 0 & 0\\
			0 & 1 & 0 & 0\\
			0 & 0 & 0 & 1\\
			0 & 0 & 1 & 0
		\end{array}\right] = \left[\begin{array}{cc}
		I & 0 \\
		0 & X
		\end{array}\right]
\end{align*}
\noindent Here, the $Z$, $S$, $T$ gates are applied to single qubit, while the \textit{CNOT} gate is applied to a pair of qubits. The \textit{CNOT} gate is a kind of \textit{controlled gate}, which has different actions on the second qubit according to the content of the first qubit: it applies the $X$ operation to the second qubit if the first qubit is in the state $\ket{1}$, and does nothing (equivalent to applying an identity gate $I$) if the first qubit is in the state $\ket{0}$:
\begin{equation*}
CNOT\ket{00}=\ket{00},\quad CNOT\ket{01}=\ket{01},\quad CNOT\ket{10}=\ket{11},\quad CNOT\ket{11}=\ket{10}
\end{equation*}
Quantum gates can also include classical parameters. For example, the \textit{rotation gates} $R_x$, $R_y$ and $R_z$ respectively take a specified rotation angle $\theta$:
\begin{align}
\label{equ:rxrz}
	&R_z(\theta) \triangleq \left[\begin{array}{cc}
			e^{-\frac{i\theta}{2}} & 0\\
			0 & e^{\frac{i\theta}{2}}
		\end{array}
		\right],\quad
	R_y(\theta) \triangleq \left[\begin{array}{cc}
		\cos\frac{\theta}{2} & -\sin\frac{\theta}{2}\\
		\sin\frac{\theta}{2} & \cos\frac{\theta}{2}
	\end{array}
	\right], \notag\\
    &R_x(\theta) \triangleq \left[\begin{array}{cc}
		\cos\frac{\theta}{2} & -i\sin\frac{\theta}{2}\\
		-i\sin\frac{\theta}{2} & \cos\frac{\theta}{2}
	\end{array}
	\right] = HR_z(\theta)H
\end{align}

\noindent Note that different $\theta$ represents different gate. Eq. (\ref{equ:rxrz}) $R_x(\theta) = HR_z(\theta)H$ means that an $R_x$ gate is \textit{equivalent} to sequentially applying three gates: $H$ gate, $R_z$ gate, and $H$ gate.

\vspace{2mm}
\noindent $\bullet$ \textbf{Global phase and relative phase.}\hspace*{1mm}
A quantum state $\ket{\phi}$ can be multiplied by a complex number of unit magnitude, i.e., $e^{i\theta}\ket{\phi}$, where the argument $\theta$ is the \textit{phase angle}. This state, $e^{i\theta}\ket{\phi}$, can be considered equivalent to $\ket{\phi}$ up to a \textit{global phase} $e^{i\theta}$, as this global phase cannot be distinguished by measurement. In contrast, \textit{relative phases} arise within the components of superposition states and can be distinguished by measurement. This relative phase can influence the behavior of quantum programs and is often critical in many quantum algorithms.

For example, compared to the state $a\ket{0} + b\ket{1}$, the state $a\ket{0} + be^{i\theta}\ket{1}$ exhibits an additional relative phase $e^{i\theta}$ between the components $\ket{0}$ and $\ket{1}$. In fact, the relative phase can be generated by an $R_z(\theta)$ gate (ignoring a global phase):
\begin{equation}
\label{equ:relative_phase}
R_z(\theta)\left(a\ket{0} + b\ket{1}\right)=
\left[\begin{array}{cc}
	e^{-\frac{i\theta}{2}} & 0\\
	0 & e^{\frac{i\theta}{2}}
\end{array}\right]
\left[\begin{array}{c}
	a\\
	b
\end{array}\right]=
e^{-\frac{i\theta}{2}}
\left[\begin{array}{c}
	a \\
	b e^{i\theta}
\end{array}\right]=
e^{-\frac{i\theta}{2}}\left(a\ket{0} + be^{i\theta}\ket{1}\right)
\end{equation}

\vspace{2mm}
\noindent $\bullet$ \textbf{Entanglement.}\hspace*{1mm}
An important property of qubits is that qubits can be \textit{entangled}. When two qubits are entangled, they cannot be treated as independent of each other, and measuring one qubit can interfere with the other, which makes entanglement a quintessential feature of quantum computation. The ingenious application of entanglement has led to the development of many powerful quantum algorithms~\cite{shor1999polynomial,grover1996fast}.

A typical entangled state is \textit{Bell state} $\ket{\beta}=\frac{1}{\sqrt 2}(\ket{00}+\ket{11})$, which has an entanglement between the first qubit and the second qubit. When measuring the first qubit of $\ket{\beta}$, we will obtain the result 0 with probability 1/2 and the result 1 with probability 1/2. Similarly, when measuring the second qubit of $\ket{\beta}$, we will also obtain result 0 with probability 1/2 and result 1 with probability 1/2. However, if we measure the first qubit and obtain a result of 0 (or 1), and then immediately measure the second qubit, then the result of the second measurement will always be 0 (or 1). In other words, the result of the second measurement will be determined by the result of the first measurement.

\begin{figure*}
	\centering
	\includegraphics[scale=0.65]{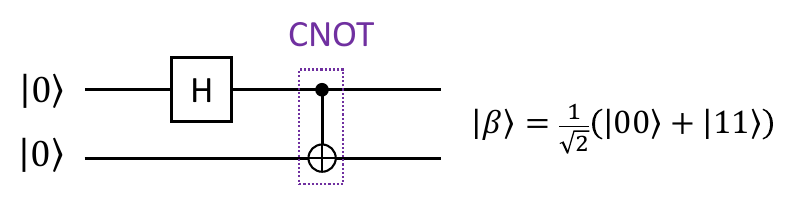}
	\caption{The quantum circuit to generate the Bell state.}
	\label{fig:Bell}
\end{figure*}

\vspace{2mm}
\noindent $\bullet$ \textbf{Quantum circuit.}\hspace*{1mm}
A quantum circuit is a widely used representation model to illustrate the quantum computation process. Each line corresponds to a qubit, and a sequence of operations, including quantum gates and measurements, is applied from left to right. Fig.~\ref{fig:Bell} shows a quantum circuit to generate Bell state$\ket{\beta}=\frac{1}{\sqrt 2}(\ket{00}+\ket{11})$ from the initial state $\ket{00}$. To generate it, we need to apply an $H$ gate on the first qubit, and then apply a \textit{CNOT} gate between these two qubits, with the first qubit as controlling and the second qubit as the target.

\subsection{Oracle Quantum Programs}
\label{subsec:oracle}

\textit{Oracle} is a common type of subroutine in many quantum computation algorithms. An oracle quantum program $\mathcal{O}$ is associated with a classical function $f$, and its behavior is determined by $f$. To some extent, oracle quantum programs serve as bridges that enable quantum computation to address classical problems. This section presents three typical types of oracle quantum programs, followed by a summary.

\subsubsection{Oracle in Grover's Algorithm: Boolean Function}
\label{subsubsec:boolfunc}

Grover's Algorithm~\cite{grover1996fast} is a typical quantum algorithm for solving the search problem in unstructured databases, which has been proven to be faster than classical search algorithms~\cite {10.5555/870802}. Consider a Boolean function $f: \{0,1\}^n \rightarrow \{0,1\}$ associated with the target search problem, which identifies the solutions of the problem.
\begin{equation*}
	f(x) = \left\{\begin{array}{ll}
		1 & \mathrm{if\:} x  \mathrm{\: is \: a \: solution} \\
		0 & \mathrm{if\:} x  \mathrm{\: is \: not \: a \: solution}
	\end{array}\right.
\end{equation*}

\noindent As a quantum algorithm, Grover's algorithm requires a quantum subroutine to evaluate the function $f$ on qubits to identify the solutions. This quantum subroutine, denoted as $\mathcal{O}_f$, takes an input $x$ represented by an $n$-qubit quantum state $\ket{x}$ based on its binary representation. There are two ways to implement the oracle. The first way introduces an additional qubit to record the result, denoted as $\mathcal{O}_f^{\mathrm{(qubit)}}$:
\begin{equation}
	\label{equ:Ofqubit}
	\ket{x}\ket{q} \xrightarrow{\mathcal{O}_f^{\mathrm{(qubit)} }} \ket{x}\ket{q \oplus f(x)}
\end{equation}
\noindent where $\ket{q}$ is a single qubit, and $\oplus$ represents the XOR operation (i.e., if $q = 0$, then $0 \oplus f(x) = f(x)$; if $q = 1$, then $1 \oplus f(x) = \overline{f(x)}$, where $\overline{f(x)}$ is the negation of $f(x)$). The use of XOR ensures that the map is unitary. The second way encodes the result into phase angles, denoted as $\mathcal{O}_f^{\mathrm{(phase)}}$:
\begin{equation}
	\label{equ:Ofphase}
	\ket{x} \xrightarrow{\mathcal{O}_f^{\mathrm{(phase)} }} (-1)^{f(x)}\ket{x}
\end{equation}

\noindent That is, if $x$ is a solution, a phase factor of $-1$ is applied to the term $\ket{x}$. These two implementation ways can be converted to each other. For example, let $\ket{q} = \frac{1}{\sqrt{2}}(\ket{0} - \ket{1})$, then
\begin{equation*}
	\ket{x}\frac{\ket{0} - \ket{1}}{\sqrt 2} \xrightarrow{\mathcal{O}_f^{\mathrm{(qubit)} }} \ket{x}\frac{\ket{f(x)} - \ket{\overline{f(x)}}}{\sqrt 2}
\end{equation*}
\noindent Note that if $f(x) = 0$, the right-hand side is $\ket{x}\frac{\ket{0} - \ket{1}}{\sqrt{2}}$; and if $f(x) = 1$, the right-hand side becomes $\ket{x}\frac{\ket{1} - \ket{0}}{\sqrt{2}} = -\ket{x}\frac{\ket{0} - \ket{1}}{\sqrt{2}}$. Thus, the mapping can also be expressed as:
\begin{equation}
	\ket{x}\frac{\ket{0} - \ket{1}}{\sqrt 2} \xrightarrow{\mathcal{O}_f^{\mathrm{(qubit)} }} (-1)^{f(x)} \ket{x}\frac{\ket{0} - \ket{1}}{\sqrt 2}
\end{equation}
\noindent It is equivalent to $\mathcal{O}_f^{\mathrm{(phase)}}$ if the unchanged second part is disregarded.

The oracle in Grover's algorithm demonstrates the fundamental structure of oracle quantum programs. An oracle quantum program associates an input quantum state $\ket{x}$ with a classical function $f$. It computes $f(x)$ and encodes the results either into the quantum phase angles or additional qubits.

\subsubsection{Quantum Arithmetic}
\label{subsubsec:qarith}

Quantum arithmetic consists of a suite of quantum algorithms designed to perform arithmetic operations on qubits~\cite{Vedral_1996,Sahin2020QAonFourier}. Unlike classical programming, which operates on "bytes" or "words," current quantum programming directly manipulates qubits~\cite{long2024testing}. As a result, basic arithmetic operations such as \textit{adders} and \textit{multipliers} need to be implemented as subroutines rather than provided as fundamental device instructions.

Generally, an arithmetic operation is associated with a function $f: \{0,1\}^m \rightarrow \{0,1\}^n$, where the input is an $m$-bit binary string (or equivalently an integer) and the output is an $n$-bit binary string. Since $m$ does not necessarily equal $n$ and $f$ is not always reversible, quantum arithmetic operations (denoted as $\mathcal{O}$) are typically implemented as follows:
\begin{equation*}
    \ket{x}\ket{0} \xrightarrow{\mathcal{O}} \ket{x}\ket{f(x)}
\end{equation*}
\noindent Here, the first part ($\ket{x}$) consists of $m$ qubits, while the second part ($\ket{0}$ and $\ket{f(x)}$—where $\ket{0}$ represents the all-zero state $\ket{0\dots0}$) consists of $n$ qubits. In this operation, $\mathcal{O}$ takes $\ket{x}$ as input, retains it, and writes the result $f(x)$ into the second part. Furthermore, the initial value of the second part can also be arbitrary, represented as $\ket{y}$ with $y \in \{0,1\}^n$, and $f(x)$ can be written into $\ket{y}$ using \textit{bitwise-XOR}:
\begin{equation}
    \label{equ:yoplusfx}
    \ket{x}\ket{y} \xrightarrow{\mathcal{O}} \ket{x}\ket{y \oplus f(x)}
\end{equation}

In this context, $\ket{x}$ can represent multiple variables. For example, consider the following \textit{triple adder}:
\begin{equation*}
    \ket{a_1}\ket{a_2}\ket{a_3}\ket{0} \xrightarrow{\texttt{TriAdder}} \ket{a_1}\ket{a_2}\ket{a_3}\ket{a_1+a_2+a_3}
\end{equation*}
\noindent Here, each $\ket{\cdot}$ consists of $n$ qubits. If we treat $\ket{a_1}\ket{a_2}\ket{a_3}$ as a single entity (i.e., let $x = a_1 a_2 a_3$, where the expression $a_1 a_2 a_3$ denotes the concatenation of the binary strings $a_1$, $a_2$, and $a_3$), the corresponding function $f$ maps $\{0,1\}^{3n} \rightarrow \{0,1\}^n$. Using this representation, Eq.~(\ref{equ:yoplusfx}) can describe multivariate and vector or matrix functions. Notably, bitwise-XOR is just one way to record the results. More generally, quantum arithmetic programs are represented as follows:
\begin{equation}
    \label{equ:QArithmetic}
    \ket{x}\ket{y} \xrightarrow{\mathcal{O}} \ket{x}\ket{\mathcal{F}(x,y)}
\end{equation}
\noindent Here, $\mathcal{F}(x,y) : \{0,1\}^m \times \{0,1\}^n \rightarrow \{0,1\}^n$ is a one-to-one mapping on $y$ for any fixed $x$, ensuring the reversibility of $\mathcal{O}$. Quantum arithmetic reduces to computing a Boolean function if $n=1$ (as in Eq.~(\ref{equ:Ofqubit})).

The implementation of quantum arithmetic programs is analogous to classical arithmetic operations. Essential operation components, such as quantum versions of "half-adders," "full-adders," and "shifters," are first prepared on qubits. Arithmetic operations are constructed using these components, akin to building an Arithmetic Logic Unit (ALU) in classical CPUs. However, unlike classical ALUs, quantum operations must be unitary and therefore reversible. Consequently, all essential components (e.g., half-adders, full-adders, and shifters) must also be reversible.


\subsubsection{Evolution of Diagonal Hamiltonian}

Hamiltonian simulation is a critical application of quantum computing~\cite{lloyd1996universal}. To simulate an $n$-qubit quantum system, the system's \textit{Hamiltonian} $H$ must be represented as a $2^n \times 2^n$ Hermitian matrix (i.e., $H^{\dagger} = H$, where $H^{\dagger}$ is the conjugate transpose of $H$). Suppose the system's initial state is $\ket{\psi_0}$. According to the Schr\"{o}dinger Equation, given an evolution time $t \in \mathbb{R}$, the final state of the system is $\ket{\psi_t} = e^{-iHt}\ket{\psi_0}$, i.e., it is equivalent to applying the unitary transform $e^{-iHt}$ on the initial state.

A significant case arises when the Hamiltonian is diagonal: $H = \mathrm{diag}\{h_0, \dots, h_{2^n-1}\}$. In this scenario, the unitary transform simplifies to $e^{-iHt} = \mathrm{diag}\{e^{-ih_0t}, \dots, e^{-ih_{2^n-1}t}\}$, and its action on a basis state is described by:
\begin{equation*}
    \ket{x} \xrightarrow{e^{-i \mathrm{diag}\{h_0, \dots, h_{2^n-1}\} t}} e^{-ih_xt}\ket{x}, \qquad x = 0, 1, \dots, 2^n-1
\end{equation*}
\noindent In this process, an additional phase factor $e^{-ih_xt}$ is applied to the component $\ket{x}$. From the perspective of oracle quantum programs, let the function $h: \{0,1\}^n \rightarrow \mathbb{R}$, where $h(x) = h_x$ ($x = 0, \dots, 2^n-1$). This transform effectively computes the function $h(x)$ on the input $\ket{x}$ and encodes the result as the corresponding phase $e^{-ih(x)t}$.

An example of diagonal Hamiltonian evolution is the \textit{Ising model}~\footnote{Strictly speaking, this Ising model is referred to as a \textit{classical Ising model} because both the spins and the external magnetic field are parallel to the z-axis. Consequently, the spins can be represented in the computational basis, and their evolution can be described by classical arithmetic operations. For a general Ising model, where the external magnetic field may point in an arbitrary direction, the evolution cannot be described by classical arithmetic operations.}, a fundamental model in statistical physics. In this model, a $\pm1$ string $\sigma$ represents a series of particles, where $-1$ and $+1$ denote two spin states. The Hamiltonian consists of two parts: the interaction energy of adjacent particles and the external magnetic field. The interaction energy is given by $H_1 = -J \sum_{<ij>} \sigma_i \sigma_j$, where the sum runs over all adjacent particle pairs (the adjacency depends on the particle arrangement). Pairs with the same spin contribute energy $-J$, while pairs with opposite spins contribute energy $J$. The effect of the external magnetic field is described by $H_2 = -B \sum_i \sigma_i$. The total Hamiltonian for the Ising model is:
\begin{equation}
    \label{equ:ising}
    H = H_1 + H_2 = -J \sum_{<ij>} \sigma_i \sigma_j - B \sum_i \sigma_i
\end{equation}
\noindent To simulate the Ising model, a qubit array $\ket{x}$ is used to represent the $\pm1$ string $\sigma$, where $\ket{0}$ corresponds to $+1$ spin and $\ket{1}$ to $-1$ spin. Specifically, let $x_i$ satisfy $\sigma_i = (-1)^{x_i}$. The evolution is then represented as:
\begin{equation}
    \ket{x} \xrightarrow{\mathrm{Ising}} e^{i\left[ J \sum_{<ij>} (-1)^{x_i}(-1)^{x_j} + B \sum_i (-1)^{x_i} \right]t}\ket{x}, \qquad x = 0, 1, \dots, 2^n-1
\end{equation}

For a general Hamiltonian $H$, note that every Hermitian matrix is diagonalizable. There exists a unitary matrix $U$ such that $H = U \mathrm{diag}\{h_0, \dots, h_{2^n-1}\} U^{-1}$, where $h_0, \dots, h_{2^n-1} \in \mathbb{R}$ are the eigenvalues of $H$. Consequently, $e^{-iHt} = U e^{-i \mathrm{diag}\{h_0, \dots, h_{2^n-1}\}t} U^{-1}$, which is equivalent to the evolution of a diagonal Hamiltonian with a pair of conjugated unitary transforms.

\subsubsection{Summary of Oracle Quantum Programs}

We have discussed three typical forms of oracle quantum programs. An oracle quantum program is fundamentally associated with one or more classical functions. It computes these functions based on qubit inputs and encodes the results into either qubits or quantum phases. As such, oracle quantum programs serve as critical bridges between quantum and classical computing.

Some newly developed quantum programming languages have started to provide explicit support for oracle quantum programs. For example, the isQ language offers mechanisms for defining quantum oracles~\cite{Guo2022isQ}. It provides two primary methods for defining quantum oracles: (1) directly specifying the value table of a function $f: \{0,1\}^m \rightarrow \{0,1\}^n$, and (2) constructing oracles using logical operations, similar to the approach used in Verilog~\cite{Verilog} for hardware design.

\subsection{Concepts in Software Testing}
\label{subsec:testing}


Software testing serves as a methodological basis for ensuring software reliability and correctness. To provide necessary background, we briefly recall several classical testing concepts that are relevant to our proposed framework, including \textit{program specification}, \textit{equivalence class partitioning}, and \textit{boundary value analysis}. These notions originate from classical software testing techniques~\cite{ammann2016introduction,myers2011art} and are adapted for use in the context of quantum software testing.

\vspace{2mm}
\noindent $\bullet$ \textbf{Software Testing.}\hspace*{1mm}
Software testing is an essential process in software engineering, which aims to determine whether software contains errors and satisfies its specifications by executing the system under specific testing procedures~\cite{ammann2016introduction}. Software testing tasks can be categorized in various ways. Depending on whether the internal structure of the program under test is known, testing can be classified as \textit{white-box testing} or \textit{black-box testing}. According to the testing level, it can be divided into \textit{unit testing}, \textit{integration testing}, and \textit{system testing}.

\vspace{2mm}
\noindent $\bullet$ \textbf{Program Specification.}\hspace*{1mm}
A program specification defines the expected behavior of a program and forms the basis of test execution~\cite{meyer1992applying,claessen2000quickcheck}. The purpose of testing is to verify whether a program produces the expected output for a given input, as defined by its specification.

\vspace{2mm}
\noindent $\bullet$ \textbf{Equivalence Class Partitioning.}\hspace*{1mm}
Equivalence class partitioning is an important testing strategy~\cite{beizer1990software,offutt1992dataflow}. It divides the input domain of a program into disjoint classes based on specific rules, and test cases are selected from each class accordingly. It assumes that test cases within the same class have a similar ability to detect potential errors in the program. There are several methods to perform equivalence class partitioning. In white-box testing, equivalence classes are typically derived from the execution paths of the program. In black-box testing, they are usually derived from the program specification of the system under test.

\vspace{2mm}
\noindent $\bullet$ \textbf{Boundary Value Analysis.}\hspace*{1mm}
Practical experience has shown that program errors are more likely to occur at or near the boundaries of input domains~\cite{myers2011art,hamlet1977testing}. Therefore, during the testing process, attention should be paid to the program's behavior when the input values are close to boundary conditions. For example, if the input domain is the interval $[a,b]$, the test inputs should include $a$, $b$, and values slightly greater than $a$ and slightly less than $b$.

\section{Formal Definitions and Research Problem}
\label{sec:question}

Oracle quantum programs are fundamental components of many quantum algorithms, bridging the gap between classical and quantum computing. Understanding their functional behavior and ensuring their correctness is critical for developing reliable quantum software. In this section, we formally define the structure and behavior of general oracle quantum programs, providing a unified representation that encompasses existing forms of these programs. Additionally, we introduce the central research problem addressed in this paper: testing oracle quantum programs in a structured and systematic way.

\subsection{Formal Definition of Oracle Quantum Programs}
\label{subsec:def}

As discussed in Section~\ref{subsec:oracle}, oracle quantum programs commonly represent the results of corresponding classical functions $g(x)$ or $h(x)$ in two ways:

\begin{enumerate}
    \item Representing the result as a phase factor:
    \begin{equation}
        \label{equ:factor}
        \ket{x} \mapsto e^{i g(x)}\ket{x}
    \end{equation}
    
    \item Overwriting the input states:
    \begin{equation}
        \label{equ:overwrite}
        \ket{x}\ket{y} \mapsto \ket{x}\ket{y \oplus h(x)}
    \end{equation}
    Here, $\oplus$ denotes the bitwise XOR operation. Note that if $y = 0$, then $0 \oplus h(x) = h(x)$.
\end{enumerate}

To generalize these two cases, we require a representation that incorporates both the phase factor (as shown in Eq.~(\ref{equ:factor})) and the ability to handle two input qubit arrays $\ket{x}$ and $\ket{y}$ (as shown in Eq.~(\ref{equ:overwrite})). Since this representation involves two input arrays, using bi-variate functions for clarity and generality is better. Consequently, we propose the following unified definition of oracle quantum programs:

\vspace{1mm}
\begin{definition}
    \label{def:oracle}
    \textit{Oracle Quantum Program}
    
    Let $\mathcal{F}(x,y)$ and $\mathcal{G}(x,y)$ be two computable classical functions:
    \begin{align*}
        & \mathcal{F} : \{0,1\}^m \times \{0,1\}^n \rightarrow \{0,1\}^n, \\
        & \mathcal{G} : \{0,1\}^m \times \{0,1\}^n \rightarrow \mathbb{R},
    \end{align*}
    where $m, n \in \mathbb{N}$ with $m \geq 0$ and $n \geq 1$. Furthermore, assume that $\mathcal{F}(x,y)$ is a one-to-one map on $y$ for any fixed $x$.
    
    Suppose that there are two input qubit arrays $\ket{x}$ and $\ket{y}$ with $m$ and $n$ qubits, respectively, where $x \in \{0,1\}^m$ and $y \in \{0,1\}^n$. The \textit{oracle quantum program} $\mathcal{P}$ corresponding to these two classical functions performs the following transform:
    \begin{equation}
        \label{equ:PS}
        \mathcal{P}: \ket{x}\ket{y} \mapsto e^{i\mathcal{G}(x,y)}\ket{x}\ket{\mathcal{F}(x,y)}
    \end{equation}
    on these qubit arrays.
\end{definition}
\vspace{1mm}

Intuitively, given input qubits $\ket{x}$ and $\ket{y}$, $\mathcal{P}$ computes the functions $\mathcal{F}(x,y)$ and $\mathcal{G}(x,y)$ simultaneously. The result of $\mathcal{F}(x,y)$ is encoded in the second qubit array as $\ket{\mathcal{F}(x,y)}$, overwriting the original $\ket{y}$. The result of $\mathcal{G}(x,y)$ is represented as an additional phase factor $e^{i\mathcal{G}(x,y)}$. 
When $m=0$, $\mathcal{F}(x,y)=y$ and $\mathcal{G}(x,y) = g(y)$, Eq.~(\ref{equ:PS}) reduces to Eq.~(\ref{equ:factor}). Similarly, when $\mathcal{G}(x,y) \equiv 0$ and $\mathcal{F}(x,y) = y \oplus h(x)$ (noting that $\oplus$ satisfies the one-to-one property), Eq.~(\ref{equ:PS}) reduces to Eq.~(\ref{equ:overwrite}).

Here we provide supplementary explanations of the definition's details. The first concerns why the definition assumes there are two inputs, $\ket{x}$ and $\ket{y}$, and that one of them, $\ket{x}$, remains unchanged during the computation. In fact, this restriction stems from the fact that all quantum transforms are unitary. Suppose we adopt the "\textit{inplace operation}" such as $\ket{x}\mapsto \ket{f(x)}$, the function $f$ must be a one-to-one map to satisfy the reversibility requirement of a unitary transform. To represent the most general transforms, we conventionally write the result into an extra qubit array, such as $\ket{x}\ket{y}\mapsto \ket{x}\ket{y \oplus f(x)}$, as Section~\ref{subsubsec:qarith} shows. The existence of the first part $\ket{x}$ ensures that although $f(x)$ is not a one-to-one map, the total map is always a one-to-one map and can be represented as a unitary transform. Furthermore, the function of the second part $y\oplus f(x)$ can be extended to a general two-variable function $\mathcal{F}(x, y)$.

Second is about why $\mathcal{F}(x,y)$ is assumed to be a one-to-one map on $y$ with any fixed $x$. This restriction also originates from the unitarity. May-to-one maps cannot be represented as a unitary transform.

Third is about why the range of function $\mathcal{G}(x,y)$ is defined as real numbers $\mathbb{R}$. This definition is based on the fact that the APIs for rotation gates (such as $R_x(\theta)$ or $R_z(\theta)$), which allow arbitrary floating-point values for the rotation angle $\theta$, are supported in most quantum software stacks. As a result, the phase factor $e^{i\theta}$ with the floating-point parameter $\theta$ is easily implemented in quantum programs by developers. We use a real number $\mathbb{R}$ to model the floating-point parameters.

Eq.~(\ref{equ:PS}) only describes the functional behavior on computational-basis input states. Using the linearity of the unitary transform, the behavior of $\mathcal{P}$ can be extended to superposition states. For instance, when the input state is a superposition of two values, such as:
\begin{equation*}
    \frac{1}{\sqrt{2}} \left(\ket{x_1}\ket{y_1} + \ket{x_2}\ket{y_2}\right),
\end{equation*}
the oracle quantum program $\mathcal{P}$ applies the transform as follows:
\begin{equation*}
    \mathcal{P}: \frac{1}{\sqrt{2}} \left(\ket{x_1}\ket{y_1} + \ket{x_2}\ket{y_2}\right) \mapsto \frac{1}{\sqrt{2}} \left( e^{i\mathcal{G}(x_1,y_1)}\ket{x_1}\ket{\mathcal{F}(x_1,y_1)} + e^{i\mathcal{G}(x_2,y_2)}\ket{x_2}\ket{\mathcal{F}(x_2,y_2)} \right).
\end{equation*}
This result illustrates how $\mathcal{P}$ processes each component of the superposition independently, applying the corresponding phase factor $e^{i\mathcal{G}(x,y)}$ and updating the second qubit array using $\mathcal{F}(x,y)$.

Previous work~\cite{abreu2022metamorphic} defines oracle quantum programs as follows:
\begin{equation}
    \label{equ:garbage}
    \mathcal{P}: \ket{x}\ket{y}\ket{0} \mapsto \ket{x}\ket{\mathcal{F}(x,y)}\ket{\alpha},
\end{equation}
where the third qubit array represents auxiliary qubits and $\ket{\alpha}$ is an unknown garbage state. It is important to note that the presence of a garbage state will eliminate the entire phase factor, as the garbage state may introduce an unknown phase. 
As a result, allowing the simultaneous presence of additional phases and garbage qubits is contradictory. Since this paper focuses on quantum phases, we adopt the definition (\ref{equ:PS}) that incorporates phases rather than the definition (\ref{equ:garbage}), which allows garbage qubits.
%

\begin{example}
\label{example:representFG}
Concrete forms of $\mathcal{F}(x,y)$ and $\mathcal{G}(x,y)$ for practical oracle quantum programs.

\vspace{1mm}
\noindent (1) Quantum programs to calculate the parity function

The parity function $\mathrm{PARITY}(x)$ is a Boolean function that returns 0 or 1 based on the number of '1's in the binary representation of the input $x$:
$$\mathrm{PARITY}(x) = \left\{ \begin{array}{ll}
    1 & \text{if there is an odd number of '1's in } x, \\
    0 & \text{if there is an even number of '1's in } x.
\end{array} \right.$$

\noindent
As discussed in Section~\ref{subsubsec:boolfunc}, quantum programs that calculate the parity function can be implemented in two forms: a phase version (P) or a qubit version (Q): 
\vspace*{2mm}

\hspace{3cm} $\texttt{Parity\_P} : $\quad$ \ket{y} \mapsto (-1)^{\mathrm{PARITY}(y)}\ket{y} = e^{i\mathrm{PARITY}(y)\cdot \pi}\ket{y}$

\hspace{3cm} $\texttt{Parity\_Q} : $\quad$ \ket{x}\ket{y} \mapsto \ket{x}\ket{y \oplus \mathrm{PARITY}(x)}$

\vspace*{2mm}
\noindent The corresponding classical functions $\mathcal{F}$ and $\mathcal{G}$ can be expressed as:

\vspace*{2mm}
\hspace{3cm} \texttt{Parity\_P} : \quad $\mathcal{F}(x,y) = y, \quad \mathcal{G}(x,y) = \mathrm{PARITY}(x) \cdot \pi$

\hspace{3cm} \texttt{Parity\_Q} : \quad $\mathcal{F}(x,y) = y \oplus \mathrm{PARITY}(x), \quad \mathcal{G}(x,y) = 0$

\vspace*{2mm}
\noindent (2) Quantum adder

Given two $n$-bit integers $x$ and $y$, represented as qubit arrays $\ket{x}$ and $\ket{y}$, the program adds $x$ to $y$ and discards the high bits if $x + y$ overflows. Formally, the transform is as follows:

\vspace*{2mm}
\hspace{3cm} $\texttt{QAdder} : $\quad$ \ket{x}\ket{y} \mapsto \ket{x}\ket{x + y \,(\mathrm{mod}\, 2^n)}$

\vspace*{2mm}
\noindent The corresponding classical functions $\mathcal{F}$ and $\mathcal{G}$ are:

\vspace*{2mm}
\hspace{3cm} $\texttt{QAdder} : \quad \mathcal{F}(x,y) = x + y \,(\mathrm{mod}\, 2^n), \quad \mathcal{G}(x,y) = 0$
\vspace*{2mm}

\end{example}

\subsection{Research Problem: Testing Oracle Quantum Programs}
\label{subsec:question}

Testing Oracle quantum programs in a structured, systematic way is a significant challenge in quantum software testing. This section defines the central research problem addressed in this paper.

\vspace{3mm}
\noindent 
\textbf{Problem Statement:} \textit{Testing Oracle Quantum Programs} 

Suppose that we have developed an oracle quantum program $\mathcal{P}$ that performs the transform described in Eq.~(\ref{equ:PS}), with an input space $U \subseteq \{0,1\}^m \times \{0,1\}^n$, where $m \geq 0$ and $n \geq 1$. Based on the classical functions $\mathcal{F}(x,y)$ and $\mathcal{G}(x,y)$, the input space $U$ can be partitioned into equivalence classes. Let this partition be represented as a family of sets $\mathcal{C} = \{C_i\}$, where each $C_i \subseteq U$ is a set of input binary strings. These equivalence classes are pairwise disjoint, i.e., for $i \neq j$, $C_i \cap C_j = \emptyset$. The research question is to systematically test the program $\mathcal{P}$ using the equivalence class partition $\mathcal{C}$, ensuring correctness and functionality with respect to $\mathcal{F}(x,y)$ and $\mathcal{G}(x,y)$.

\vspace{3mm}
\noindent $\bullet$ \textbf{Testing Assumptions.}\hspace*{0.8mm}
To address the above problem, we adopt the following assumptions and constraints in this paper:

\begin{itemize}
    \item[(i)] \textbf{Black-box testing:} The testing task is conducted as a black-box testing process, which means the internal implementation details of $\mathcal{P}$ are unknown.
    
    \item[(ii)] \textbf{Available classical subroutines:} It is assumed that classical subroutines for computing $\mathcal{F}(x,y)$ and $\mathcal{G}(x,y)$ are available before testing. These subroutines can be treated as black-box functions during testing.
    
    \item[(iii)] \textbf{Pre-given equivalence class partition:} The equivalence class partition $\mathcal{C}$ of the input space is assumed to be pre-given. This paper does not focus on the concrete construction of $\mathcal{C}$, as it is a purely classical process. Existing classical testing methods, such as boundary value analysis, can be used to perform this partitioning.
    
    \item[(iv)] \textbf{No inverse transform:} The inverse (adjoint) transform of $\mathcal{P}$ is not required. In other words, the testing framework does not depend on the availability or use of $\mathcal{P}^{-1}$.
\end{itemize}

\vspace{2mm}
\noindent $\bullet$ \textbf{Discussion and Goals.}\hspace*{0.8mm}
The above problem statement and assumptions aim to provide a structured framework for testing oracle quantum programs in practical scenarios. By leveraging classical equivalence class partitioning and black-box testing principles, we target the following objectives:

\begin{itemize}
\item Verify the correctness of $\mathcal{P}$ in the inputs in each equivalence class in $\mathcal{C}$.
\item Provide a systematic framework applicable to various oracle quantum programs.
\item Address practical constraints, such as the unavailability of internal details and adjoint transforms, to enhance the applicability of the framework.
\end{itemize}

This research problem lays the foundation for designing a comprehensive testing framework for oracle quantum programs, which will be elaborated upon in the subsequent sections.

\section{Testing Frameworks}
\label{sec:methods}

In this section, we propose a testing framework to address the problem outlined in Section~\ref{subsec:question}, i.e., to test general oracle quantum programs defined by Eq.~(\ref{equ:PS}). First, we present the theoretical foundations of the testing algorithms in Section~\ref{subsec:theoretical}. Next, we describe the construction of equivalence class pairings for superposition input states based on predefined equivalence classes in Section~\ref{subsec:EqclsPairing}. Then, we discuss the details of the execution of the testing in Sections~\ref{subsec:RunTesting} and~\ref{subsec:RepeatTimes}. Finally, we integrate these components into a comprehensive testing framework in Section~\ref{subsec:OverallFramework}.

\subsection{Theoretical Foundations}
\label{subsec:theoretical}

The purpose of the testing is to verify whether the target program $\mathcal{P}$ correctly performs the specified transform:
\begin{equation}
	\label{equ:classical}
	\ket{x}\ket{y} \xrightarrow{\mathcal{P}} e^{i\mathcal{G}(x,y)}\ket{x}\ket{\mathcal{F}(x,y)},
\end{equation}

\noindent which involves checking both the output state $\ket{\mathcal{F}(x,y)}$ and the output phase $e^{i\mathcal{G}(x,y)}$ for a given input state $\ket{x}\ket{y}$. Several approaches exist for verifying the output state, such as (a) running multiple times and analyzing the distribution of the results; (b) performing a swap test~\cite{ekert2002direct} with an expected state; (c) computing the state vectors; and (d) applying an appropriate inverse transform followed by measurement~\cite{Andriy2025QUnitTest}. Previous studies have shown both theoretically~\cite{long2024testing} and experimentally~\cite{Andriy2025QUnitTest} that approach (d) offers better performance than the others. Therefore, in this paper, we adopt approach (d) to verify output states by constructing the corresponding inverse transform.

\subsubsection{Checking the Output State $\ket{\mathcal{F}(x,y)}$}
\label{subsubsec:Fxy}

Given binary strings $x$ and $y$, $\mathcal{F}(x,y)$ can be computed classically and is therefore a known binary string (as stated in assumption (ii) in Section~\ref{subsec:question}). The classical state $\ket{\mathcal{F}(x,y)}$ can be easily prepared from the all-zero state $\ket{0}$ by applying $X$ gates to the qubits corresponding to the '1'-bits in $\mathcal{F}(x,y)$. Denoting this transform as $U_s$, we have:
\begin{equation}
	\label{equ:U}
	U_s \ket{0} = \ket{s},
\end{equation}

\noindent where $s$ represents the target binary string. The inverse transform, $U_s^{-1}\ket{s} = \ket{0}$, recovers state $\ket{s}$ back to $\ket{0}$. Since $X^{-1} = X$, it follows that $U_s^{-1} = U_s$. If the target state is $\ket{\mathcal{F}(x,y)}$, applying $U_{\mathcal{F}(x,y)}$ to it will return it to $\ket{0}$.

Measuring state $\ket{0}$, we will always obtain the classical result 0. Conversely, if the target state is not $\ket{\mathcal{F}(x,y)}$, applying the transform $U_{\mathcal{F}(x,y)}$ will not recover it to $\ket{0}$, and a non-zero measurement result may occur. A non-zero result indicates a bug in the program. However, it is possible for a wrong program to yield a zero result, leading to a misjudgment. To address this, the process may be repeated multiple times, and we conclude that the program is correct only if all measurements yield zero. The probability of misjudging a wrong program decreases exponentially with repeated trials.

This procedure can be further simplified. Note that directly measuring the expected output state $\ket{\mathcal{F}(x,y)}$ always yields the binary string $\mathcal{F}(x,y)$. Therefore, we can omit the inverse transform and directly measure the output state: if any measurement result differs from $\mathcal{F}(x,y)$, some bugs exist in the program. In fact, the following proposition shows that the "directly measure" (DM) procedure is equivalent to the "inverse and measure" (IM) procedure for checking computational-basis states.

\begin{proposition}
\label{prop:directmeasure}
Let $\ket{s}$ denote a computational-basis quantum state with binary string $s$, and let $U_s$ be the procedure that generates $\ket{s}$ from $\ket{0}$. Suppose $\ket{\psi}$ is the output state to be checked, and we want to verify whether $\ket{\psi} = \ket{s}$. The DM approach is to directly measure the state and check whether the result equals the binary string $s$:
\begin{equation*}
\ket{\psi}  \xrightarrow{M}
\begin{cases}
	=s & \text{(with probability $p_1$)}, \\
	\neq s & \text{(with probability $p_2$)}.
\end{cases}
\end{equation*}
\noindent The IM approach is to apply the inverse transform $U_s$, then measure and check whether the result is $0$:
\begin{equation*}
\ket{\psi}  \xrightarrow{U_s} \xrightarrow{M}
\begin{cases}
	=0 & \text{(with probability $p_3$)}, \\
	\neq 0 & \text{(with probability $p_4$)}.
\end{cases}
\end{equation*}
The conclusion is that these two checking procedures are equivalent; that is, the corresponding probabilities satisfy $p_1 = p_3$ and $p_2 = p_4$.
\end{proposition}

The proof of Proposition~\ref{prop:directmeasure} is provided in Appendix~\ref{subappd:directmeasure}. However, it should be noted that this equivalence holds only for checking computational-basis states. As discussed in the following section, the DM and IM procedures are not equivalent when distinguishing relative phases. Furthermore, there may be differences in execution efficiency (running time) between these two procedures. We experimentally evaluate their execution efficiency in Section~\ref{sec:evaluation}.

\subsubsection{Checking the Output Phase $e^{i\mathcal{G}(x,y)}$}
\label{subsubsec:eiGxy}

Since the global phase cannot be distinguished by measurement~\cite{nielsen2002quantum}, the output phase can only be evaluated by the relative phase. Consider a superposition input state involving two sets of variables, $(x_1, y_1)$ and $(x_2, y_2)$: 
$\frac{1}{\sqrt{2}}(\ket{x_1}\ket{y_1} + \ket{x_2}\ket{y_2})$.
When $\mathcal{P}$ is applied, this state transforms as follows:
\begin{align}
	\label{equ:super}
	\frac{1}{\sqrt{2}} \left(\ket{x_1}\ket{y_1} + \ket{x_2}\ket{y_2}\right) &\xrightarrow{\mathcal{P}} \frac{1}{\sqrt{2}} \left\{ e^{i\mathcal{G}(x_1,y_1)}\ket{x_1}\ket{\mathcal{F}(x_1,y_1)} + e^{i\mathcal{G}(x_2,y_2)}\ket{x_2}\ket{\mathcal{F}(x_2,y_2)} \right\} \notag \\
	&= \frac{1}{\sqrt{2}} e^{i\mathcal{G}(x_1,y_1)} \left\{ \ket{x_1}\ket{\mathcal{F}(x_1,y_1)} + e^{i[\mathcal{G}(x_2,y_2) - \mathcal{G}(x_1,y_1)]}\ket{x_2}\ket{\mathcal{F}(x_2,y_2)} \right\}.
\end{align}

\noindent Ignoring the global phase, the output state takes the form:
$\frac{1}{\sqrt{2}}(\ket{s_1} + e^{i\theta}\ket{s_2})$, where $\ket{s_1} = \ket{x_1}\ket{\mathcal{F}(x_1,y_1)}$, $\ket{s_2} = \ket{x_2}\ket{\mathcal{F}(x_2,y_2)}$, and $\theta = \mathcal{G}(x_2,y_2) - \mathcal{G}(x_1,y_1)$. This is referred to as a \textit{two-value superposition state}.
This state can also be generated by applying a proper unitary transform $V_{s_1, s_2, \theta}$ on the all-zero state:

\begin{equation}
	\label{equ:V}
	V_{s_1,s_2,\theta}\ket{0} = \frac{1}{\sqrt{2}}\left( \ket{s_1} + e^{i\theta}\ket{s_2} \right).
\end{equation}

\noindent
Similarly, applying $V^{-1}_{x_1\mathcal{F}(x_1,y_1), x_2\mathcal{F}(x_2,y_2), \mathcal{G}(x_2,y_2) - \mathcal{G}(x_1,y_1)}$ will return the target state to the all-zero state, and a measurement will yield a result of 0. If a non-zero measurement result occurs, it indicates a bug in the program. As in $\ket{\mathcal{F}(x,y)}$, repeating this process can reduce the probability of misjudgment.

Note that if we measure the state in Eq.~(\ref{equ:V}), the result will always be $s_1$ with probability 1/2 and $s_2$ with probability 1/2, which is irrelevant to the phase $e^{i\theta}$. As a result, unlike the checking of the output state $\ket{\mathcal{F}(x,y)}$, the DM procedure cannot be used to check the output phase $e^{i\mathcal{G}(x,y)}$. Therefore, we must adopt the IM procedure to verify the output phase $e^{i\mathcal{G}(x,y)}$. In other words, the DM and IM procedures are not equivalent for this purpose.

\subsubsection{Key Questions for the Testing Framework}
\label{subsubsec:keyquestion}

Based on the above discussion, the testing framework involves verifying the output state $\ket{\mathcal{F}(x,y)}$ using the computational basis input $\ket{x}\ket{y}$ and checking the phase $e^{i[\mathcal{G}(x_2,y_2) - \mathcal{G}(x_1,y_1)]}$ using a two-value superposition input $\frac{1}{\sqrt{2}}(\ket{x_1}\ket{y_1} + \ket{x_2}\ket{y_2})$. To achieve this, the following questions must be addressed:

\begin{itemize}
	\item[(1)] Which quantum input states should be selected?
	\item[(2)] Which steps should be executed in the testing process?
	\item[(3)] To limit the probability of misjudgment, how many repetitions should be performed?
\end{itemize}

The first question (1) will be addressed in Section~\ref{subsec:EqclsPairing}, the second (2) in Section~\ref{subsec:RunTesting}, and the third (3) through theoretical analysis in Section~\ref{subsec:RepeatTimes}.

\subsection{Equivalence Classes for Quantum Input}
\label{subsec:EqclsPairing}

Equivalence class partitioning is a fundamental concept in program testing~\cite{ammann2016introduction}. For classical programs, the input space is first partitioned into several equivalence classes. A representative input is selected for each class to form input-output pairs, which are then used to verify whether the program produces the expected output under each input.
Similarly, testing quantum programs also requires equivalence classes for quantum inputs. As described in assumption (iii) in Section~\ref{subsec:question}, we assume that the equivalence class partition $\mathcal{C} = \{C_i\}$ is given for the classical functions $\mathcal{F}(x, y)$ and $\mathcal{G}(x, y)$. The testing framework should construct proper quantum equivalence classes $\mathcal{Q}$ for quantum inputs based on equivalence classes $\mathcal{C}$.

\subsubsection{Types of Input States}

The direct construction of quantum equivalence classes involves converting binary strings in each $C_i$ into computational-basis states. Specifically, the equivalence class $Q_i$ corresponding to $C_i$ is defined as:
\begin{equation}
	Q_i = \left\{ \ket{x}\ket{y} \mid xy \in C_i \subseteq \{0,1\}^m \times \{0,1\}^n \right\},
\end{equation}
\noindent where $xy$ denotes the concatenation of two binary strings $x \in \{0,1\}^m$ and $y \in \{0,1\}^n$. However, these computational-basis states $\{Q_i\}$ can only check the qubits $\ket{\mathcal{F}(x,y)}$. As discussed in Section~\ref{subsec:theoretical}, to verify the phase factor $e^{i\mathcal{G}(x, y)}$, it is necessary to check two-value superposition states.

To construct equivalence classes for two-value superposition states, we select two distinct binary strings, $s_1$ and $s_2$, from $\{C_i\}$ to form the input state $\frac{1}{\sqrt{2}}(\ket{s_1} + \ket{s_2})$. Two cases arise:

\begin{itemize}
\item[(1)] Both values are selected from the same $C_i$ (if $|C_i| \geq 2$). In this case, the equivalence class $S_{ii}$ for two-value superposition states is:
\begin{equation}
	S_{ii} = \left\{ \begin{array}{cl}
		\left\{ \frac{1}{\sqrt{2}}\left(\ket{s_1} + \ket{s_2}\right) \mid s_1, s_2 \in C_i, s_1 \neq s_2 \right\} & \quad \mathrm{if} \: |C_i| \geq 2, \\
		\emptyset & \quad \mathrm{if} \: |C_i| \leq 1.
	\end{array} \right.
\end{equation}

\item[(2)] The two values are selected from different equivalence classes, $C_i$ and $C_j$ ($i \neq j$). The equivalence class $S_{ij}$ is defined as
\begin{equation}
	S_{ij} = \left\{ \frac{1}{\sqrt{2}}\left(\ket{s_1} + \ket{s_2}\right) \mid s_1 \in C_i, s_2 \in C_j \right\}, \qquad i \neq j.
\end{equation}
\end{itemize}

The complete equivalence classes for quantum inputs are the union of computational-basis states $\{Q_i\}$ and two-value superposition states $\{S_{ij}\}$:
\begin{equation}
	\mathcal{Q} = \{Q_i\} \cup \{S_{ij}\}.
\end{equation}
\noindent
We summarize this discussion in the following testing criterion:

\begin{criterion}
	\label{crit:types}
	Types of Quantum Input States for Testing Oracle Quantum Programs

	Given a set of equivalence class partitions $\{C_i\}$ for functions $\mathcal{F}(x, y)$ and $\mathcal{G}(x, y)$ associated with the target oracle quantum program, the corresponding equivalence classes for quantum inputs should include the following three types of input states:
	\begin{itemize}
		\item[(1)] Computational-basis states $Q_i$ corresponding to each $C_i$.
		\item[(2)] Two-value superposition states $S_{ii}$ constructed from pairs of values in the same $C_i$ (if $|C_i| \geq 2$).
		\item[(3)] Two-value superposition states $S_{ij}$ constructed from values in different $C_i$ and $C_j$ ($i \neq j$).
	\end{itemize}
\end{criterion}

Previous work~\cite{long2024testing} highlighted the importance of covering both computational basis and superposition states in testing quantum programs. The Criterion~\ref{crit:types} details the selection of superposition states for oracle quantum programs.

\subsubsection{Pairing of Different $C_i$}

Suppose that there are $k$ equivalence classes for $\mathcal{F}(x, y)$ and $\mathcal{G}(x, y)$, i.e., $|\mathcal{C}| = k$. Each class $C_i$ derives a set $Q_i$ and a set $S_{ii}$ (if $|C_i|\geq 2$), while $S_{ij} (i \ne j) $ is derived from pairs of classes $(C_i, C_j)$. There are $\frac{k(k-1)}{2}$ possible pairs $(i, j)$, leading to a quadratic increase in the number of $S_{ij}$ as $k$ increases. To reduce the number of pairs, we can select a subset of all pairs.

\begin{figure}
	\centering
	\subfigure[All-Coverage Pairing]{\includegraphics[scale=0.6]{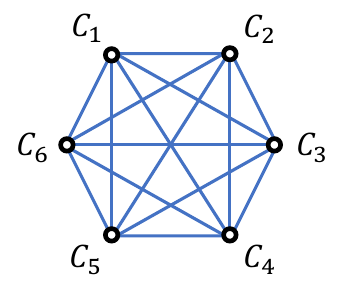}}
	\qquad
	\subfigure[Tree-Coverage Pairing]{\includegraphics[scale=0.6]{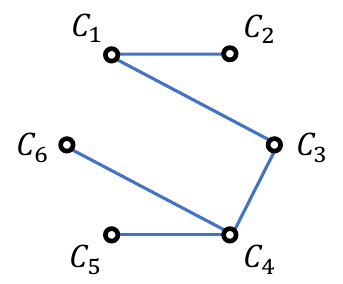}}
	\qquad
	\subfigure[Each-Choice Pairing]{\includegraphics[scale=0.6]{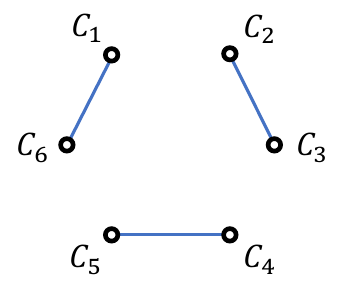}}
	\caption{Pairing graphs illustrating the all-coverage, tree-coverage, and each-choice pairing criteria.}
	\label{fig:pairing}
\end{figure}

We use a \textit{pairing graph} to represent the selection process, where each vertex corresponds to a class $C_i$. If the pair $(C_i, C_j)$ is selected, an edge is drawn connecting the two vertices $C_i$ and $C_j$. 
On the one hand, selecting all possible pairs corresponds to constructing a complete graph, as illustrated in Fig.~\ref{fig:pairing}(a). This approach represents the \textit{maximum pairing}, with $\frac{k(k-1)}{2} = O(k^2)$ pairs. 
On the other hand, the \textit{minimum pairing} ensures that each vertex is included in at least one pair. This is equivalent to constructing a graph where every vertex is connected to at least one other vertex and no vertex is isolated, as shown in Fig.~\ref{fig:pairing}(c). In this case, the number of pairs is at least $\lceil\frac{k}{2}\rceil = \Omega(k)$. 
We summarize these two extreme cases with the following testing criteria:

\begin{criterion}
	\label{crit:max}
	All-Coverage Pairing (Maximum)

	Select the pairs $\{(C_i, C_j)\}$ such that they form the edges of a complete graph over the vertex set $\{C_i\}$.
\end{criterion}

\begin{criterion}
	\label{crit:min}
	Each-Choice Pairing (Minimum)

	Select the pairs $\{(C_i, C_j)\}$ such that each $C_i$ is included in at least one pair.
\end{criterion}

In addition to these two extreme cases, intermediate criteria can also be employed. An efficient criterion should involve a linear number of pairs, i.e., $\Theta(k)$. For example, we can select pairs that form a tree in the pairing graph, as described in Fig.~\ref{fig:pairing}(b), and define it as follows:

\begin{criterion}
	\label{crit:tree}
	Tree-Coverage Pairing

	Select the pairs $\{(C_i, C_j)\}$ such that they form the edges of a tree over the vertex set $\{C_i\}$.
\end{criterion}

\noindent The rationale behind adopting this criterion is to ensure that the pairing graph remains connected. In a connected graph, each edge has at least one neighboring edge, enabling better propagation of testing information. For example, if a bug is detected in a superposition input state corresponding to a pair $(C_i, C_j)$, the testing results on neighboring edges—those involving $C_i$ or $C_j$—can provide additional insight into the location or nature of the bug. As the minimally connected graph on the $k$ vertices, a tree contains $k-1$ edges, which makes the number of pairs $\Theta(k)$.
In Section~\ref{sec:evaluation}, we will experimentally evaluate the effectiveness of these three pairing criteria.

\subsection{Testing Execution and Verification}
\label{subsec:RunTesting}

Once the equivalence classes for the quantum input states are established, the next step is to prepare the input states, run the target program, and verify the corresponding output states, similarly to testing classical programs.

\subsubsection{Preparation of Quantum Input States} 
To prepare quantum input states, we should implement the transform $U_s$, $V_{s_1, s_2, \theta}$, as described in Section~\ref{subsec:theoretical}. Previous work~\cite{long2024testing} has detailed the state generation process for computational-basis states $\ket{x}$ and two-value superposition states $\frac{1}{\sqrt{2}}(\ket{x} + e^{i\theta}\ket{y})$, which implement the transform $U_s$ and $V_{s_1, s_2, \theta}$. These algorithms can be adapted for this context. While testing for oracle quantum programs highlights the importance of two-value superposition states, the algorithms themselves are not the contributions of this paper. Thus, the implementation details of transform $U_s$ and $V_{s_1, s_2, \theta}$ are shown in Appendix~\ref{appd:inputstate}.

In the context of testing oracle quantum programs, the input consists of two binary strings, $x$ and $y$. These can be concatenated into a single binary string $s=xy$ to facilitate state generation algorithms, as shown below:
\begin{align*}
& \ket{0}\ket{0} \xrightarrow{U_{xy}} \ket{x}\ket{y}, \\
& \ket{0}\ket{0} \xrightarrow{V_{x_1 y_1, x_2 y_2, 0}} \frac{1}{\sqrt{2}}\left( \ket{x_1}\ket{y_1} + \ket{x_2}\ket{y_2} \right).
\end{align*}

\subsubsection{Run the Target Program}
Having the input states, we can run the target program on them, as shown in Eqs.~(\ref{equ:classical}) and (\ref{equ:super}), restated below for clarity:
\begin{align*}
	& \ket{x}\ket{y} \xrightarrow{\mathcal{P}} e^{i\mathcal{G}(x,y)}\ket{x}\ket{\mathcal{F}(x,y)},\\
	& \frac{1}{\sqrt{2}} \left(\ket{x_1}\ket{y_1} + \ket{x_2}\ket{y_2}\right) \xrightarrow{\mathcal{P}} \frac{1}{\sqrt{2}} e^{i\mathcal{G}(x_1,y_1)} \left\{ \ket{x_1}\ket{\mathcal{F}(x_1,y_1)} + e^{i[\mathcal{G}(x_2,y_2) - \mathcal{G}(x_1,y_1)]}\ket{x_2}\ket{\mathcal{F}(x_2,y_2)} \right\}.
\end{align*}

\noindent
Note that in the first case, the global phase $e^{i\mathcal{G}(x_1,y_1)}$ cannot be observed through measurement. Therefore, in the second case, the relative phase $e^{i[\mathcal{G}(x_2,y_2) - \mathcal{G}(x_1,y_1)]}$ is used to verify the correctness of the computed $\mathcal{G}(x,y)$.


\subsubsection{Uncomputing Output States}
According to the discussion in Section~\ref{subsubsec:Fxy}, to verify the correctness of the computational-basis state given in Eq.~(\ref{equ:classical}), we can adopt either the "directly measure" (DM) or the "inverse and measure" (IM) procedure:
\begin{align}
	&  \text{Directly measure:\qquad} \ket{x}\ket{\mathcal{F}(x, y)} \xrightarrow{M} x, \mathcal{F}(x,y) ?  \notag \\
	& \text{Inverse and measure:\quad} \ket{x}\ket{\mathcal{F}(x, y)} \xrightarrow{U_{x\mathcal{F}(x, y)}} \ket{0}\ket{0}? \xrightarrow{M} 0? \label{equ:uncompute_cb}
\end{align}

\noindent If we adopt the IM procedure, the inverse transform $U_{x\mathcal{F}(x,y)}$ should be applied, followed by a measurement to check whether the result is $0$. If we adopt the DM procedure, we simply measure the state and check whether the result equals $x$ concatenated with $\mathcal{F}(x,y)$.

According to the discussion in Section~\ref{subsubsec:eiGxy}, to verify the correctness of the two-value superposition state given in Eq.~(\ref{equ:super}), we need to apply the inverse transform $V_{x_1\mathcal{F}(x_1,y_1),\, x_2\mathcal{F}(x_2,y_2),\, \mathcal{G}(x_2,y_2) - \mathcal{G}(x_1,y_1)}^{-1}$ and then measure the resulting state.
\begin{align}
	\frac{1}{\sqrt{2}} \left\{ \ket{x_1}\ket{\mathcal{F}(x_1, y_1)} + e^{i[\mathcal{G}(x_2, y_2) - \mathcal{G}(x_1, y_1)]}\ket{x_2}\ket{\mathcal{F}(x_2, y_2)} \right\} \notag \\ \xrightarrow{V^{-1}_{x_1\mathcal{F}(x_1, y_1), x_2\mathcal{F}(x_2, y_2), \mathcal{G}(x_2, y_2) - \mathcal{G}(x_1, y_1)}} \ket{0}\ket{0}? \xrightarrow{M} 0? \label{equ:uncompute_tv}
\end{align}

\noindent If the output state matches the expected one, the measurement always yields 0. Any nonzero result indicates a deviation from the expected state and suggests a potential bug.

Note that Eqs.~(\ref{equ:uncompute_cb}) and~(\ref{equ:uncompute_tv}) share the same pattern: applying an appropriate inverse transform maps the state to an all-zero state, after which a subsequent measurement yields a result of $0$. Therefore, we refer to this process as \textit{uncomputing} the output states.


%


\subsubsection{The PRUM Process}
\begin{figure*}
	\centering
	\includegraphics[scale=0.75]{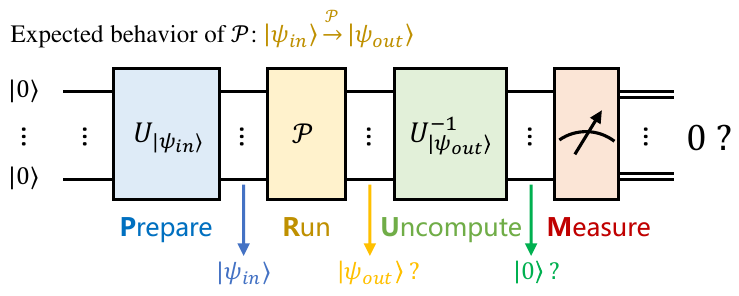}
	\caption{The general pattern of the PRUM process.}
	\label{fig:PRUM}
\end{figure*}

Proposition~\ref{prop:directmeasure} shows the equivalence between the DM and IM procedures. Therefore, in the following discussion, we always adopt the IM procedure. In this way, the testing process for both computational-basis states and two-value superposition states can be represented as four steps: (1) \textbf{P}reparing input states, (2) \textbf{R}unning the target program, (3) \textbf{U}ncomputing the corresponding output states, and (4) \textbf{M}easuring the states. We refer to this process as the \textit{Prepare-Run-Uncompute-Measure (PRUM)} process. Figure~\ref{fig:PRUM} illustrates the overall structure of the PRUM process.

Suppose we plan to test a target quantum program $\mathcal{P}$ with the input state $\ket{\psi_{in}}$, and the expected output state is $\ket{\psi_{out}}$. In the "Prepare" step, a unitary transform $U_{\ket{\psi_{in}}}$ is applied to an all-zero initial state to prepare the input state $\ket{\psi_{in}}$. In the "Uncompute" step, a unitary transform $U_{\ket{\psi_{out}}}^{-1}$ is applied to uncompute the expected output state $\ket{\psi_{out}}$ back to the all-zero state. If the actual output of the target program matches the expected output, the uncompute step restores the state to the all-zero state, and the measurement in the "Measure" step will always yield result 0. Thus, the occurrence of a nonzero result indicates that the output of the target program does not match the expected result.

However, as discussed in Section~\ref{subsec:theoretical}, a zero measurement result does not guarantee the absence of bugs; thus, the PRUM process must be repeated several times to minimize the risk of misjudgment. We conclude that the target program is correct only when all measurement results are zero. The following section will discuss the determination of repetition times.

\subsection{Number of Repetitions for the PRUM Process}
\label{subsec:RepeatTimes}

As discussed earlier, repeating the PRUM process multiple times is crucial to reducing the probability of misjudgment. The input states for the testing are categorized into computational base states and two-value superposition states. This section presents a theoretical analysis of the probability of misjudgment and the required repetition times for the two input types. Let $N_{cb}$ and $N_{tv}$ denote the repetition times for the computational-basis input states and the input states of the two-value superposition, respectively, and let $\alpha$ ($0 < \alpha < 1$) represent an acceptable probability of misjudgment. The question is how to choose $N_{cb}$ and $N_{tv}$ such that the probability of misjudgment does not exceed $\alpha$. This section provides a theoretical analysis of this question.

\subsubsection{$N_{cb}$: Number of Repetitions for Computational-Basis Input States}
\label{subsubsec:Ncb}

To derive a concrete bound for $N_{cb}$, some additional information is required. According to Eq.~(\ref{equ:uncompute_cb}), the PRUM process always reverts the expected output state $\ket{x}\ket{\mathcal{F}(x,y)}$ to the all-zero state, ensuring that the measurement result is always $0$. Conversely, if the output state deviates from the expected one, the PRUM process may yield both zero and nonzero results. When a nonzero result occurs, the unexpected output state is detected. Intuitively, the more the output state deviates from the expected one, the higher the probability that a nonzero result will occur. We need a parameter to quantitatively describe this deviation.

Here we use the orthogonal decomposition, i.e., we decompose the output state $\ket{\psi_{\mathrm{output}}}$ into a linear combination of the expected state $\ket{\psi_{\mathrm{expected}}}$ and its orthogonal complement:
\begin{equation}
\label{equ:output_decomp}
	\ket{\psi_{\mathrm{output}}} = a\ket{\psi_{\mathrm{expected}}} + b\ket{\psi_{\mathrm{expected}}^{\bot}},
\end{equation}
\noindent where the complex number $a$ is the amplitude of the expected component, satisfying $0 \le |a| \le 1$ and $|a|^2 + |b|^2 = 1$. In fact, $a$ is just the \textit{inner product} between the output state and the expected state,\footnote{$\left< \psi_{\mathrm{expected}} \big| \psi_{\mathrm{output}} \right> = a\left< \psi_{\mathrm{expected}} \big| \psi_{\mathrm{expected}} \right> + b\left< \psi_{\mathrm{expected}} \Big| \psi_{\mathrm{expected}}^{\bot} \right> = a\cdot 1 + b \cdot 0 = a$.} and therefore, it can be used to describe the deviation between the erroneous output state and the expected one: the larger $|a|$ is, the closer the output state is to the expected state. Given $a$, which represents the degree of deviation of the output state $\ket{\psi_{\mathrm{output}}}$ to be distinguished, our goal is to determine $N_{cb}$ such that the probability of misjudgment does not exceed $\alpha$.

The following proposition provides the criterion for selecting $N_{cb}$:

\begin{proposition}
\label{prop:Ncb}
For the repetition times for computational-basis input states, if we choose  $N_{cb}$ satisfying:
\begin{equation}
	\label{equ:Ncb}
	N_{cb} \geq \frac{\ln \alpha}{\ln |a|^2},
\end{equation}
\noindent where $0 < \alpha < 1$ and $0 < a \leq 1$. Then the probability of misjudgment will not be over $\alpha$.
\end{proposition}

The proof of \ref{prop:Ncb} is provided in Appendix~\ref{subappd:Ncb}. Note that $\ln \alpha < 0$, so $\frac{\ln \alpha}{\ln |a|^2}$ is monotonically increasing with respect to $|a|$. This indicates that the closer the erroneous output state to be distinguished is to the correct state, the more repetitions of the PRUM process are required to avoid incorrect discrimination. The following example illustrates several typical cases.

\begin{example}
\label{example:Ncb}
Several typical cases are discussed below:
\begin{itemize}
	\item[(1)] $a = 0$, and thus $|b| = 1$:
	
	In this case, the unexpected output state contains no component of the expected state. Therefore, misjudgment is impossible even if the PRUM process is executed only once. A typical example is when the unexpected output state is a computational-basis state associated with a wrong integer. Let $N_{cb} = 1$ be sufficient to detect the error with certainty.

	\item[(2)] $|a|^2 = |b|^2 = \frac{1}{2}$ and $\alpha = 0.01$:
	\begin{equation*}
		N_{cb} \geq \frac{\ln 0.01}{\ln 0.5} \approx 6.6.
	\end{equation*}
	This case represents an output state that corresponds to half of the expected state. We can see that only a few repetitions are required to ensure that the probability of misjudgment remains below the desired threshold.

    \item[(3)] $|a|^2=\frac{1}{4},|b|^2=\frac{3}{4}$ and $\alpha=0.01$ \quad $\Rightarrow$ \quad $N_{cb} \geq \frac{\ln 0.01}{\ln 0.25} \approx 3.3$.
    \item[(4)] $|a|^2=\frac{3}{4},|b|^2=\frac{1}{4}$ and $\alpha=0.01$ \quad $\Rightarrow$ \quad $N_{cb} \geq \frac{\ln 0.01}{\ln 0.75} \approx 16$.
\end{itemize}
\end{example}

Note that the above analysis considers only a single input state. However, in practice, there may exist multiple equivalence classes of computational-basis input states. Sampling across these classes can potentially reduce the number of repeated PRUM executions required for a single input. This issue will be further examined in the experimental evaluation.

\subsubsection{$N_{tv}$: Number of Repetitions for Two-Value Superposition Input States}
\label{subsubsec:Ntv}

Similar to the parameter $a$ used for $N_{cb}$, to derive a concrete bound for $N_{tv}$, we also need an additional parameter to quantitatively describe the deviation of the output state from the expected one.
As discussed in Section~\ref{subsec:theoretical}, two-value superposition input states are used to verify the phase factor $e^{i\mathcal{G}(x,y)}$. Here, we assume that the target program correctly handles qubit arrays, i.e., the qubit array $\ket{\mathcal{F}(x,y)}$ is always correct, and errors occur only in the phase factor $e^{i\mathcal{G}(x,y)}$. This assumption is reasonable because errors in $\ket{\mathcal{F}(x,y)}$ can be detected using computational-basis input states, allowing two-value superposition states to focus exclusively on phase errors in $e^{i\mathcal{G}(x,y)}$.

Based on the above assumption, the phase-angle difference $\Delta\theta$ can be used to characterize the deviation between the erroneous output state and the expected state.
\begin{equation*}
\Delta\theta = \left| \mathcal{G}_{\mathrm{output}}(x,y) - \mathcal{G}_{\mathrm{expected}}(x,y) \right|.
\end{equation*}
\noindent Here, $\mathcal{G}_{\mathrm{expected}}(x,y)$ denotes the expected phase angle, and $\mathcal{G}_{\mathrm{output}}(x,y)$ denotes the actual output phase angle. Obviously, the smaller the $\Delta\theta$ is, the closer the output state is to the expected one. Given $\Delta\theta$, which characterizes the phase deviation of the output state to be distinguished, our goal is to determine $N_{tv}$ such that the probability of misjudgment does not exceed $\alpha$.

The following proposition provides the criterion for selecting $N_{tv}$:

\begin{proposition}
\label{prop:Ntv}
For the selection of the number of repetitions for two-value superposition states, suppose we aim to distinguish a phase-angle difference $\Delta\theta$. If we choose $N_{tv}$ such that
\begin{equation}
	\label{equ:Ntv}
	N_{tv} \geq \frac{\ln \alpha}{\ln \left(\cos^2\frac{\Delta\theta}{2}\right)},
\end{equation}
\noindent then the probability of misjudgment will not exceed $\alpha$. Furthermore, if $N_{tv}$ is chosen based on the lower bound in (\ref{equ:Ntv}), we have the following asymptotic relation:
\begin{equation}
	\label{equ:NtvAsy}
	N_{tv} = O\left(\frac{\log\left(1/\alpha\right)}{(\Delta\theta)^2}\right).
\end{equation}
\end{proposition}

The proof of Proposition~\ref{prop:Ntv} is provided in Appendix~\ref{subappd:Ntv}.
 Eq. (\ref{equ:NtvAsy}) indicates that repetition times are inversely proportional to the square of the distinguishable phase angle difference $\Delta\theta$. To ensure the ability to distinguish all possible values of $\mathcal{G}(x,y)$, the setting of $\Delta\theta$ should not be greater than the minimum non-zero phase difference between two distinct function values of $\mathcal{G}(x,y)$, i.e.,
\begin{equation}
	\label{equ:minDTheta}
	\Delta\theta \leq \min_{ \mathcal{G}(x_1,y_1) \neq \mathcal{G}(x_2,y_2) } \left\{ |\mathcal{G}(x_1,y_1) - \mathcal{G}(x_2,y_2)| \right\}.
\end{equation}

\vspace*{.5mm}
\begin{example}
\label{example:Ntv}
We discuss two typical cases to illustrate the repetition times required for two-value superposition input states.

\begin{itemize}
	\item[(1)] Oracle quantum program $\mathcal{O}_f^{\mathrm{(phase)}}$:

	This program calculates a Boolean function on the phase (refer to Eq.~(\ref{equ:Ofphase})). In this case, the output phase factor is either $1 = e^{i \cdot 0}$ or $-1 = e^{i\pi}$, resulting in $\Delta\theta = \pi$. When applying $R_x(\pi)$, the transform is $R_x(\pi)\ket{0} = -i\ket{1}$, meaning the measurement result will never be 0. Thus, executing the PRUM process once suffices to distinguish the phase factor $1$ from $-1$ with certainty.

	\item[(2)] Phase angle difference $\Delta\theta = \frac{\pi}{2}, \frac{\pi}{4}, \frac{\pi}{8}, \frac{\pi}{16}, \frac{\pi}{32}$ with $\alpha = 0.01$:
	\begin{align*}
		&\frac{\pi}{2}: N_{tv} \geq \frac{\ln 0.01}{\ln(\cos^2 (\pi/4))} \approx 7; &&\frac{\pi}{4}: N_{tv} \geq \frac{\ln 0.01}{\ln(\cos^2 (\pi/8))} \approx 30; \\
        &\frac{\pi}{8}: N_{tv} \geq \frac{\ln 0.01}{\ln(\cos^2 (\pi/16))} \approx 119; && \frac{\pi}{16}: N_{tv} \geq \frac{\ln 0.01}{\ln(\cos^2 (\pi/32))} \approx 477; \\
        &\frac{\pi}{32}: N_{tv} \geq \frac{\ln 0.01}{\ln(\cos^2 (\pi/64))} \approx 1910.
	\end{align*}
	This result shows that reliably detecting a phase difference of a small fraction of $\pi$ requires hundreds to thousands of repetitions.

\end{itemize}
\end{example}

\subsubsection{Summary of the number of repetitions}
The number of repetitions required in the PRUM process depends on the deviation that must be distinguished between the erroneous state and the expected state, as well as the allowable probability of misjudgment. Here we summarize the used symbols as follows:
\begin{itemize}
    \item[$\alpha$ :] the allowable probability of misjudgment.
    \item[$a$ :] the amplitude of the component of the expected state within the erroneous state, which characterizes the deviation in computational-basis output states.
    \item[$N_{cb}$:] the number of repetitions for computational-basis input states.
    \item[$\Delta\theta$:] the phase-angle difference between the erroneous output state and the expected output state, which characterizes the deviation in two-value superposition output states.
    \item[$N_{tv}$:] the number of repetitions for two-value superposition input states.
\end{itemize}

The conclusion is that if we select the numbers of repetitions satisfying:
\begin{equation*}
N_{cb} \geq \frac{\ln \alpha}{\ln |a|^2}, \qquad N_{tv} \geq \frac{\ln \alpha}{\ln \left(\cos^2\frac{\Delta\theta}{2}\right)},
\end{equation*}
\noindent then the probability of misjudgment will not exceed $\alpha$.

\subsection{Overall Testing Framework}
\label{subsec:OverallFramework}

\begin{figure*}[p]
	\centering
	\includegraphics[scale=0.7]{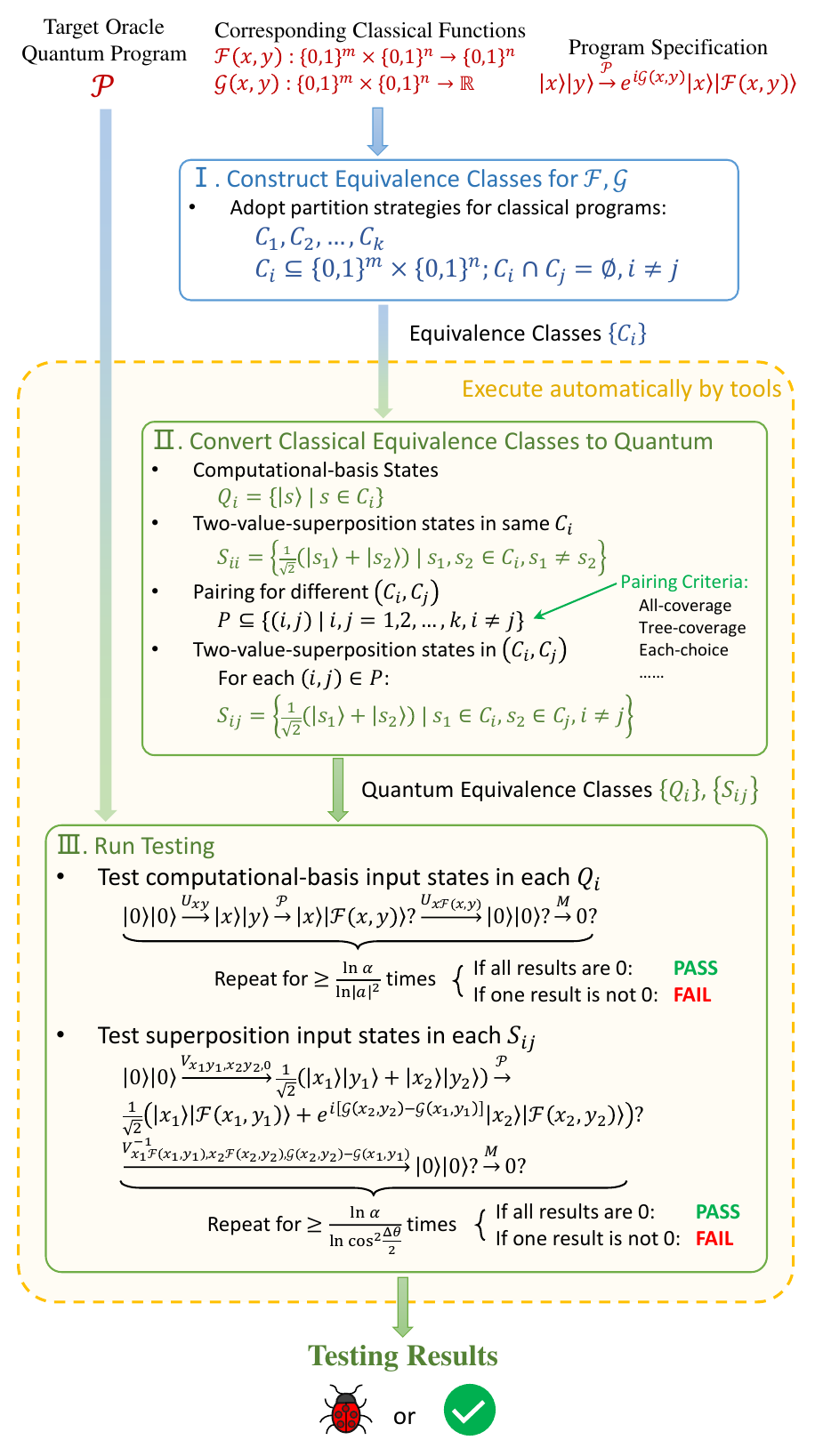}
	\caption{The overall testing framework for oracle quantum programs.}
	\label{fig:framework}
\end{figure*}

Based on the above discussions in this section, we propose a comprehensive testing framework for oracle quantum programs. Fig.~\ref{fig:framework} illustrates the overall testing framework. The testing framework takes the target oracle quantum program $\mathcal{P}$ along with its corresponding classical functions $\mathcal{F}$ and $\mathcal{G}$ as inputs. The goal is to check whether $\mathcal{P}$ conforms to its program specification. The framework consists of three primary steps: 
(1) construct equivalence classes for $\mathcal{F}$ and $\mathcal{G}$; 
(2) convert classical equivalence classes to quantum equivalence classes, and (3) perform a testing process based on quantum equivalence classes.

\begin{itemize}
\setlength{\itemsep}{2pt}
\item \textbf{Step I:} \textit{Construct equivalence classes for $\mathcal{F}$ and $\mathcal{G}$.}

This step involves generating equivalence classes based on the specific properties of the classical functions $\mathcal{F}(x,y)$ and $\mathcal{G}(x,y)$.
As this is a purely classical process, many existing strategies can be applied, such as boundary value analysis~\cite{ammann2016introduction}.
This step results in a set of equivalence classes $\{C_i\}$, which will be used in subsequent steps.

\item \textbf{Step II:} \textit{Convert classical equivalence classes to quantum equivalence classes.}

This step converts the equivalence classes $\{C_i\}$ into quantum input equivalence classes.
Each $C_i$ corresponds directly to an equivalence class $Q_i$ consisting of computational-basis states. For any $C_i$ containing at least two elements, we derive $S_{ii}$, the equivalence class for the two-value-superposition states formed by pairs of different values within $C_i$. We also construct $S_{ij}$ ($i \neq j$), which includes superposition states formed by pairs of values from different equivalence classes $C_i$ and $C_j$. To reduce complexity, we can select a subset of all possible pairs $(C_i, C_j)$ using suitable pairing criteria.
Upon completing this step, we obtain the quantum equivalence classes $\{Q_i\}$ and $\{S_{ij}\}$.

\item \textbf{Step III:} \textit{Run testing based on quantum equivalence classes.}
 
This step involves testing the computational-basis input states in each $Q_i$ and the superposition input states in each $S_{ij}$.
To check the execution result for each quantum input state, a PRUM process is executed.
Each quantum input state yields a testing result (PASS or FAIL). Aggregating these results across all input states forms the final testing result.
\end{itemize}

Steps II and III are well-suited for automation in this framework, as their procedures follow clearly defined execution rules. Step I, which involves partitioning equivalence classes for classical functions, can leverage existing testing tools. The following example demonstrates the workflow of the proposed testing framework for a practical oracle quantum program.

\begin{example}
\label{example:QAdd}
Testing the Quantum Adder Program

In Example~\ref{example:representFG}(2), we discuss the quantum adder program \texttt{QAdder}:

\hspace{3cm} $\texttt{QAdder} : \ket{x}\ket{y} \mapsto \ket{x}\ket{x+y\,(\mathrm{mod}\, 2^n)}$

\noindent and its corresponding classical functions $\mathcal{F}$ and $\mathcal{G}$:

\hspace{3cm} $\mathcal{F}(x,y) = x+y\;(\mathrm{mod}\;2^n), \quad \mathcal{G}(x,y) = 0$

\noindent
Suppose that the length of each qubit array is set to $n=5$. The testing process for this program follows the three steps outlined in Fig.~\ref{fig:framework}.

\vspace*{1mm}
\noindent \textbf{$\bullet$ Step I:} Construct equivalence classes for $\mathcal{F}$ and $\mathcal{G}$

Since $\mathcal{G}$ is constant, we need only to focus on $\mathcal{F}$. The test cases need to cover scenarios where $x+y$ either overflows or does not overflow. Additionally, 0 is a boundary value for the integers $x$ and $y$. Let $\mathrm{MAX} = 2^n-1 = 31$. We partition the input ranges for $x$ and $y$ as follows:
	
\qquad $x,y=0$ \quad $1\leq x,y\leq\mathrm{MAX} / 2$ \quad $\mathrm{MAX}/2+1 \leq x,y \leq\mathrm{MAX}$
	
\noindent For $n=5$, these ranges become:
	
\qquad $x,y=0$ \quad $1\leq x,y\leq 15$ \quad $16\leq x,y\leq 31$
	
We combine the partitions for $x$ and $y$ to derive equivalence classes. If both $x$ and $y$ belong to the second range $[1,15]$, their sum will not overflow. Conversely, their sum will overflow if both belong to the third range $[16,31]$. The equivalence classes that cover these cases are:

\qquad EC1 : \quad $x \in [1,15], \; y \in [1,15]$ \qquad EC2 : \quad $x \in [1,15], \; y \in [16,31]$

\qquad EC3 : \quad $x \in [16,31], \; y \in [1,15]$ \qquad EC4 : \quad $x \in [16,31], \; y \in [16,31]$
	
Next, we address the boundary case where $x = 0$ or $y = 0$. Note that if either variable is 0, there will be no overflow. So we can merge the second and third ranges for one variable when the other is 0. Then we obtain additional equivalence classes:

\qquad EC5 : \quad $x=0, \; y=0$ \qquad EC6 : \quad $x=0, \; y \in [1,31]$ \qquad EC7 : \quad $x \in [1,31], \; y=0$

Finally, the equivalence classes for classical functions $\mathcal{F}$ and $\mathcal{G}$ are EC1 through EC7.

\vspace*{1mm}
\noindent \textbf{$\bullet$ Step II:} Convert classical equivalence classes to quantum equivalence classes

Constructing the corresponding quantum equivalence classes for computational-basis input states and two-value-superposition states from the same equivalence class (EC*) is straightforward:
\begin{align*}
&Q_i = \left\{ \ket{x}\ket{y} \;|\; (x,y) \in \mathrm{EC}i \right\}, \quad i = 1, 2, 3, 4, 5, 6, 7 \\
&S_{ii} = \left\{ \frac{1}{\sqrt{2}} \left( \ket{x_1}\ket{y_1} + \ket{x_2}\ket{y_2} \right) \;|\; (x_1, y_1), (x_2, y_2) \in \mathrm{EC}i \right\}, \quad i = 1, 2, 3, 4, 6, 7
\end{align*}
\noindent For example, the state $\ket{7}\ket{20}$ belongs to equivalence class $Q_2$, and the state $\frac{1}{\sqrt{2}} \left( \ket{0}\ket{10} + \ket{0}\ket{21} \right)$ belongs to equivalence class $S_{66}$. Note that there are 7 $Q_i$ classes and 6 $S_{ii}$ classes because EC5 contains only one state, and thus, it cannot generate a two-value-superposition state.

Next, we construct the quantum equivalence classes for two-value-superposition states involving pairs from different equivalence classes. Assuming the adoption of the all-coverage criterion, we combine $EC1 \sim EC7$ pairwise to obtain:
\begin{equation*}
S_{ij} = \left\{ \frac{1}{\sqrt{2}} \left( \ket{x_1}\ket{y_1} + \ket{x_2}\ket{y_2} \right) \;|\; (x_1, y_1) \in \mathrm{EC}i, (x_2, y_2) \in \mathrm{EC}j \right\}, \quad i, j = 1, \dots, 7, \; i < j
\end{equation*}
\noindent For example, the state $\frac{1}{\sqrt{2}} \left( \ket{7}\ket{20} + \ket{0}\ket{21} \right)$ is included in equivalence class $S_{26}$. There are 21 $S_{ij}$ classes.

\vspace*{1mm}
\noindent \textbf{$\bullet$ Step III:} Run Testing

The testing task is performed based on the classical functions $\mathcal{F}$, $\mathcal{G}$, and the equivalence classes $Q_i$, $S_{ij}$ generated in Step II. Sample input states from every equivalence class, execute the PRUM process for each input state to check whether the corresponding output meets the expected value. The repetition times for computational-basis input states $N_{cb}$ and for two-value superposition input states $N_{tv}$ can be selected by Eqs.~(\ref{equ:Ncb}) and (\ref{equ:Ntv}).


\end{example}



\section{Evaluation}
\label{sec:evaluation}

This section evaluates the proposed testing framework for oracle quantum programs. The experiments aim to assess its effectiveness in detecting bugs and ensuring correctness. We analyze the framework's robustness and efficiency under different configurations by testing a range of benchmark programs with adjustable parameters.

\subsection{Experiment Design}

A testing algorithm yields a PASS or FAIL result for the given target program. To ensure a comprehensive evaluation, it is essential to prepare two categories of target programs: \textit{expected-pass} and \textit{expected-fail}. An effective algorithm should return PASS for \textit{expected-pass} targets and FAIL for \textit{expected-fail} targets. Therefore, our benchmark programs include both types of target programs.

As outlined in Section~\ref{sec:methods}, given a target oracle quantum program and the equivalence classes for its corresponding classical functions, several adjustable parameters and strategies exist within our testing framework. These include the repetition times for the PRUM process with computational-basis states ($N_{cb}$) and two-value-superposition states ($N_{tv}$), as well as the pairing strategies for generating quantum equivalence classes from classical equivalence classes. To guide the evaluation, we address the following research questions (RQs):

\begin{itemize}
\setlength{\itemsep}{2pt}
\item \textbf{RQ1:} How does the parameter $N_{cb}$ affect the bug detection capability for errors in qubits?

\textbf{RQ2:} How different are the execution efficiencies between the "directly measure" procedure and the "inverse and measure" procedure when checking computational-basis output states?

\item \textbf{RQ3:} How well do the three equivalence class pairing criteria for superposition states perform?

\item \textbf{RQ4:} How does the parameter $N_{tv}$ affect the bug detection capability for errors in phase?

\item \textbf{RQ5:} How effective are our checking methods on the benchmark programs?
\end{itemize}

To facilitate our experiments, we developed a software project with a dedicated prototype tool for testing Q\# oracle quantum programs and a series of testing scripts for evaluating these RQs. The prototype tool and testing scripts are implemented in Q\#, while Python controls the testing processes and analyzes results. The software structure of our project is designed to maximize module reuse by leveraging Q\# features, such as function factories and functions as arguments\footnote{The source code for the prototype tool and experimental evaluation is available at \\ \href{https://github.com/MgcosA/Code_of_Testing_Oracle_Quantum_Program_Article}{https://github.com/MgcosA/Code\_of\_Testing\_Oracle\_Quantum\_Program\_Article}}.

Our experiments were conducted using the Azure Quantum Development Kit (version 1.12.1) and its simulator on a personal computer equipped with an Intel Core i7-13700 CPU and 48 GB RAM. We executed the testing tasks sequentially to ensure accuracy and minimize measurement errors caused by CPU cooling limitations.

\subsection{Benchmark Programs}
\label{subsec:benchmark}

To evaluate the proposed testing framework, we constructed a set of benchmark programs representing various oracle quantum program types. These programs encompass Boolean functions, arithmetic operations, diagonal Hamiltonian evolutions, and mixed behaviors, providing a comprehensive test suite for assessing the effectiveness and generality of our framework.

\subsubsection{Original: Expected-Pass Programs}

\begin{table*}
\setlength{\tabcolsep}{1.4mm}
\scriptsize
\caption{The corresponding classical functions $\mathcal{F}$, $\mathcal{G}$, and parameter settings for our benchmark programs.}
\label{table:original_programs}

\begin{tabular}{c|cc|cc|c}
	\toprule
	\textbf{Program} & $m$ & $n$ & $\mathcal{F}(x,y)$ & $\mathcal{G}(x,y)$ & \textbf{Other Parameter Settings}\\
	\midrule
	\texttt{Parity\_P} & 0 & 6 & $y$ & $\mathrm{PARITY}(y)\cdot\pi$ & - \\
	\texttt{Is2Power\_P} & 0 & 6 & $y$ & $\mathrm{IS2POWER}(y)\cdot\pi$ & - \\
	\texttt{LessThan\_P} & 0 & 5 & $y$ & $\mathrm{LESSTHAN}_{k}(y)\cdot\pi$ & $k=10$ \\
	\texttt{Parity\_Q} & 6 & 1 & $y\oplus\mathrm{PARITY}(x)$ & $0$ & - \\
	\texttt{Is2Power\_Q} & 6 & 1 & $y\oplus\mathrm{IS2POWER}(x)$ & $0$ & - \\
	\texttt{LessThan\_Q} & 5 & 1 & $y\oplus\mathrm{LESSTHAN}_{k}(x)$ & $0$ & $k=10$ \\
	\texttt{QAdder} & 5 & 5 & $x+y\,(\mathrm{mod}\, 2^n)$ & $0$ & - \\
	\texttt{HamiltonX} & 0 & 3 & $y$ & $yt$ & $t=0.2$ \\
	\texttt{Ising} & 0 & 7 & $y$ & $\left[ J\sum_{i=0}^{n-1}(-1)^{y_i}(-1)^{y_{i+1(\mathrm{mod}\;n)}} + B\sum_{i=0}^{n-1}(-1)^{y_i} \right] \cdot t$ & $t=0.2, J=1, B=1$ \\
	\texttt{Mixed\_Proc} & 5 & 5 & $x+y\,(\mathrm{mod}\, 2^n)$ & $yt$ & $t=0.2$ \\
	\bottomrule
\end{tabular}
\end{table*}

We selected several representative oracle quantum programs as benchmark programs to evaluate the testing framework. As discussed in Section~\ref{subsec:oracle}, the benchmark includes the following categories: (1) oracle programs for Boolean functions using phase implementation; (2) oracle programs for Boolean functions using qubit implementation; (3) quantum arithmetic programs; (4) programs for diagonal Hamiltonian evolution; (5) general oracle quantum programs with non-trivial $\mathcal{F}$ and $\mathcal{G}$. The selected programs and their corresponding classical functions are detailed in Table~\ref{table:original_programs}. 

\begin{figure}
	\centering
	\includegraphics[scale=0.5]{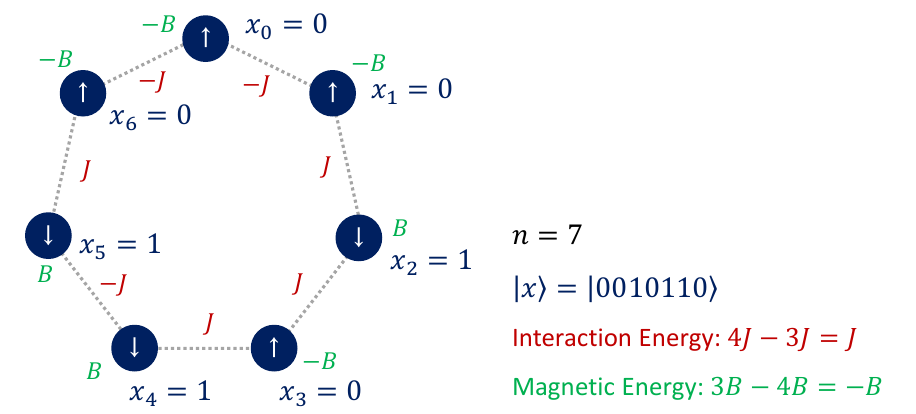}
	\caption{The particle configuration of program \texttt{Ising}. Seven particles are arranged in a circle, with spin interactions between adjacent pairs. The diagram shows the spin state $\ket{0010110}$.}
	\label{fig:Ising}
\end{figure}

Table~\ref{table:original_programs} also lists the selected classical parameters for these programs, such as the number of qubits $m$ and $n$. For programs requiring additional parameters, like the evolution time $t$ in Hamiltonian simulation, these are also included. Each program is implemented correctly to serve as an expected-pass program. Additionally, we adopt the \textit{little-endian mode} to represent integers in binary format in the following experiments, with indices starting from 0 (i.e., $x_0$ denotes the least significant bit and $x_{n-1}$ the most significant bit). Below is a brief description of these selected programs:

\begin{itemize}
\setlength{\itemsep}{3pt}
\item \texttt{Parity\_P, Parity\_Q}
	
	These two programs compute the Boolean parity function $\mathrm{PARITY}(x)$, which returns 0 or 1 based on the number of '1' bits in the binary representation of input $x$:
	$$\mathrm{PARITY}(x) = \left\{ \begin{array}{ll}
		1 & \text{if there are odd number of '1' bits in } x \\
		0 & \text{if there are even number of '1' bits in } x
	\end{array} \right.$$
	\texttt{Parity\_P} encodes the result in the phase, while \texttt{Parity\_Q} encodes the result in an auxiliary qubit:
	
	\qquad \texttt{Parity\_P} : $\ket{y} \mapsto (-1)^{\mathrm{PARITY}(y)}\ket{y} =  e^{i\mathrm{PARITY}(y)\cdot \pi}\ket{y}$
	
	\qquad \texttt{Parity\_Q} : $\ket{x}\ket{y} \mapsto \ket{x}\ket{y\oplus \mathrm{PARITY}(x)}$
	
\item \texttt{Is2Power\_P, Is2Power\_Q}
	
	These two programs compute the Boolean function $\mathrm{IS2POWER}(x)$, which identifies whether $x$ is a power of 2:
	$$\mathrm{IS2POWER}(x) = \left\{ \begin{array}{ll}
		1 & \text{if } \exists k\in\mathbb{N}, x=2^k \\
		0 & \text{otherwise}
	\end{array} \right.$$
	
	\qquad \texttt{Is2Power\_P} : $\ket{y} \mapsto (-1)^{\mathrm{IS2POWER}(y)}\ket{y} =  e^{i\mathrm{IS2POWER}(y)\cdot \pi}\ket{y}$

	\qquad \texttt{Is2Power\_Q} : $\ket{x}\ket{y} \mapsto \ket{x}\ket{y\oplus \mathrm{IS2POWER}(x)}$
	
\item \texttt{LessThan\_P, LessThan\_Q}
	
	These two programs compute the Boolean function $\mathrm{LESSTHAN}_{k}(x)$, which compares $x$ to a given integer     $k$:
	$$\mathrm{LESSTHAN}_{k}(x) = \left\{ \begin{array}{ll}
	1 & \text{if } x<k \\
	0 & \text{otherwise}
	\end{array} \right.$$
	
	\qquad \texttt{LessThan\_P} : $\ket{y} \mapsto (-1)^{\mathrm{LESSTHAN}_{k}(y)}\ket{y} =  e^{i\mathrm{LESSTHAN}_{k}(y)\cdot \pi}\ket{y}$

	\qquad \texttt{LessThan\_Q} : $\ket{x}\ket{y} \mapsto \ket{x}\ket{y\oplus \mathrm{LESSTHAN}_{k}(x)}$
	
	We referenced~\cite{Javier2025LessThan} for implementing these programs. In our experiments, $k$ is fixed to 10.
	
\item \texttt{QAdder}

	This program implements the addition for two $n$-bit qubit arrays $\ket{x}$ and $\ket{y}$. It computes $x+y$ and discards the high bits (equivalent to modulo $2^n$) when overflowing:

    \qquad \texttt{QAdder} : $\ket{x}\ket{y} \mapsto \ket{x}\ket{x+y\,(\mathrm{mod}\, 2^n)}$

\item \texttt{HamiltonX}

This program implements the Hamiltonian evolution for the position operator $\hat{x}$ in the coordinate representation:

\qquad \texttt{HamiltonX} : $\ket{x} \mapsto e^{ixt}\ket{x}$

In our experiment, we fix the evolution time $t=0.2$.


\item \texttt{Ising}

This program implements the Hamiltonian evolution of the Ising model:
$$H = -J\sum_{<ij>}(-1)^{x_i}(-1)^{x_j} - B\sum_{i}(-1)^{x_i}$$
In our experiment, we arrange seven particles in a circular configuration, as shown in Fig.~\ref{fig:Ising}. Thus, we use $n=7$ qubits, where each qubit represents the spin state of a particle. The spin interactions occur between adjacent pairs of particles, i.e., between the $i$-th particle and the $[i+1 (\mathrm{mod}\;n)]$-th particle:

\qquad \texttt{Ising} : $\ket{y} \mapsto e^{i\left[ J\sum_{i=0}^{n-1}(-1)^{y_i}(-1)^{y_{i+1(\mathrm{mod}\;n)}} + B\sum_{i=0}^{n-1}(-1)^{y_i} \right]t}\ket{y}$

We fix the parameters: time $t=0.2$, interaction energy $J=1$, and external magnetic field $B=1$.

\item \texttt{Mixed\_Proc}

We construct a general oracle quantum program (qubit-phase-mixed) by combining the behaviors of \texttt{QAdder} and \texttt{HamiltonX}. Specifically, it performs the following transform:

\qquad \texttt{Mixed\_Proc} : $\ket{x}\ket{y} \mapsto e^{iyt}\ket{x}\ket{x+y\,(\mathrm{mod}\, 2^n)}$

This program features both non-trivial $\mathcal{F}$ and $\mathcal{G}$.
\end{itemize}

\subsubsection{Classical Equivalence Classes}

Since our testing framework requires equivalence classes for the classical functions $\mathcal{F}$ and $\mathcal{G}$ as input, we partition the input space for each original program based on its $\mathcal{F}$ and $\mathcal{G}$ using existing equivalence class partitioning strategies. These strategies include partitioning by boundary values, special values, or output results~\cite{ammann2016introduction}. The equivalence classes for each program are summarized in Table~\ref{table:equivalence_classes}. The column "\#EC" indicates the total number of equivalence classes, while the column "Partition Guidelines" lists the strategies used for partitioning.

A special case is the \texttt{HamiltonX} program. Since we fixed the number of qubits to 3, the total number of inputs is limited to 8. Consequently, each input is treated as a separate equivalence class.

\begin{table*}
\footnotesize
\caption{Selected equivalence classes for each program and corresponding partition guidelines.}
\label{table:equivalence_classes}
	
\begin{tabular}{c|l|c|l}
	\toprule
	\textbf{Program} & \textbf{Equivalence Classes} & \textbf{\#EC} & \textbf{Partition Guidelines} \\
	\midrule
	\makecell{ \texttt{Parity\_P} \\ \texttt{Parity\_Q} } & \makecell[l]{\textcircled{1} All-zero input \\ \textcircled{2} All-one input \\ \textcircled{3} Input is parity \\ \textcircled{4} Input is not parity} & 4 & \makecell[l]{Boundary values \\ Output results} \\
	\midrule
	\makecell{ \texttt{Is2Power\_P} \\ \texttt{Is2Power\_Q} } & \makecell[l]{\textcircled{1} All-zero input \\ \textcircled{2} All-one input \\ \textcircled{3} Input is the power of 2 \\ \textcircled{4} Input is not the power of 2} & 4 & \makecell[l]{Boundary values \\ Output results} \\
	\midrule
	\makecell{\texttt{LessThan\_P} \\ \texttt{LessThan\_Q} } & \makecell[l]{\textcircled{1} $x=0$ \\ \textcircled{2} $0<x<k-1$ \\ \textcircled{3} $x=k-1$ \\ \textcircled{4} $x=k$ \\ \textcircled{5} $x=k+1$ \\ \textcircled{6} $k+1<x<2^n-1$ \\ \textcircled{7} $x=2^n-1$ } & 7 & \makecell[l]{Boundary values \\ Special values} \\
	\midrule
	\makecell{ \texttt{QAdder} \\ \texttt{Mixed\_Proc} } & \makecell[l]{$\mathrm{MAX} = 2^n - 1$ \\ \textcircled{1} $x=0,\quad y=0$ \\ \textcircled{2} $x=0,\quad 1\leq y\leq \mathrm{MAX}$ \\ \textcircled{3} $1\leq x\leq \mathrm{MAX},\quad y=0$ \\ \textcircled{4} $1\leq x\leq \mathrm{MAX}/2,\quad 1\leq y\leq \mathrm{MAX}/2$ \\ \textcircled{5} $1\leq x\leq \mathrm{MAX}/2,\quad \mathrm{MAX}/2+1\leq y\leq \mathrm{MAX}$ \\ \textcircled{6} $\mathrm{MAX}/2+1\leq x\leq \mathrm{MAX},\quad 1\leq y\leq \mathrm{MAX}/2$ \\ \textcircled{7} $\mathrm{MAX}/2+1\leq x\leq \mathrm{MAX},\quad \mathrm{MAX}/2+1\leq y\leq \mathrm{MAX}$} & 7 & \makecell[l]{ Boundary values \\ Whether output overflows \\ Combination of two variables } \\
	\midrule
	\texttt{HamiltonX} & \textcircled{0}$\sim$\textcircled{7}: $y=0,\dots, 7$ & 8 & Enumeration \\
	\midrule
	\texttt{Ising} & \makecell[l]{\textcircled{1} State with all '0' \\ \textcircled{2} State with one '1' \\ \textcircled{3} State with two adjacent  '1's \\ \textcircled{4} State with two separate '1's \\ \textcircled{5} State with more than two '1's} & 5 & \makecell[l]{Boundary values \\ Special values} \\
	\bottomrule
\end{tabular}
\end{table*}

\subsubsection{Expected-Fail Programs}

We construct expected-fail programs by introducing specific errors into the expected-pass programs. 
In general, we consider the following three types of expected-fail programs: 
(1) To evaluate the selection of $N_{cb}$ and $N_{tv}$ (RQ1 and RQ4), we use expected-fail programs that differ from their corresponding correct versions by a given amplitude or phase difference; 
(2) To evaluate the effectiveness of quantum equivalence class construction methods (RQ3), we use expected-fail programs generated by altering the program behavior according to its equivalence classes; 
and (3) to evaluate the performance of testing methods (RQ5), in addition to the above two types, we also include more mutant programs that simulate typical coding errors.

\vspace{2mm}
\noindent $\bullet$ \textit{Expected-fail programs for evaluating the selection of $N_{cb}$ (RQ1).}\hspace*{1mm}
According to Eq.~(\ref{equ:Ncb}), the number of repetitions for computational-basis input states, $N_{cb}$, is related to $a$, the amplitude of the expected-state component in the following decomposition: 
$\ket{\psi_{\mathrm{output}}} = a\ket{\psi_{\mathrm{expected}}} + b\ket{\psi_{\mathrm{expected}}^{\bot}}$. 
The following proposition explains how to generate an erroneous output state $\ket{\psi_{\mathrm{output}}}$ from the correct state $\ket{\psi_{\mathrm{expected}}}$.

\begin{proposition}
\label{prop:aExp_bNExp}
To convert the expected computational-basis state $\ket{\psi_{\mathrm{expected}}}$ into an erroneous output state $\ket{\psi_{\mathrm{output}}}$ of the following form,\footnote{Note that a global phase factor $e^{i\delta}$ can be extracted. Without loss of generality, we assume $a \in \mathbb{R}$.}
\begin{equation*}
\ket{\psi_{\mathrm{output}}} = a\ket{\psi_{\mathrm{expected}}} + b\ket{\psi_{\mathrm{expected}}^{\bot}}, 
\quad 0 \le a \le 1,\; |a|^2 + |b|^2 = 1,
\end{equation*}
\noindent we can apply an $R_y(\theta)$ gate to any qubit of $\ket{\psi_{\mathrm{expected}}}$, where the rotation angle $\theta = 2\arccos(a)$, as shown below:
\begin{equation*}
\ket{\psi_{\mathrm{expected}}} 
\xrightarrow{\;R_y(2\arccos(a))\text{ on any qubit}\;}
\ket{\psi_{\mathrm{output}}}.
\end{equation*}
\end{proposition}

The proof of Proposition~\ref{prop:aExp_bNExp} is provided in Appendix~\ref{appd:aExp_bNExp}. Accordingly, we construct the expected-fail programs for evaluating the selection of $N_{cb}$ by adding an $R_y(\theta)$ gate to the first qubit at the end of each expected-pass program. Specifically, we select $a = \frac{\sqrt{3}}{2}$, $\frac{\sqrt{2}}{2}$, $\frac{1}{2}$, and $0$, with the corresponding $\theta$ values of $\frac{\pi}{3}$, $\frac{\pi}{2}$, $\frac{2\pi}{3}$, and $\pi$, respectively. We use the suffixes "\texttt{\_AddRyPiDiv3}", "\texttt{\_AddRyPiDiv2}", "\texttt{\_AddRy2PiDiv3}", and "\texttt{\_AddRyPi}" to denote these mutant programs, as shown in Table~\ref{table:mutant_programs}. For example, the program \texttt{Parity\_P\_AddRyPiDiv3} is generated by adding an $R_y(\frac{\pi}{3})$ gate at the end of the program \texttt{Parity\_P}.

\begin{table*}
\footnotesize
\centering
\caption{Mutant programs and their corresponding outputs for evaluating the choice of $N_{cb}$ and $N_{tv}$.}
\label{table:mutant_programs}
\begin{tabular}{c|c|c|c|c}
\toprule
\textbf{Evaluate} & \textbf{Suffix} & \textbf{Added Gate} & \textbf{Original Output} & \textbf{Mutant Output} \\
\midrule
\multirow{5}{*}{$N_{cb}$} & \texttt{\_AddRyPiDiv3} & $R_y(\frac{\pi}{3})$ & \multirow{5}{*}{$\ket{\psi_{\mathrm{expected}}}$} & $\frac{\sqrt 3}{2}\ket{\psi_{\mathrm{expected}}} \pm \frac{1}{2}\ket{\psi_{\mathrm{expected}}^{\bot}}$ \\
& \texttt{\_AddRyPiDiv2} & $R_y(\frac{\pi}{2})$ && $\frac{\sqrt 2}{2}\ket{\psi_{\mathrm{expected}}} \pm \frac{\sqrt 2}{2}\ket{\psi_{\mathrm{expected}}^{\bot}}$ \\
& \texttt{\_AddRy2PiDiv3} & $R_y(\frac{2\pi}{3})$ && $\frac{1}{2}\ket{\psi_{\mathrm{expected}}} \pm \frac{\sqrt 3}{2}\ket{\psi_{\mathrm{expected}}^{\bot}}$ \\
& \texttt{\_AddRyPi} & $R_y(\pi)$ && $\pm \ket{\psi_{\mathrm{expected}}^{\bot}}$ \\
\midrule
\multirow{8}{*}{$N_{tv}$} & \texttt{\_AddZ} & $Z$ & \multirow{8}{*}{$\frac{1}{\sqrt 2}\left( \ket{\psi_1} + e^{i\Delta\mathcal{G}_{\mathrm{expected}}}\ket{\psi_2} \right)$} & $\frac{1}{\sqrt 2}\left( \ket{\psi_1} + e^{i(\Delta\mathcal{G}_{\mathrm{expected}} + \pi)}\ket{\psi_2} \right)$ \\
& \texttt{\_AddS} & $S$ && $\frac{1}{\sqrt 2}\left( \ket{\psi_1} + e^{i(\Delta\mathcal{G}_{\mathrm{expected}} + \pi/2)}\ket{\psi_2} \right)$ \\
& \texttt{\_AddT} & $T$ && $\frac{1}{\sqrt 2}\left( \ket{\psi_1} + e^{i(\Delta\mathcal{G}_{\mathrm{expected}} + \pi/4)}\ket{\psi_2} \right)$ \\
& \texttt{\_AddRz8} & $R_z(\frac{\pi}{8})$ && $\frac{1}{\sqrt 2}\left( \ket{\psi_1} + e^{i(\Delta\mathcal{G}_{\mathrm{expected}} + \pi/8)}\ket{\psi_2} \right)$ \\
& \texttt{\_AddRz16} & $R_z(\frac{\pi}{16})$ && $\frac{1}{\sqrt 2}\left( \ket{\psi_1} + e^{i(\Delta\mathcal{G}_{\mathrm{expected}} + \pi/16)}\ket{\psi_2} \right)$ \\
& \texttt{\_AddRz32} & $R_z(\frac{\pi}{32})$ && $\frac{1}{\sqrt 2}\left( \ket{\psi_1} + e^{i(\Delta\mathcal{G}_{\mathrm{expected}} + \pi/32)}\ket{\psi_2} \right)$ \\
\bottomrule
\end{tabular}
\end{table*}

\vspace{2mm}
\noindent $\bullet$ \textit{Expected-fail programs for evaluating the selection of $N_{tv}$ (RQ4).}\hspace*{1mm}
Equation~(\ref{equ:Ntv}) indicates that the number of repetitions for two-value superposition input states, $N_{tv}$, is related to the phase difference $\Delta\theta$. According to Eq. (\ref{equ:relative_phase}), applying an $R_z(\theta)$ gate to a qubit in superposition introduces a relative phase factor $e^{i\theta}$. However, a fixed gate cannot be used for generating the desired phase difference because some qubits may not be in superposition. For example, consider the two-value superposition state $\frac{1}{\sqrt{2}}\left(\ket{00} + \ket{10}\right)$, where the first qubit is in superposition but the second qubit is not. Consequently, the phase difference is generated only when the $R_z(\theta)$ gate is applied to the superposed first qubit, as shown below:
\begin{equation*}
\frac{1}{\sqrt 2}\left(\ket{00}+\ket{10}\right) \left\{
\begin{array}{l}
\xrightarrow{\text{apply $R_z(\theta)$ gate on 1st qubit}} \frac{1}{\sqrt 2}(\ket{00} + e^{i\theta}\ket{10}) \\
\xrightarrow{\text{apply $R_z(\theta)$ gate on 2nd qubit}} \frac{1}{\sqrt 2}(\ket{00}+\ket{10})
\end{array}
\right.
\end{equation*}
\noindent Unless we know which qubits are in superposition, we cannot use fixed mutant programs to generate a phase difference between the two components of a two-value superposition state.

Thus, in our experiments, we use fixed inputs rather than random ones, and the phase difference is introduced by adding an $R_z(\theta)$ gate to a pre-identified superposed qubit.
 We select the phase differences $\pi$, $\frac{\pi}{2}$, $\frac{\pi}{4}$, $\frac{\pi}{8}$, $\frac{\pi}{16}$, and $\frac{\pi}{32}$ respectively, and the corresponding mutation gates are $Z$, $S$, $T$, $R_z(\frac{\pi}{8})$, $R_z(\frac{\pi}{16})$, and $R_z(\frac{\pi}{32})$\footnote{$Z$, $S$, and $T$ gates are equivalent to $R_z(\pi)$, $R_z(\frac{\pi}{2})$ and $R_z(\frac{\pi}{4})$ respectively, ignoring a global phase.}. We use the suffixes "\texttt{\_AddZ}", "\texttt{\_AddS}", "\texttt{\_AddT}", "\texttt{\_AddRz8}", "\texttt{\_AddRz16}", and "\texttt{\_AddRz32}" to denote these mutant programs, as shown in Table~\ref{table:mutant_programs}. For example, the program \texttt{Parity\_P\_AddRz8} is generated by adding an $R_z(\frac{\pi}{8})$ gate to the end of the program \texttt{Parity\_P}.

\vspace{2mm}
\noindent $\bullet$ \textit{Expected-fail programs for evaluating the effectiveness of quantum equivalence class construction (RQ3).}\hspace*{1mm}
For the error type involving changes to program behavior according to its equivalence classes, we consider three categories of behavior modifications: (1) reversing the TRUE/FALSE outputs for Boolean functions; (2) altering the endianness mode (i.e., using big-endian mode instead of the default little-endian mode); and (3) modifying all output values within a single equivalence class. These modifications represent common implementation errors that arise from misunderstandings of the requirements. We selected several original programs and constructed the corresponding expected-failure programs by applying these three behavior changes. The resulting expected-fail programs are listed below:

\begin{itemize}
\setlength{\itemsep}{2pt}
\item Flipping the TRUE/FALSE outputs for Boolean functions:

\texttt{Parity\_P\_FlipOut},\quad \texttt{Parity\_Q\_FlipOut},\quad \texttt{Is2Power\_P\_FlipOut},\quad \texttt{Is2Power\_Q\_FlipOut}, \quad \texttt{GreaterThanEq\_P},\quad \texttt{GreaterThanEq\_Q}.

\item Changing the endianness mode:

\texttt{LessThan\_P\_BE},\quad \texttt{LessThan\_Q\_BE},\quad \texttt{QAdder\_BE},\quad \texttt{HamiltonX\_BE}.

\item Changing the output values in one equivalence class:

\texttt{Parity\_P\_FlipAll1},\quad \texttt{Parity\_Q\_FlipAll1},\quad \texttt{LessThanEq\_P},\quad \texttt{LessThanEq\_Q},\quad \\ \texttt{QAdder\_change0p}.

\end{itemize}

Similar to the expected-fail programs generated by mutations, these behavior-based expected-fail programs are named with specific suffixes for clarity. The suffix "\texttt{\_FlipOut}" denotes programs with flipped Boolean outputs. The suffix "\texttt{\_BE}" indicates that the endianness mode has been changed to \textit{big-endian}. Additionally, some programs have special naming conventions: \texttt{GreaterThanEq\_*} refers to quantum programs that compute the "$\geq$" function, equivalent to flipping the Boolean output of \texttt{LessThan\_*}. \texttt{LessThanEq\_*} denotes quantum programs that compute the "$\leq$" function, constructed by modifying the output under the input state critical value $\ket{k}$ while keeping other outputs unchanged. \texttt{Parity\_P\_FlipAll1} and \texttt{Parity\_Q\_FlipAll1} are derived by flipping the output for all-one inputs in \texttt{Parity\_P} and \texttt{Parity\_Q}. Lastly, \texttt{QAdder\_change0p} modifies its behavior by performing the transform $\ket{x}\ket{y} \mapsto \ket{x}\ket{x+y}$ and flipping the first qubit of $\ket{x+y}$ when $x=0$.

\vspace{2mm}
\noindent $\bullet$ \textit{Two-qubit-gate mutant programs.}\hspace*{1mm}
A common approach to simulating coding errors in programming is to insert mutation operations into the original program. Previous studies on quantum mutants~\cite{fortunato2022mutation,long2024testing,long2024equivalence} have shown that even a single mutation can significantly affect program outputs. Therefore, for each expected-fail program, we introduce only one mutation. The previous expected-fail programs were generated by adding an extra quantum gate at the end of the original program, but these mutants involved only single-qubit gates. Since two-qubit gates such as \texttt{CNOT} and \texttt{CZ} are also common in quantum programming, we additionally create mutant programs by adding these two-qubit gates at the end of each program. The suffixes "\texttt{\_AddCNOT}" and "\texttt{\_AddCZ}" are used to denote them.

\subsection{Experiment Configuration}
\label{subsec:ExConfig}

We describe the experimental configurations for each research question (RQ) as follows:

\vspace{2mm}
\noindent $\bullet$ \textbf{RQ1:}\hspace*{1mm}
We evaluate the bug detection capability of computational-basis state checking under different repetition counts $N_{cb}$ using expected-fail programs with a series of $R_y$ mutants: $R_y(\frac{\pi}{3})$, $R_y(\frac{\pi}{2})$, $R_y(\frac{2\pi}{3})$, and $R_y(\pi)$ (see Table~\ref{table:mutant_programs}). The testing range is set to $N_{cb} = 1 \dots 10$, following the analysis in Example~\ref{example:Ncb}. Because $N_{cb}$ is associated with computational-basis input states, we consider only quantum equivalence classes corresponding to these states, i.e., $\{Q_i\}$. For each mutant program, 100 input states are selected from each $Q_i$ under every $N_{cb}$, making the total number of tested inputs for each program equal to the product of the number of equivalence classes and 100.

According to the proof of Proposition~\ref{prop:Ncb} (see Appendix~\ref{subappd:Ncb}), when the PRUM process is repeated $N_{cb}$ times, the probability of obtaining a PASS result (i.e., the measurement yields 0) is given by $\left(|a|^2\right)^{N_{cb}}$. Specifically, $a=\frac{\sqrt{3}}{2}$, $\frac{\sqrt{2}}{2}$, $\frac{1}{2}$, and $0$ for the mutants $R_y(\frac{\pi}{3})$, $R_y(\frac{\pi}{2})$, $R_y(\frac{2\pi}{3})$, and $R_y(\pi)$, respectively. As a result, the corresponding PASS probabilities, as functions of the repetition count, should theoretically follow the exponential forms: $y=\left(\frac{3}{4}\right)^x$, $y=\left(\frac{1}{2}\right)^x$, $y=\left(\frac{1}{4}\right)^x$, and $y=0$, respectively. We record the proportion of PASS inputs for each $N_{cb}$ and compute their proportions to verify whether the experimental results align with these theoretical exponential functions.

\vspace{2mm}
\noindent $\bullet$ \textbf{RQ2:}\hspace*{1mm}
In this evaluation, we use the same experimental configurations as in RQ1, employing the four mutants $R_y(\frac{\pi}{3})$, $R_y(\frac{\pi}{2})$, $R_y(\frac{2\pi}{3})$, and $R_y(\pi)$. For each program, we execute the "inverse and measure" (IM) process and the "directly measure" (DM) process, and then compute the execution time ratio $t_{IM}/t_{DM}$. In this evaluation, we fix $N_{cb}=1$ to compare the time of a single process execution and select 500 input states from each $Q_i$ to reduce random fluctuations.

\vspace{2mm}
\noindent $\bullet$ \textbf{RQ3:}\hspace*{1mm}
To evaluate the performance of the three equivalence-class pairing criteria—\textit{all-coverage}, \textit{tree-coverage}, and \textit{each-choice}—we use expected-fail programs constructed by modifying program behavior according to equivalence classes. We consider three types of quantum equivalence classes: (1) computational-basis state classes $Q_i$, (2) two-value superposition state classes within the same classical class $S_{ii}$, and (3) two-value superposition state classes across different classical classes $S_{ij}$, where $i \neq j$. The pairing criteria are applied to construct the $S_{ij}$ classes.

We first conduct experiments using the all-coverage criterion and record the classes in which unexpected outputs are detected. Since the graphs of the tree-coverage and each-choice criteria can be regarded as spanning subgraphs of the all-coverage criterion, their effectiveness can be analyzed based on the all-coverage results. For each program, the testing process is repeated 100 times, with balanced parameter settings of $N_{cb}=1$ and $N_{tv}=100$.

\vspace{2mm}
\noindent $\bullet$ \textbf{RQ4:}\hspace*{1mm}
We evaluate the bug detection capability of two-value-superposition state checking under different repetition counts $N_{tv}$ using expected-fail programs with a series of $R_z$ mutants: $Z$, $S$, $T$, $R_z(\frac{\pi}{8})$, $R_z(\frac{\pi}{16})$, and $R_z(\frac{\pi}{32})$ (see Table~\ref{table:mutant_programs}). These mutants introduce phase angles of $\pi$, $\frac{\pi}{2}$, $\frac{\pi}{4}$, $\frac{\pi}{8}$, $\frac{\pi}{16}$, and $\frac{\pi}{32}$, respectively. Substituting these angles into Eq.~(\ref{equ:Ntv}) and setting the misjudgment probability limit $\alpha=0.01$, we obtain the corresponding $N_{tv}$ values required to distinguish the phase differences: 1, 7, 30, 119, 477, and 1910 (see Example~\ref{example:Ntv}(2)). Since the range of these values is large, the experiment does not use a continuous sequence of $N_{tv}$ values as in RQ1.

Instead, we fix the selection of $N_{tv}$ to 1, 7, 30, 119, 477, and 1910. Note that not all output states change under $R_z(\theta)$ mutations. To obtain precise results, for each program, we select a fixed input whose output state is affected by the $R_z(\theta)$ gate. Specifically, a mutation gate is added to the first qubit, and the input states are chosen as follows:

\begin{itemize}
\item \texttt{Parity\_P}, \texttt{Is2Power\_P}, \texttt{LessThan\_P}, \texttt{Parity\_Q}, \texttt{Is2Power\_Q}, \texttt{LessThan\_Q}, \texttt{HamiltonX}, and \texttt{Ising}: $\frac{1}{\sqrt{2}} \left( \ket{0\dots0} + \ket{1\dots1} \right)$, i.e., the GHZ state.
\item \texttt{QAdder} and \texttt{Mixed\_Proc}: $\frac{1}{\sqrt{2}} \left( \ket{0\dots0}\ket{0\dots0} + \ket{1\dots1}\ket{0\dots0} \right)$, where the variable $x$ is in the GHZ state and $y$ is in the zero state.
\end{itemize}

This selection is based on the fact that the output states of these programs under GHZ input are superposed across all qubits, allowing an $R_z$ gate to generate a detectable phase difference. The testing process is repeated 100 times for each input state at $N_{tv}=1$, 7, 30, 119, 477, and 1910, and the proportion of PASS results is recorded for each execution.

\vspace{2mm}
\noindent $\bullet$ \textbf{RQ5:}\hspace*{1mm}
We evaluate the overall effectiveness of our testing framework on all benchmark programs, including both expected-pass and expected-fail programs. For each program, the testing process is repeated 100 rounds. Balanced parameter settings for $N_{cb}$ and $N_{tv}$, obtained from RQ1$\sim$RQ4, are used. We record the number of PASS rounds and the running time for each shot. The framework is considered effective if expected-pass programs consistently achieve nearly 100 PASS rounds, while expected-fail programs achieve nearly 0 PASS rounds.

\vspace{2mm}
\noindent In summary, the experimental configurations for each research question (RQ) have been established, with appropriate parameters selected for evaluation. These configurations form the basis for assessing the performance and effectiveness of the proposed testing framework, and the results are presented in the following section.

\subsection{Result Analysis}
\label{subsec:results}

In this section, we analyze the experimental results presented in Section~\ref{subsec:ExConfig}. We evaluate the impact of different parameters and configurations, including the number of repetitions for computational-basis and two-value-superposition state checks, as well as various equivalence class pairing strategies. Through this analysis, we aim to assess the performance of the testing framework and its effectiveness in detecting bugs across different test scenarios.

\subsubsection{RQ1: Parameter $N_{cb}$}
\begin{figure*}
	\centering
	\includegraphics[scale=0.54]{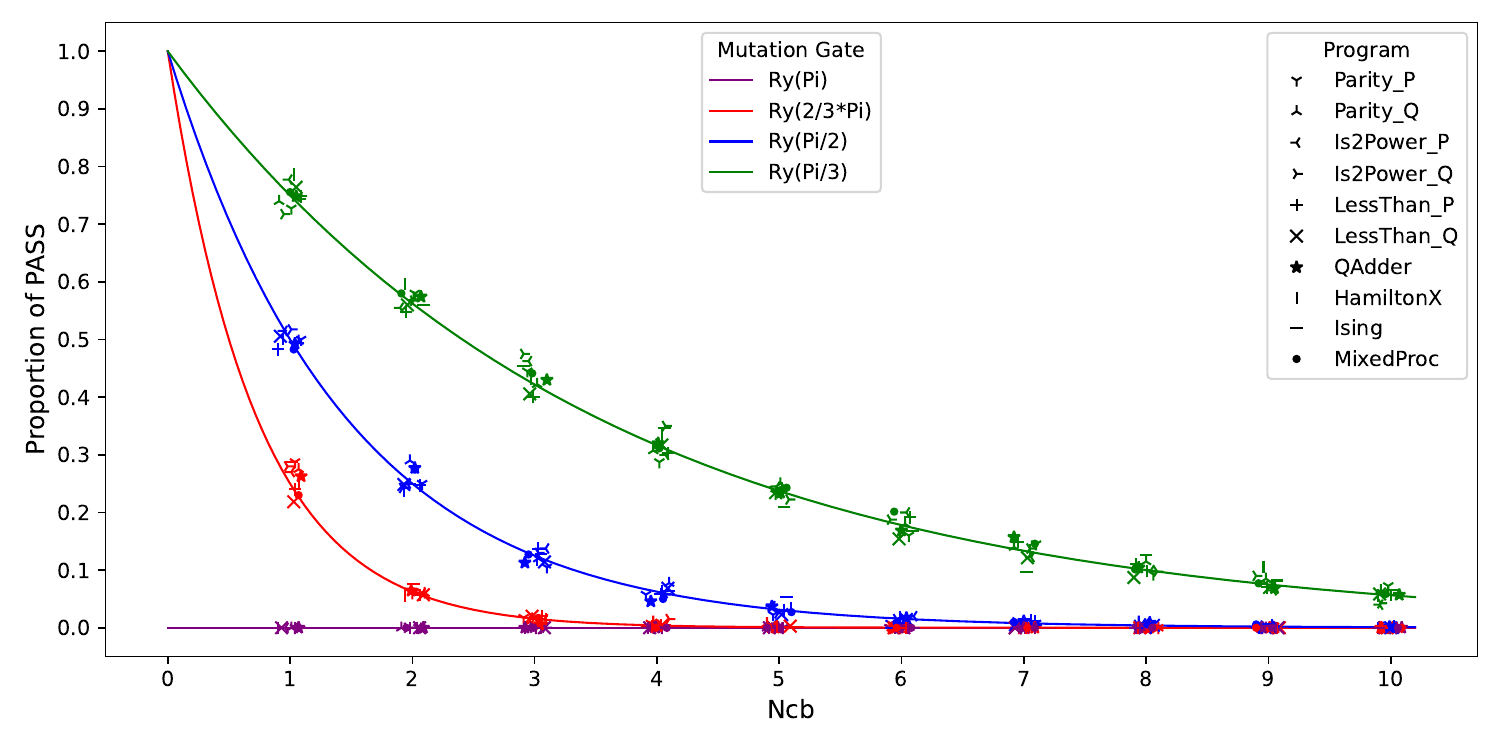}
    \caption{Results for RQ1: the proportion of PASS inputs for each expected-fail program with different $R_y$ mutants. The four curves represent the theoretical functions of PASS proportion versus $N_{cb}$, and the marked points indicate the experimental results. Different markers correspond to different programs, and different colors correspond to different mutation gates.}

	\label{fig:RQ1Result}
\end{figure*}

Fig.~\ref{fig:RQ1Result} presents the experimental results showing the relationship between the proportion of PASS inputs and $N_{cb}$. The experimental results agree well with the theoretical predictions. Moreover, the PASS proportion is independent of specific programs and inputs, which is consistent with the analysis in Proposition~\ref{prop:aExp_bNExp} and Appendix~\ref{appd:aExp_bNExp}. These observations indicate that the analytical framework used in Proposition~\ref{prop:Ncb}—namely, decomposing the output state into its expected component and orthogonal complement (Eq.~(\ref{equ:output_decomp}))—is valid. Therefore, the correctness of Proposition~\ref{prop:Ncb} regarding the selection of $N_{cb}$ is confirmed.

We also recorded the overall testing results for each expected-fail program. For each $N_{cb}$, $N_{cb}$ input states were selected from each $Q_i$, and the outcomes were aggregated according to the framework in Fig.~\ref{fig:framework}. Interestingly, the results show that misjudgment is nearly zero even when $N_{cb}=1$. This can be explained by the existence of multiple equivalence classes—when $N_{cb}=1$, the number of tested inputs equals the number of classes. In the worst case, where there are four equivalence classes and $|a|^2=\frac{3}{4}$, the misjudgment probability is $\left(\frac{3}{4}\right)^4 \approx 0.316$. Combined with the conclusion that the outcomes are independent of specific programs and inputs, this suggests that sampling across different equivalence classes can effectively substitute for multiple repetitions of the PRUM process on a single input. Consequently, selecting $N_{cb}=1$ or $2$ is sufficient for most practical testing scenarios.

\subsubsection{RQ2: Execution Efficiency of DM and IM}

\begin{figure*}
	\centering
	\includegraphics[scale=0.45]{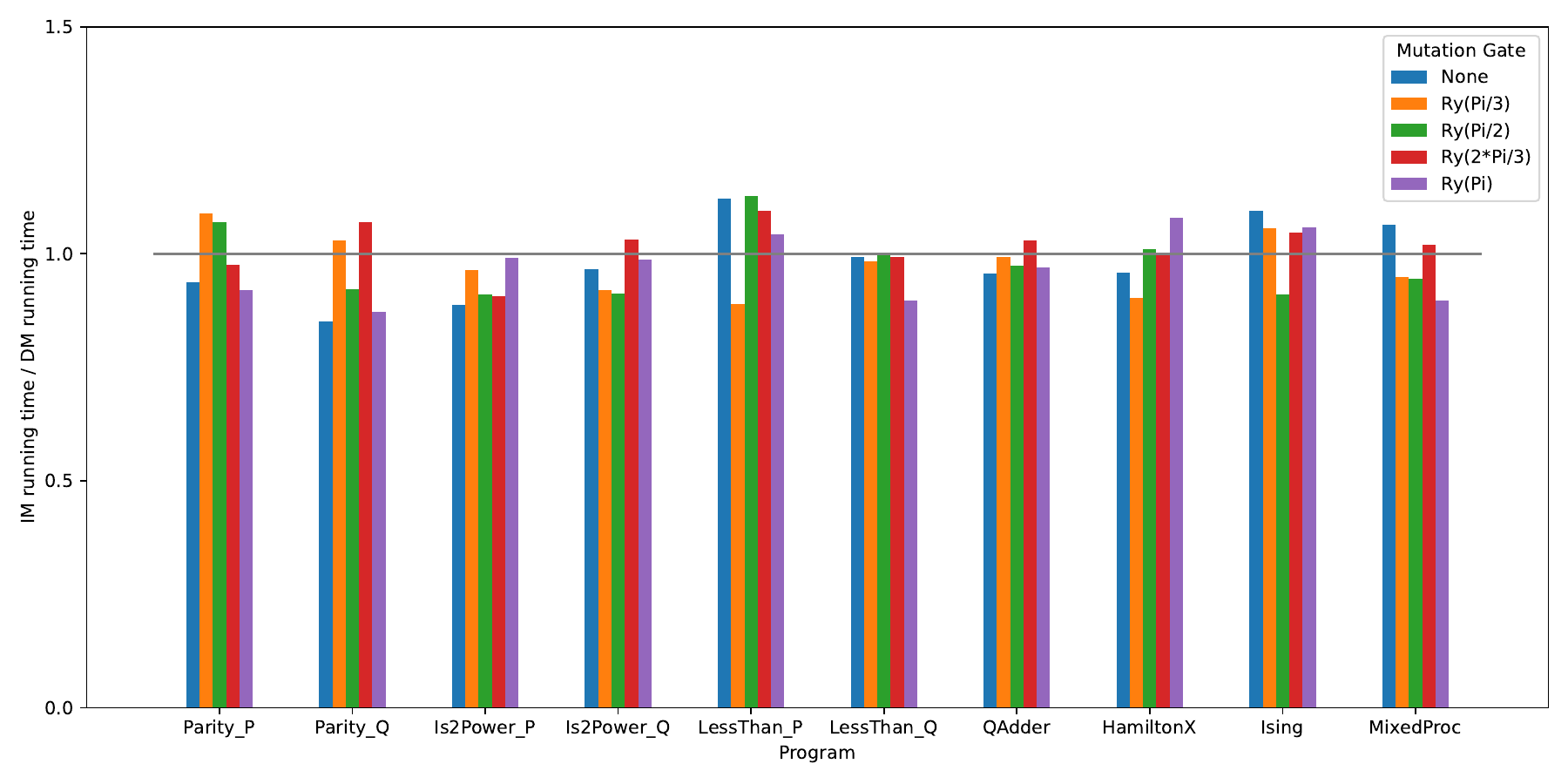}
    \caption{Results for RQ2: the time ratio $t_{IM}/t_{DM}$ for each original program and its corresponding $R_y$ mutants. Each group of bars represents an original program and its mutants, with different bar colors indicating different mutation gates. The gray horizontal line denotes $t_{IM}/t_{DM}=1$, where the two processes have equal execution time.}
	\label{fig:RQ2Result}
\end{figure*}

The results are shown in Fig.~\ref{fig:RQ2Result}. For all programs, the DM and IM processes exhibit similar execution times, with differences within 20\%. For some programs, IM is slightly faster, while for others, DM performs better, showing no clear trend.

We further analyze the components of execution time for both processes. The IM process includes an inverse unitary transform, which corresponds to a layer of $X$ gates when using computational-basis inputs. The measurement result is then compared against the integer 0. In contrast, the DM process omits the inverse transform, but its measurement result must be compared with an integer variable, requiring additional preparation time. This observation is noteworthy: although DM appears to reduce one layer of circuit depth and thus might seem faster, it also increases the complexity of classical computation. Therefore, the relative efficiency between IM and DM depends on the execution time of a layer of $X$ gates versus the time required to prepare a classical integer. In practice, a typical layer of quantum gates executes in approximately $10^{-9}\!\sim\!10^{-7}$\,s~\cite{barends2014superconducting,kjaergaard2020superconducting}, while a classical CPU operating at around $10^9$\, Hz executes one instruction in about $10^{-9}$\,s. Hence, the difference in execution efficiency between IM and DM seems negligible. In our next experiments, we adopt the IM process to maintain consistency within the PRUM procedure.

\subsubsection{RQ3: Bug Behaviors Across Equivalence Classes}

\begin{figure*}[p]
	\centering
	\includegraphics[scale=0.7]{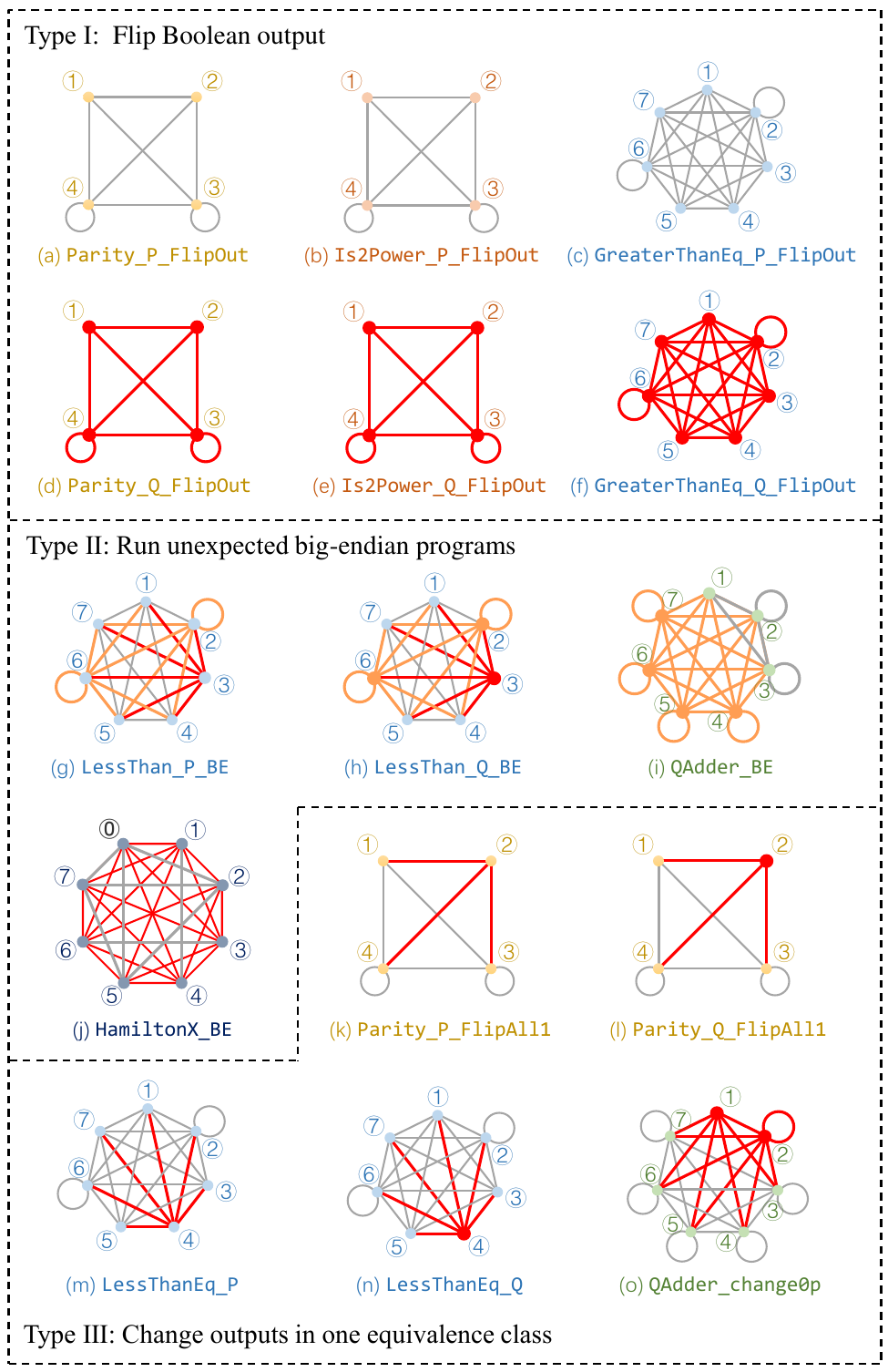}
    \caption{Results for RQ3: each graph corresponds to one program, where vertices represent $Q_i$ classes, self-loops represent $S_{ii}$ classes, and edges between two different vertices represent $S_{ij}$ classes. If bugs are detected by all inputs in a class, the corresponding edge or vertex is highlighted in red. If bugs are detected by only some inputs, it is highlighted in orange. If no bugs are detected in a class, no highlight is applied.}

	\label{fig:RQ3Result}
\end{figure*}

We represent the experimental results of each program using a graph, where vertices correspond to $Q_i$ classes (the circled labels correspond to Table~\ref{table:equivalence_classes}), self-loops represent $S_{ii}$ classes, and edges between different vertices represent $S_{ij}$ classes. If bugs are detected by all inputs in a class, the corresponding edge or vertex is highlighted in red. If bugs are detected by only some inputs, it is highlighted in orange. If no bugs are detected in a class, it remains unhighlighted. The experimental results are shown in Fig.~\ref{fig:RQ3Result}, where programs are grouped according to their types of behavioral change.

We next analyze the results for each type. In the "Flip Boolean output" type, an interesting observation is that no bugs are detected in the "phase" version programs, while all inputs report bugs in the "qubit" version programs. Ideally, flipping the Boolean output should lead to FAIL under all inputs. The results show that the "qubit" version behaves as expected, whereas the "phase" version does not. This can be explained by the fact that the phase difference between $1$ and $-1$ is $\pi$: flipping the output changes $1$ to $-1$ and $-1$ to $1$, preserving the relative phase. Consequently, such a change cannot be detected through phase-based comparison. In other words, for Boolean oracles expressed in phase form, $\ket{x} \rightarrow (-1)^{f(x)}\ket{x}$, the specific TRUE or FALSE values are indistinguishable; only the relative output differences between inputs can be observed. Therefore, flipping Boolean outputs encoded as phases does not introduce detectable bugs, which is consistent with the experimental results.

Next, we discuss the "Run unexpected big-endian programs" type. For the \texttt{LessThan} program (cases (g) and (h)), note that we set $k=10=01010_{(2)}$, and the classes \textcircled{1}, \textcircled{3}, \textcircled{4}, \textcircled{5}, and \textcircled{7} correspond to the inputs $0=00000_{(2)}$, $9=01001_{(2)}$, $10=01010_{(2)}$, $11=01011_{(2)}$, and $31=11111_{(2)}$, respectively. After reversing the endianness, they become $00000$, $10010$, $01010$, $11010$, and $11111$. We can see that class \textcircled{3} is no longer less than 10 after the endianness reversal. Thus, in case (h), vertex \textcircled{3} and its connected edges are highlighted in red. Classes \textcircled{2} and \textcircled{6} each contain multiple inputs; some of them change their order relations with 10 after the endianness reversal, while others remain unchanged. As a result, these vertices and their related edges are highlighted in orange.

Comparing cases (g) and (h), we find that no vertices are highlighted in case (g). This is because using only computational-basis inputs cannot detect bugs related to phase behavior. For \texttt{QAdder\_BE} (case (i)), note that if either input $x$ or $y$ equals 0, the output remains $y$ or $x$, respectively. Therefore, classes \textcircled{1}, \textcircled{2}, and \textcircled{3}, as well as the edges among them, are not highlighted. For \texttt{HamiltonX\_BE} (case (j)), since the program behavior depends on phase and each equivalence class contains only one input, the vertices are not highlighted, while the highlighted edges are red. The unhighlighted edges correspond to classes that remain unchanged after the endianness reversal. Overall, because some integers remain the same after reversing the endianness, the testing results for programs using unexpected endianness appear irregular across equivalence classes.

Next, we discuss the "Change outputs in one equivalence class" type. For these programs, the affected classes can be clearly identified in the graphs. If all edges connected to a vertex are highlighted, it indicates that bugs occur in the class associated with that vertex. In case (o), modifying the output for $x=0$ affects two classes, \textcircled{1} and \textcircled{2}.

According to the above analysis, we now examine the effectiveness of different pairing criteria. Due to the limitations of the computational-basis input classes $\{Q_i\}$ (see cases (g), (j), (k), and (m) in Fig.~\ref{fig:RQ3Result}), the superposition classes $\{S_{ij}\}$ play an important role in testing. From the results under the all-coverage criterion, we observe that if a bug appears in a computational-basis input class $Q_i$, it can also be detected in the corresponding two-value superposition classes $\{S_{ij}\,|\,\forall j\}$. Therefore, to ensure that bugs in any equivalence class can be detected, a rational pairing criterion should cover all classes, with the minimal configuration being the \textit{each-choice} pairing. In fact, if the goal is only to determine whether a program contains bugs, the each-choice pairing is sufficient. However, if further bug localization is required, broader coverage criteria such as \textit{tree-coverage} or \textit{all-coverage} should be adopted, as the bug locations can then be inferred from the graphs of the testing results.

\subsubsection{RQ4: Parameter $N_{tv}$}

\begin{figure*}
	\centering
	\includegraphics[scale=0.45]{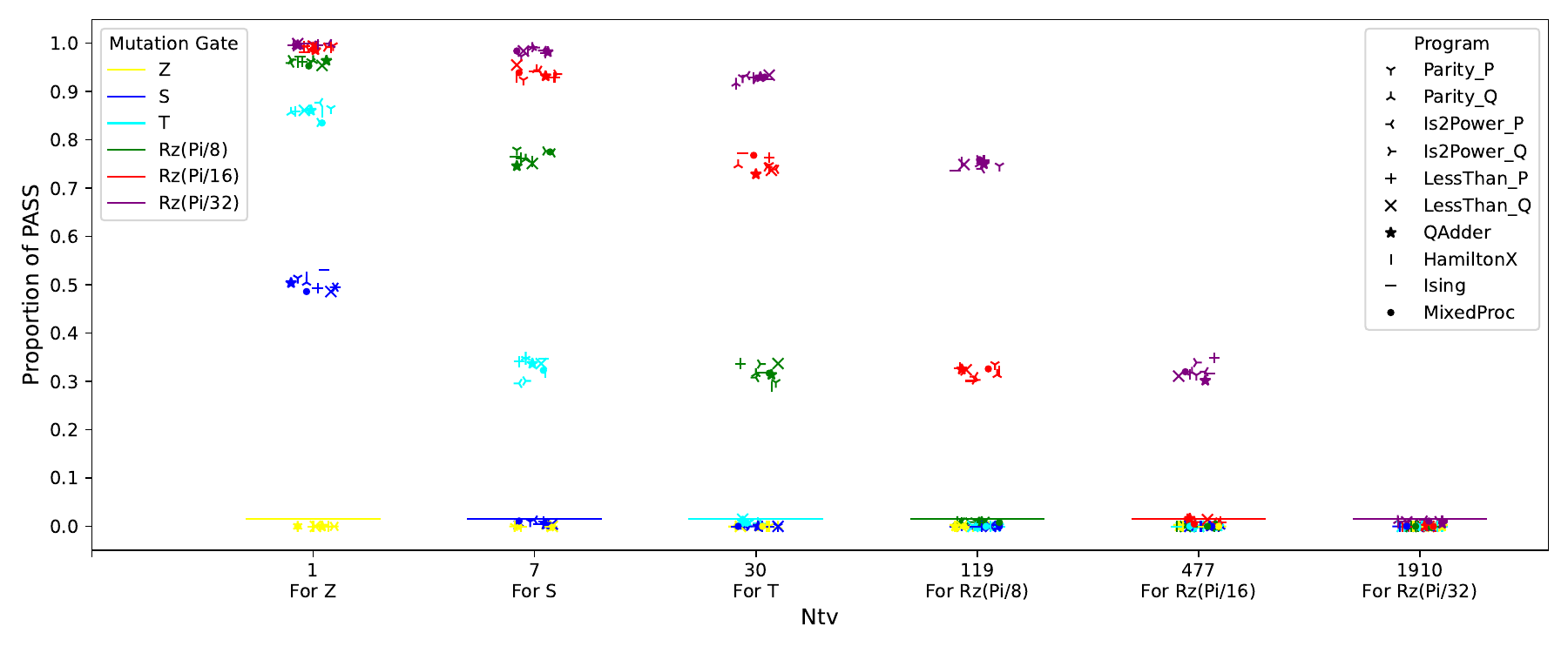}
    \caption{Results for RQ4: the proportion of PASS inputs for each expected-fail program with different $R_z$ mutants. The horizontal axis denotes the selected $N_{tv}$ values for each mutation gate. The colored horizontal short lines indicate the allowable misjudgment probability $\alpha=0.01$. Different markers correspond to different programs, and different colors correspond to different mutation gates.}
	\label{fig:RQ4Result}
\end{figure*}

Fig.~\ref{fig:RQ4Result} shows the relationship between the proportion of PASS inputs for expected-fail programs and $N_{tv}$. In this figure, data points of different colors represent different mutation gates. The ordinate indicates the proportion of PASS inputs, which corresponds to the estimated misjudgment probability, while the colored horizontal lines denote the allowable misjudgment probability $\alpha=0.01$. The selected $N_{tv}$ values (on the horizontal axis) are derived from Proposition~\ref{prop:Ntv} and Eq.~(\ref{equ:Ntv}). We can see that all data points lie below their corresponding (same-colored) horizontal lines. The experimental results confirm the effectiveness of the parameter selection method for $N_{tv}$ described in Proposition~\ref{prop:Ntv} and Eq.~(\ref{equ:Ntv}). As $N_{tv}$ increases, the misjudgment proportion decreases, allowing $R_z$ mutants with smaller phase angles to be distinguished. Moreover, if a given $N_{tv}$ value is sufficient to distinguish an $R_z(\theta)$ mutation, it is also effective for distinguishing all $R_z(\theta')$ mutations with $\theta' < \theta$.

\subsubsection{RQ5: Overall Performance of Benchmark Programs}

According to the evaluation of RQ1$\sim$RQ4, we select a balanced set of parameters, $N_{cb}=1$ and $N_{tv}=100$, to assess the overall performance of all benchmark programs. The results are shown in Table~\ref{table:RQ4result}. Our testing framework performs well for all benchmarks, as the number of PASS rounds (column "\# of PASS rounds") is close to 100 for expected-pass programs and close to 0 for expected-fail programs. In addition, each testing task can be completed within a reasonable time (column "Average running time for each round"), and programs using more qubits naturally require longer testing time.

We also compute the ratio of the running time of each expected-fail program to that of its corresponding expected-pass program (column "\% of correct"). The results show that expected-fail programs generally run faster because the testing process terminates once any nonzero result occurs, whereas a PASS result requires all input states to be tested. Furthermore, even under the same parameter settings, different expected-fail programs exhibit different time ratios: (1) as the amplitude $a$ in the output state $a\ket{\psi_{\mathrm{expected}}} + b\ket{\psi_{\mathrm{expected}}^{\bot}}$ decreases (from "\texttt{*\_AddRyPiDiv3}" to "\texttt{*\_AddRyPi}"), the time ratio decreases; (2) as the introduced phase decreases (from "\texttt{*\_AddZ}" to "\texttt{*\_AddRz32}"), the time ratio increases; and (3) \texttt{QAdder} and \texttt{Mixed\_Proc} exhibit longer execution times. Observations (1) and (2) arise from the fact that, for an expected-fail program closer to the correct one, the probability of obtaining a PASS result for each input is higher. Observation (3) arises because \texttt{QAdder} and \texttt{Mixed\_Proc} are more complex and involve more qubits than the other programs. Moreover, it is noteworthy that programs "\texttt{*\_AddRyPiDiv2}" and "\texttt{*\_AddS}" involve similar rotation angles of $\frac{\pi}{2}$ for the $R_y$ and $R_z$ gates (noting that $R_z(\frac{\pi}{2})$ is equivalent to $S$ up to a global phase). However, the running time of "\texttt{*\_AddS}" is significantly longer than that of "\texttt{*\_AddRyPiDiv2}". This may be because bugs caused by $R_z$ mutations cannot be detected by computational-basis input states, whereas those caused by $R_y$ mutations can. Consequently, buggy programs with $R_z$ mutations require more input states to be executed during testing.

\begin{table*}
\tiny
\caption{The testing results of benchmark programs using balanced parameter selection.}
\label{table:RQ4result}
\begin{tabular}{p{9mm}<{\centering}|p{25mm}<{\centering}|p{6mm}<{\centering}p{14mm}<{\centering}|p{20mm}<{\centering}|p{14mm}<{\centering}p{10mm}<{\centering}}
	\toprule
	&& \multicolumn{2}{c|}{$N_{cb} = 1, N_{tv} = 100$} & \# of total rounds $=100$ & \multicolumn{2}{c}{\textbf{Average running time}} \\
	\textbf{Expected} & \textbf{Program} & \textbf{\#Qubits} & \textbf{\# of Eq Classes} & \textbf{\# of PASS rounds} & for each round & \% of correct \\
	\midrule
	\multirow{10}{*}{PASS} & \texttt{Parity\_P} & $0+6$ & 12 & 100 & 37.70ms & \multirow{10}{*}{-} \\
	& \texttt{Is2Power\_P} & $0+6$ & 12 & 100 & 128.6ms &\\
	& \texttt{LessThan\_P} & $0+5$ & 30 & 100 & 144.4ms &\\
	& \texttt{Parity\_Q} & $6+1$ & 12 & 100 & 39.10ms &\\
	& \texttt{Is2Power\_Q} & $6+1$ & 12 & 100 & 138.5ms &\\
	& \texttt{LessThan\_Q} & $5+1$ & 30 & 100 & 160.7ms &\\
	& \texttt{QAdder} & $5+5$ & 34 & 100 & 1506ms &\\
	& \texttt{HamiltonX} & $0+3$ & 36 & 100 & 101.7ms & \\
	& \texttt{Ising} & $0+7$ & 19 & 100 & 85.19ms & \\
	& \texttt{Mixed\_Proc} & $5+5$ & 34 & 100 & 1513ms & \\
	\midrule
	\multirow{90}{*}{FAIL} & \texttt{Parity\_P\_AddRyPiDiv3} & $0+6$ & 12 & 0 & 2.073ms & 5.50\% \\
	& \texttt{Is2Power\_P\_AddRyPiDiv3} & $0+6$ & 12 & 0 & 6.266ms & 4.87\% \\
	& \texttt{LessThan\_P\_AddRyPiDiv3} & $0+5$ & 30 & 0 & 7.137ms & 4.94\% \\
	& \texttt{Parity\_Q\_AddRyPiDiv3} & $6+1$ & 12 & 0 & 2.302ms & 7.91\% \\
	& \texttt{Is2Power\_Q\_AddRyPiDiv3} & $6+1$ & 12 & 0 & 6.796ms & 4.91\% \\
	& \texttt{LessThan\_Q\_AddRyPiDiv3} & $5+1$ & 30 & 0 & 8.016ms & 4.99\% \\
	& \texttt{QAdder\_AddRyPiDiv3} & $5+5$ & 34 & 0 & 66.80ms & 4.44\% \\
	& \texttt{HamiltonX\_AddRyPiDiv3} & $0+3$ & 36 & 0 & 5.055ms & 4.97\% \\
	& \texttt{Ising\_AddRyPiDiv3} & $0+7$ & 19 & 0 & 4.777ms & 5.61\% \\
	& \texttt{Mixed\_Proc\_AddRyPiDiv3} & $5+5$ & 34 & 0 & 64.90ms & 4.29\% \\
	\cmidrule{2-7}
	& \texttt{Parity\_P\_AddRyPiDiv2} & $0+6$ & 12 & 0 & 1.334ms & 3.54\% \\
	& \texttt{Is2Power\_P\_AddRyPiDiv2} & $0+6$ & 12 & 0 & 3.568ms & 2.77\% \\
	& \texttt{LessThan\_P\_AddRyPiDiv2} & $0+5$ & 30 & 0 & 4.273ms & 2.96\% \\
	& \texttt{Parity\_Q\_AddRyPiDiv2} & $6+1$ & 12 & 0 & 1.334ms & 4.58\% \\
	& \texttt{Is2Power\_Q\_AddRyPiDiv2} & $6+1$ & 12 & 0 & 3.999ms & 2.89\% \\
	& \texttt{LessThan\_Q\_AddRyPiDiv2} & $5+1$ & 30 & 0 & 4.721ms & 2.94\% \\
	& \texttt{QAdder\_AddRyPiDiv2} & $5+5$ & 34 & 0 & 34.94ms & 2.32\% \\
	& \texttt{HamiltonX\_AddRyPiDiv2} & $0+3$ & 36 & 0 & 3.119ms & 3.07\% \\
	& \texttt{Ising\_AddRyPiDiv2} & $0+7$ & 19 & 0 & 3.158ms & 3.71\% \\
	& \texttt{Mixed\_Proc\_AddRyPiDiv2} & $5+5$ & 34 & 0 & 35.28ms & 2.33\% \\
	\cmidrule{2-7}
	& \texttt{Parity\_P\_AddRy2PiDiv3} & $0+6$ & 12 & 0 & 1.205ms & 3.20\% \\
	& \texttt{Is2Power\_P\_AddRy2PiDiv3} & $0+6$ & 12 & 0 & 2.682ms & 2.09\% \\
	& \texttt{LessThan\_P\_AddRy2PiDiv3} & $0+5$ & 30 & 0 & 3.247ms & 2.25\% \\
	& \texttt{Parity\_Q\_AddRy2PiDiv3} & $6+1$ & 12 & 0 & 1.167ms & 4.01\% \\
	& \texttt{Is2Power\_Q\_AddRy2PiDiv3} & $6+1$ & 12 & 0 & 3.003ms & 2.17\% \\
	& \texttt{LessThan\_Q\_AddRy2PiDiv3} & $5+1$ & 30 & 0 & 3.573ms & 2.22\% \\
	& \texttt{QAdder\_AddRy2PiDiv3} & $5+5$ & 34 & 0 & 25.13ms & 1.67\% \\
	& \texttt{HamiltonX\_AddRy2PiDiv3} & $0+3$ & 36 & 0 & 2.551ms & 2.51\% \\
	& \texttt{Ising\_AddRy2PiDiv3} & $0+7$ & 19 & 0 & 2.603ms & 3.06\% \\
	& \texttt{Mixed\_Proc\_AddRy2PiDiv3} & $5+5$ & 34 & 0 & 25.32ms & 1.67\% \\
	\cmidrule{2-7}
	& \texttt{Parity\_P\_AddRyPi} & $0+6$ & 12 & 0 & 1.047ms & 2.78\% \\
	& \texttt{Is2Power\_P\_AddRyPi} & $0+6$ & 12 & 0 & 2.515ms & 1.96\% \\
	& \texttt{LessThan\_P\_AddRyPi} & $0+5$ & 30 & 0 & 2.720ms & 1.88\% \\
	& \texttt{Parity\_Q\_AddRyPi} & $6+1$ & 12 & 0 & 1.015ms & 3.49\% \\
	& \texttt{Is2Power\_Q\_AddRyPi} & $6+1$ & 12 & 0 & 2.567ms & 1.85\% \\
	& \texttt{LessThan\_Q\_AddRyPi} & $5+1$ & 30 & 0 & 3.039ms & 1.89\% \\
	& \texttt{QAdder\_AddRyPi} & $5+5$ & 34 & 0 & 20.13ms & 1.34\% \\
	& \texttt{HamiltonX\_AddRyPi} & $0+3$ & 36 & 0 & 2.024ms & 1.99\% \\
	& \texttt{Ising\_AddRyPi} & $0+7$ & 19 & 0 & 2.054ms & 2.41\% \\
	& \texttt{Mixed\_Proc\_AddRyPi} & $5+5$ & 34 & 0 & 20.14ms & 1.33\% \\
	\cmidrule{2-7}
	& \texttt{Parity\_P\_AddCNOT} & $0+6$ & 12 & 0 & 8.731ms & 23.2\% \\
	& \texttt{Is2Power\_P\_AddCNOT} & $0+6$ & 12 & 0 & 42.51ms & 33.1\% \\
	& \texttt{LessThan\_P\_AddCNOT} & $0+5$ & 30 & 0 & 33.25ms & 23.0\% \\
	& \texttt{Parity\_Q\_AddCNOT} & $6+1$ & 12 & 0 & 8.646ms & 22.1\% \\
	& \texttt{Is2Power\_Q\_AddCNOT} & $6+1$ & 12 & 0 & 47.97ms & 34.6\% \\
	& \texttt{LessThan\_Q\_AddCNOT} & $5+1$ & 30 & 0 & 28.92ms & 18.0\% \\
	& \texttt{QAdder\_AddCNOT} & $5+5$ & 34 & 0 & 593.3ms & 39.4\% \\
	& \texttt{HamiltonX\_AddCNOT} & $0+3$ & 36 & 0 & 23.09ms & 22.7\% \\
	& \texttt{Ising\_AddCNOT} & $0+7$ & 19 & 0 & 46.30ms & 54.3\% \\
	& \texttt{Mixed\_Proc\_AddCNOT} & $5+5$ & 34 & 0 & 592.7ms & 39.2\% \\
	\cmidrule{2-7}
	& \texttt{Parity\_P\_AddCZ} & $0+6$ & 12 & 0 & 19.09ms & 50.6\% \\
	& \texttt{Is2Power\_P\_AddCZ} & $0+6$ & 12 & 0 & 71.72ms & 55.8\% \\
	& \texttt{LessThan\_P\_AddCZ} & $0+5$ & 30 & 0 & 74.06ms & 51.3\% \\
	& \texttt{Parity\_Q\_AddCZ} & $6+1$ & 12 & 0 & 20.34ms & 52.0\% \\
	& \texttt{Is2Power\_Q\_AddCZ} & $6+1$ & 12 & 0 & 74.97ms & 54.1\% \\
	& \texttt{LessThan\_Q\_AddCZ} & $5+1$ & 30 & 0 & 82.38ms & 51.3\% \\
	& \texttt{QAdder\_AddCZ} & $5+5$ & 34 & 0 & 1125ms & 74.7\% \\
	& \texttt{HamiltonX\_AddCZ} & $0+3$ & 36 & 0 & 58.40ms & 57.4\% \\
	& \texttt{Ising\_AddCZ} & $0+7$ & 19 & 0 & 72.50ms & 85.1\% \\
	& \texttt{Mixed\_Proc\_AddCZ} & $5+5$ & 34 & 0 & 1015ms & 67.1\% \\
	\cmidrule{2-7}
\end{tabular}
\end{table*}

\begin{table*}
\tiny
Continued from Table~\ref{table:RQ4result}

\begin{tabular}{p{9mm}<{\centering}|p{25mm}<{\centering}|p{6mm}<{\centering}p{14mm}<{\centering}|p{20mm}<{\centering}|p{14mm}<{\centering}p{10mm}<{\centering}}
	\toprule
	&& \multicolumn{2}{c|}{$N_{cb} = 1, N_{tv} = 100$} & \# of total rounds $=100$ & \multicolumn{2}{c}{\textbf{Average running time}} \\
	\textbf{Expected} & \textbf{Program} & \textbf{\#Qubits} & \textbf{\# of Eq Classes} & \textbf{\# of PASS rounds} & for each round & \% of correct \\
	\midrule
	\multirow{30}{*}{FAIL} & \texttt{Parity\_P\_AddZ} & $0+6$ & 12 & 0 & 16.64ms & 44.1\% \\
	& \texttt{Is2Power\_P\_AddZ} & $0+6$ & 12 & 0 & 59.53ms & 46.3\% \\
	& \texttt{LessThan\_P\_AddZ} & $0+5$ & 30 & 0 & 68.61ms & 47.5\% \\
	& \texttt{Parity\_Q\_AddZ} & $6+1$ & 12 & 0 & 17.04ms & 43.6\% \\
	& \texttt{Is2Power\_Q\_AddZ} & $6+1$ & 12 & 0 & 64.07ms & 46.3\% \\
	& \texttt{LessThan\_Q\_AddZ} & $5+1$ & 30 & 0 & 75.95ms & 47.3\% \\
	& \texttt{QAdder\_AddZ} & $5+5$ & 34 & 0 & 846.3ms & 56.2\% \\
	& \texttt{HamiltonX\_AddZ} & $0+3$ & 36 & 0 & 45.81ms & 45.0\% \\
	& \texttt{Ising\_AddZ} & $0+7$ & 19 & 0 & 56.52ms & 66.3\% \\
	& \texttt{Mixed\_Proc\_AddZ} & $5+5$ & 34 & 0 & 797.1ms & 52.7\% \\
	\cmidrule{2-7}
	& \texttt{Parity\_P\_AddS} & $0+6$ & 12 & 0 & 16.61ms & 44.1\% \\
	& \texttt{Is2Power\_P\_AddS} & $0+6$ & 12 & 0 & 60.80ms & 47.3\% \\
	& \texttt{LessThan\_P\_AddS} & $0+5$ & 30 & 0 & 68.92ms & 47.7\% \\
	& \texttt{Parity\_Q\_AddS} & $6+1$ & 12 & 0 & 17.84ms & 45.6\% \\
	& \texttt{Is2Power\_Q\_AddS} & $6+1$ & 12 & 0 & 67.97ms & 49.1\% \\
	& \texttt{LessThan\_Q\_AddS} & $5+1$ & 30 & 0 & 77.38ms & 48.2\% \\
	& \texttt{QAdder\_AddS} & $5+5$ & 34 & 0 & 815.8ms & 54.2\% \\
	& \texttt{HamiltonX\_AddS} & $0+3$ & 36 & 0 & 45.49ms & 44.7\% \\
	& \texttt{Ising\_AddS} & $0+7$ & 19 & 0 & 53.36ms & 62.6\% \\
	& \texttt{Mixed\_Proc\_AddS} & $5+5$ & 34 & 0 & 886.8ms & 58.6\% \\
	\cmidrule{2-7}
	& \texttt{Parity\_P\_AddT} & $0+6$ & 12 & 0 & 18.60ms & 49.3\% \\
	& \texttt{Is2Power\_P\_AddT} & $0+6$ & 12 & 0 & 64.74ms & 50.3\% \\
	& \texttt{LessThan\_P\_AddT} & $0+5$ & 30 & 0 & 71.89ms & 49.8\% \\
	& \texttt{Parity\_Q\_AddT} & $6+1$ & 12 & 0 & 19.24ms & 49.2\% \\
	& \texttt{Is2Power\_Q\_AddT} & $6+1$ & 12 & 0 & 70.32ms & 50.8\% \\
	& \texttt{LessThan\_Q\_AddT} & $5+1$ & 30 & 0 & 80.97ms & 50.4\% \\
	& \texttt{QAdder\_AddT} & $5+5$ & 34 & 0 & 830.4ms & 55.1\% \\
	& \texttt{HamiltonX\_AddT} & $0+3$ & 36 & 0 & 48.75ms & 47.9\% \\
	& \texttt{Ising\_AddT} & $0+7$ & 19 & 0 & 59.17ms & 69.5\% \\
	& \texttt{Mixed\_Proc\_AddT} & $5+5$ & 34 & 0 & 825.7ms & 54.6\% \\
	\cmidrule{2-7}
	& \texttt{Parity\_P\_AddRz8} & $0+6$ & 12 & 0 & 21.89ms & 58.1\% \\
	& \texttt{Is2Power\_P\_AddRz8} & $0+6$ & 12 & 0 & 78.11ms & 60.7\% \\
	& \texttt{LessThan\_P\_AddRz8} & $0+5$ & 30 & 0 & 91.39ms & 63.3\% \\
	& \texttt{Parity\_Q\_AddRz8} & $6+1$ & 12 & 0 & 23.60ms & 60.4\% \\
	& \texttt{Is2Power\_Q\_AddRz8} & $6+1$ & 12 & 0 & 83.14ms & 60.0\% \\
	& \texttt{LessThan\_Q\_AddRz8} & $5+1$ & 30 & 0 & 96.88ms & 60.3\% \\
	& \texttt{QAdder\_AddRz8} & $5+5$ & 34 & 0 & 1031ms & 68.5\% \\
	& \texttt{HamiltonX\_AddRz8} & $0+3$ & 36 & 0 & 60.34ms & 59.3\% \\
	& \texttt{Ising\_AddRz8} & $0+7$ & 19 & 0 & 60.90ms & 71.5\% \\
	& \texttt{Mixed\_Proc\_AddRz8} & $5+5$ & 34 & 0 & 984.6ms & 65.1\% \\
	\cmidrule{2-7}
	& \texttt{Parity\_P\_AddRz16} & $0+6$ & 12 & 0 & 29.55ms & 78.4\% \\
	& \texttt{Is2Power\_P\_AddRz16} & $0+6$ & 12 & 0 & 105.7ms & 82.2\% \\
	& \texttt{LessThan\_P\_AddRz16} & $0+5$ & 30 & 0 & 118.4ms & 82.0\% \\
	& \texttt{Parity\_Q\_AddRz16} & $6+1$ & 12 & 1 & 32.08ms &  82.0\% \\
	& \texttt{Is2Power\_Q\_AddRz16} & $6+1$ & 12 & 0 & 113.1ms & 81.7\% \\
	& \texttt{LessThan\_Q\_AddRz16} & $5+1$ & 30 & 0 & 132.6ms & 82.5\% \\
	& \texttt{QAdder\_AddRz16} & $5+5$ & 34 & 0 & 1258ms & 83.5\% \\
	& \texttt{HamiltonX\_AddRz16} & $0+3$ & 36 & 0 & 81.29ms & 79.9\% \\
	& \texttt{Ising\_AddRz16} & $0+7$ & 19 & 0 & 76.99ms & 90.4\% \\
	& \texttt{Mixed\_Proc\_AddRz16} & $5+5$ & 34 & 0 & 1278ms & 84.5\% \\
	\cmidrule{2-7}
	& \texttt{Parity\_P\_AddRz32} & $0+6$ & 12 & 1 & 35.19ms & 93.3\% \\
	& \texttt{Is2Power\_P\_AddRz32} & $0+6$ & 12 & 1 & 120.5ms & 93.7\% \\
	& \texttt{LessThan\_P\_AddRz32} & $0+5$ & 30 & 0 & 140.9ms & 97.6\% \\
	& \texttt{Parity\_Q\_AddRz32} & $6+1$ & 12 & 0 & 38.48ms & 98.4\% \\
	& \texttt{Is2Power\_Q\_AddRz32} & $6+1$ & 12 & 2 & 132.1ms & 95.4\% \\
	& \texttt{LessThan\_Q\_AddRz32} & $5+1$ & 30 & 0 & 153.4ms & 95.5\% \\
	& \texttt{QAdder\_AddRz32} & $5+5$ & 34 & 0 & 1448ms & 96.1\% \\
	& \texttt{HamiltonX\_AddRz32} & $0+3$ & 36 & 0 & 94.56ms & 93.0\% \\
	& \texttt{Ising\_AddRz32} & $0+7$ & 19 & 0 & 80.30ms & 96.6\% \\
	& \texttt{Mixed\_Proc\_AddRz32} & $5+5$ & 34 & 0 & 1436ms & 94.9\% \\
	\cmidrule{2-7}
	& \texttt{Parity\_Q\_FlipOut} & $6+1$ & 12 & 0 & 0.9800ms & 2.51\% \\
	& \texttt{Is2Power\_Q\_FlipOut} & $6+1$ & 12 & 0 & 2.458ms & 1.77\% \\
	& \texttt{GreaterThanEq\_Q\_FlipOut} & $5+1$ & 30 & 0 & 2.722ms & 1.69\% \\
	& \texttt{LessThan\_P\_BE} & $0+5$ & 30 & 0 & 86.35ms & 59.8\% \\
	& \texttt{LessThan\_Q\_BE} & $5+1$ & 30 & 0 & 83.32ms & 51.8\% \\
	& \texttt{QAdder\_BE} & $5+5$ & 34 & 0 & 431.1ms & 28.6\% \\
	& \texttt{HamiltonX\_BE} & $0+3$ & 36 & 0 & 34.81ms & 34.2\% \\
	& \texttt{Parity\_P\_FlipAll1} & $0+6$ & 12 & 0 & 33.23ms & 88.1\% \\
	& \texttt{Parity\_Q\_FlipAll1} & $6+1$ & 12 & 0 & 35.68ms & 91.3\% \\
	& \texttt{LessThanEq\_P} & $0+5$ & 30 & 0 & 132.9ms & 92.0\% \\
	& \texttt{LessThanEq\_Q} & $5+1$ & 30 & 0 & 150.8ms & 93.8\% \\
	& \texttt{QAdder\_change0p} & $5+5$ & 34 & 0 & 891.9ms & 59.2\% \\
	\bottomrule
\end{tabular}
\end{table*}

\section{Threats to Validity}
\label{sec:threat}

The experimental evaluation has demonstrated the effectiveness of our testing framework for oracle quantum programs. However, as with other testing methods, specific threats to the validity of our framework should be acknowledged.

\vspace{2mm}
\noindent $\bullet$ \textbf{Scalability.}\hspace*{1mm}
Theoretically, the execution time of the PRUM process excluding the run step (i.e., the prepare (P), uncompute (U), and measure (M) steps) satisfies $T_P + T_U + T_M = O(n)$, where $n$ is the number of qubits in the target program. Therefore, the testing framework scales linearly with qubit count. In practice, however, the computational complexity of the target program itself may grow faster than linear with respect to $n$, making the run (R) step the dominant factor. Consequently, our framework remains practical when the target program is not overly complex, but for programs with exponential complexity, executing the target program as part of the checking process may become infeasible.

\vspace{2mm}
\noindent $\bullet$ \textbf{Programming Languages.}\hspace*{1mm}
To enable automation within the testing framework, the chosen quantum programming language must support advanced features, such as function factories and functions as arguments. The evaluation presented in Section~\ref{sec:evaluation} is based on the Q\# language~\cite{svore2018q}, which offers these advanced capabilities. However, many quantum programming languages and device platforms currently lack such features. This limitation poses challenges for automating the testing process, particularly when developing and testing oracle quantum programs on these platforms.

\vspace{2mm}
\noindent $\bullet$ \textbf{Quantum Noise.}\hspace*{1mm}
In this paper, the theoretical analysis of the required number of repetitions assumes the perfect execution of the quantum programs. However, quantum noise and execution errors cannot be ignored in practical scenarios in the NISQ era. A feasible approach to addressing this issue is to develop and test prototype quantum programs on simulators before executing them on real quantum hardware. Some bugs can be detected and fixed at the prototype testing stage. This approach helps mitigate the effects of hardware noise during verification.

\section{Related Work}
\label{sec:relatedwork}

Quantum software testing is a rapidly evolving research field that aims to ensure the accuracy and reliability of quantum programs~\cite{miranskyy2019testing,garcia2021quantum,paltenghi2024survey,ramalho2024testing,zhao2020quantum,ali2022software}. Various methodologies have been proposed to adapt classical software testing techniques to the quantum domain, addressing unique quantum-specific challenges such as superposition, entanglement, and measurement-induced state collapse.

Several approaches have been explored for testing quantum programs. Classical testing techniques, such as search-based testing~\cite{wang2021generating}, fuzz testing~\cite{wang2018quanfuzz}, combinatorial testing~\cite{wang2021application}, metamorphic testing~\cite{abreu2022metamorphic}. Property-based testing methods~\cite{honarvar2020property} and concolic testing~\cite{xia2024concolic} have been extended to test quantum programs systematically. Quantum-specific assertions, such as projection-based assertions~\cite{li2020projection} and statistical assertions~\cite{huang2019statistical}, enable runtime verification of quantum state properties. These techniques aim to enhance the effectiveness of test oracles, which remain a key challenge in quantum software testing.

To assess the effectiveness of test cases, various test coverage criteria have been proposed. Ali et al.~\cite{ali2021assessing} introduced input-output coverage criteria for quantum programs, while Long and Zhao~\cite{long2024testing} developed equivalence class partitioning strategies. More recently, Shao and Zhao~\cite{shao2024coverage} extended these ideas to quantum neural networks by proposing multi-granularity coverage criteria. In addition, Gate Coverage~\cite{fortunato2024gate} was introduced to evaluate the effectiveness of test cases by analyzing quantum gate-level coverage.

Mutation testing has been widely used to evaluate the quality of quantum test suites. Several mutation testing frameworks have been developed, including Muskit~\cite{mendiluze2021muskit} and QmutPy~\cite{fortunato2022qmutpy}. Recent studies, such as~\cite{usandizaga2023quantum}, have analyzed large-scale mutant datasets (more than 700{,}000 mutants) to assess the fault-detection capabilities of different testing strategies. In addition, the MutTG approach~\cite{wang2022mutation} was introduced to minimize the size of the test suite while maintaining high coverage of the mutants.

Several frameworks and tools have been proposed to support the testing of quantum software. Long and Zhao~\cite{long2024testing} developed a framework for unit and integration testing of quantum programs, emphasizing systematic testing processes and automation. Relation-based testing approaches~\cite{long2024equivalence} leverage equivalence, identity, and unitarity relations to verify the correctness of the program. Mutation testing frameworks, such as Muskit~\cite{mendiluze2021muskit} and QmutPy~\cite{fortunato2022qmutpy}, evaluate the fault-detection capabilities of test suites using quantum-specific mutation operators.

In addition to testing techniques, formal methods have been applied to verify quantum software. Approaches such as symbolic execution for quantum programs~\cite{bauer2023symqv,fang2024symbolic} and quantum Hoare logic~\cite{ying2012floyd} contribute to formal verification efforts. These methods offer rigorous correctness guarantees but often face scalability challenges due to the exponential growth of quantum state spaces.

Oracle quantum programs play a critical role in many quantum algorithms and applications, serving as fundamental building blocks for evaluating Boolean functions, performing arithmetic operations, and simulating Hamiltonians~\cite {Vedral_1996,draper2000additionquantumcomputer,lloyd1996universal}. Previous research has focused on efficient implementations and optimizations of quantum oracles~\cite{Vedral_1996,draper2000additionquantumcomputer,cuccaro2004newquantumripplecarryaddition,Sahin2020QAonFourier,Thapliyal2021Division}. Recent work~\cite{Javier2025LessThan} introduced an optimized implementation of the less-than oracle, which we use as a benchmark program in this paper. Despite extensive research on oracle implementations, systematic testing methodologies for oracle quantum programs remain underexplored.
Existing work on testing oracle quantum programs is limited. Rui Abreu et al.~\cite{abreu2022metamorphic} applied metamorphic testing to verify quantum modular adders, demonstrating the potential of test relations for oracle verification. However, a generalized framework for systematically testing oracle quantum programs is still lacking. To the best of our knowledge, this paper is the first to propose a structured black-box testing framework specifically for oracle quantum programs, providing a foundation for future research in this area.

\section{Conclusion and Future Work}
\label{sec:conclusion}

This paper proposes a black-box testing framework for oracle quantum programs, an essential type of quantum software. We started by mathematically defining oracle quantum programs and outlining the key challenges in testing them. We then developed a testing framework that includes equivalence class construction, testing execution, and parameter selection. To evaluate our framework, we implemented a prototype tool and tested it using a set of benchmark oracle quantum programs. The experimental results demonstrated that our framework can effectively identify bugs and verify the correctness of oracle quantum programs.
There are several directions for future work:

\begin{itemize}
\setlength{\itemsep}{2pt}
\item \textit{Exploring the impact of classical equivalence class partitioning on quantum testing}: In this paper, we assume that the classical equivalence classes are pre-defined (assumption (iii) in Section~\ref{subsec:question}), and that the quantum equivalence classes are generated from these classical ones, with the focus placed on quantum-specific properties. However, since the selection of classical equivalence classes can affect the effectiveness of classical software testing, it may also influence the effectiveness of bug detection in oracle quantum programs. Therefore, exploring this aspect further would be a valuable direction for future research.

\item \textit{Exploring more pairing methods for generating superposition input states}: This paper used three specific strategies for generating superposition input states. Studying other pairing methods that might work better for different types of oracle quantum programs would be helpful.

\item \textit{Testing general} Hamiltonian evolution: Our framework works well for diagonal Hamiltonians, but testing programs with general (non-diagonal) Hamiltonians is still challenging and worth further research.

\item \textit{Adapting the framework for noisy quantum devices:} Our analysis assumes ideal conditions. Real quantum devices often have noise and errors, so adapting our framework to handle these challenges effectively is essential.

\item \textit{Automating the testing framework across diverse quantum languages and platforms:} The prototype tool developed in this study relies on advanced features of the Q\# language, such as function factories. Extending the framework to support other quantum programming languages and platforms that lack these features presents a valuable direction for future exploration.
\end{itemize}

These directions can enhance the testing of oracle quantum programs and advance the field of quantum software testing.

\begin{acks}
Peixun Long is supported in part by National High Energy Physics Data Center and High Energy Physics Data Center, Chinese Academy of Science. Jianjun Zhao is supported in part by the JSPS KAKENHI Grant No. JP24K02920, No. JP23K28062, No. JP24K14908.

\end{acks}


\bibliographystyle{acm}
\bibliography{ref}

\section*{Appendix}
\appendix

\section{Details of Input State Generation}
\label{appd:inputstate}

\noindent $\bullet$ $U_s$:\hspace*{1mm}
To implement $U_s$, $X$ gates are applied to the qubits corresponding to the '1' bits in $s$. For example, consider $s = 25 = 11001_{(2)}$ (i.e., to generate the state $\ket{11001}$). Three $X$ gates are applied to the 1st, 2nd, and 5th qubits, respectively.

\begin{figure*}
	\centering
	\includegraphics[scale=0.44]{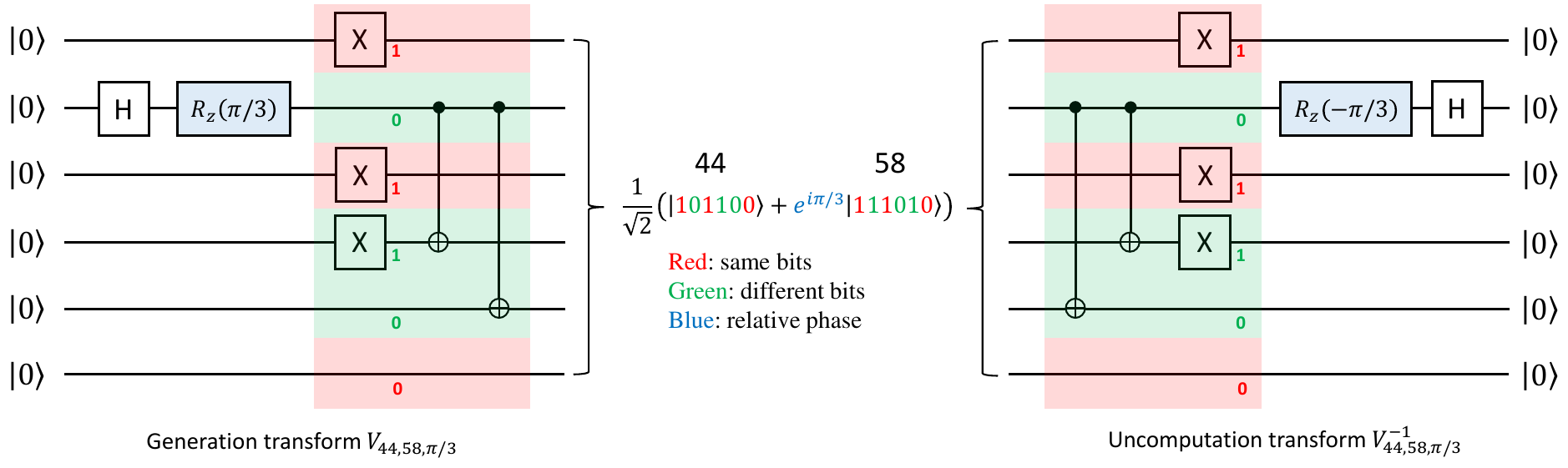}
	\caption{Quantum circuit to implement $V_{44,58,\pi/3}$ and its inverse $V_{44,58,\pi/3}^{-1}$.}
	\label{fig:GenTVS}
\end{figure*}

\vspace{2mm}
\noindent $\bullet$ $V_{s_1, s_2, \theta}$:\hspace*{1mm}
To implement $V_{s_1, s_2, \theta}$, it is necessary to identify the bits that are identical and those that differ in the binary representations of $s_1$ and $s_2$. Different operations are then applied to the qubits corresponding to these two groups of bits. Fig.~\ref{fig:GenTVS} illustrates the implementation of $V_{44, 58, \pi/3}$ and its inverse $V_{44, 58, \pi/3}^{-1}$, where the binary representations are $44 = 101100_{(2)}$ and $58 = 111010_{(2)}$. The generation process involves four steps:
\begin{enumerate}
	\item Identify the same and different bits in the binary representations of 44 and 58.
	\item Apply $X$ gates to the qubits corresponding to the '1'-bits in the same positions (i.e., 1st and 3rd bits in this example).
	\item For the differing bits (i.e., 2nd, 4th, and 5th), apply an $H$ gate, appropriate $X$ gates, and a series of \texttt{CNOT} gates to generate the superposition state.
	\item Insert an $R_z(\pi/3)$ gate\footnote{Strictly speaking, $R_1(\theta) = \mathrm{diag}\{1, e^{i\theta}\}$ should be used instead of $R_z(\theta) = \mathrm{diag}\{e^{-i\theta/2}, e^{i\theta/2}\}$, but the two differ only by a global phase factor, which cannot be distinguished by measurement.} after the $H$ gate to generate the relative phase $e^{i\pi/3}$.
\end{enumerate}

The uncomputation process $V_{44, 58, \pi/3}^{-1}$ reverses these steps, with the $R_z(\pi/3)$ gate replaced by $R_z(-\pi/3)$.

\section{Proofs of Proposition~\ref{prop:directmeasure} $\sim$ \ref{prop:Ntv}}
\label{appd:theoretical_analysis}

\subsection{Proof of Proposition~\ref{prop:directmeasure}}
\label{subappd:directmeasure}

Consider the orthogonal decomposition of the target state $\ket{\psi} = a\ket{s} + b\ket{s^\bot}$, where $\ket{s}$ is the target computational-basis state, $\ket{s^\bot}$ is its orthogonal complement, and $a, b \in \mathbb{C}$ satisfy $|a|^2 + |b|^2 = 1$. Clearly, if we measure $\ket{\psi}$ directly, we obtain the result $s$ with probability $p_1 = |a|^2$ and a non-$s$ result with probability $p_2 = |b|^2$.

Next, we consider applying $U_s$ to $\ket{\psi}$:
\begin{equation*}
U_s\ket{\psi} = U_s\left(a\ket{s} + b\ket{s^\bot}\right)
= aU_s\ket{s} + bU_s\ket{s^\bot}
= a\ket{0} + b\ket{0^\bot},
\end{equation*}
\noindent where the third equality follows from the fact that a unitary transform preserves orthogonality, and thus, $U_s\ket{s^\bot} = \ket{0^\bot}$. Measuring this state yields the result 0 with probability $p_3 = |a|^2 = p_1$, and a non-zero result with probability $p_4 = |b|^2 = p_2$.

\subsection{Proof of Proposition~\ref{prop:Ncb}}
\label{subappd:Ncb}

As shown in Eq. (\ref{equ:uncompute_cb}), if the output of the target program is correct, the uncomputation process will revert the output state to the all-zero state, ensuring that the measurement result is always 0. For simplicity, we combine the two parts of the state $\ket{x}\ket{\mathcal{F}(x,y)}$ into a single ket representation, denoting the expected output state as $\ket{\psi_{\mathrm{expected}}}$ and thus the uncomputation operation as $U_{\psi_{\mathrm{expected}}}$. Formally, the uncomputation process can be expressed as:
\begin{equation*}
    \ket{\psi_{\mathrm{expected}}}
    \xrightarrow{U_{\psi_{\mathrm{expected}}}}
    \ket{0\dots 0}
    \xrightarrow{M}
    0,
    \quad \text{with probability 1.}
\end{equation*}
\noindent Therefore, no misjudgment occurs for the expected output state.

Next, we analyze the case where the output state deviates from the expected result. Let the practical output state be $\ket{\psi_{\mathrm{output}}}$. This state can generally be decomposed into a superposition of the expected state and its orthogonal complement:
\begin{equation*}
    \ket{\psi_{\mathrm{output}}}
    = a\ket{\psi_{\mathrm{expected}}}
    + b\ket{\psi_{\mathrm{expected}}^{\bot}},
\end{equation*}
\noindent where $a, b \in \mathbb{C}$ satisfy $|a|^2 + |b|^2 = 1$. When $\ket{\psi_{\mathrm{output}}}$ matches the expected state, $|a| = 1$; otherwise, $|a| < 1$. Applying $U_{\psi_{\mathrm{expected}}}$ to $\ket{\psi_{\mathrm{output}}}$ transforms it into
\begin{equation*}
    a\ket{0\dots 0} + b\,U_{\psi_{\mathrm{expected}}}\ket{\psi_{\mathrm{expected}}^{\bot}}.
\end{equation*}
\noindent Since unitary transforms preserve orthogonality, $\ket{0\dots 0}$ is orthogonal to $U_{\psi_{\mathrm{expected}}}\ket{\psi_{\mathrm{expected}}^{\bot}}$. Consequently, measurement yields a result of $0$ with probability $|a|^2$ or a non-zero result with probability $|b|^2 = 1 - |a|^2$, as follows:
\begin{equation*}
\ket{\psi_{\mathrm{output}}}
\xrightarrow{U_{\psi_{\mathrm{expected}}}}
a\ket{0\dots 0} + b\,U_{\psi_{\mathrm{expected}}}\ket{\psi_{\mathrm{expected}}^{\bot}}
\xrightarrow{M}
\begin{cases}
    0 & \text{with probability $|a|^2$}, \\
    \neq 0 & \text{with probability $|b|^2$}.
\end{cases}
\end{equation*}
\noindent By repeating the PRUM process for $N_{cb}$ times, the probability that all measurement results are 0 (i.e., the misjudgment probability) is $(|a|^2)^{N_{cb}}$. This probability decreases exponentially with increasing $N_{cb}$. To limit the misjudgment probability to $\alpha$:
\begin{equation*}
    (|a|^2)^{N_{cb}} \leq \alpha,
\end{equation*}
the following condition must be satisfied:
\begin{equation*}
	N_{cb} \geq \frac{\ln \alpha}{\ln |a|^2}.
\end{equation*}
\noindent Eq.~(\ref{equ:Ncb}) is thus proved.

\subsection{Proof of Proposition~\ref{prop:Ntv}}
\label{subappd:Ntv}

\begin{figure*}
	\centering
	\includegraphics[scale=0.55]{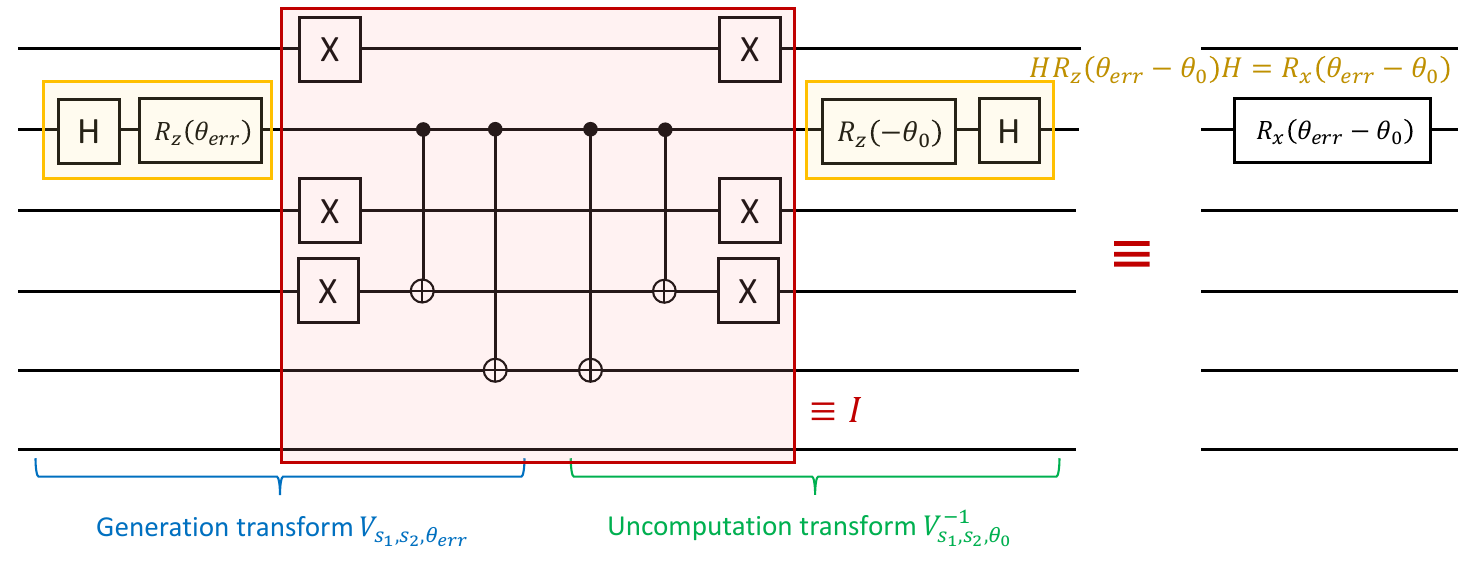}
	\caption{The execution of a pair of generation and uncomputation transforms with identical binary strings but different phase angles. The total operation is equivalent to an $R_x$ gate with a phase difference $\Delta\theta = \theta_{\mathrm{err}} - \theta_0$.}
	\label{fig:GenUncompTVS}
\end{figure*}

We examine the uncomputation process for an incorrect output state to analyze the misjudgment probability. Suppose that the expected output state is $\frac{1}{\sqrt{2}}(\ket{s_1} + e^{i\theta_0}\ket{s_2})$, which can be uncomputed to the all-zero state by $V_{s_1,s_2,\theta_0}^{-1}$. By contrast, the unexpected state takes the form $\frac{1}{\sqrt{2}}(\ket{s_1} + e^{i\theta_{\mathrm{err}}}\ket{s_2})$, which can be generated by $V_{s_1,s_2,\theta_{\mathrm{err}}}$ from the all-zero state, where $\theta_{\mathrm{err}} \neq \theta_0$. Consider the sequential execution of $V_{s_1,s_2,\theta_{\mathrm{err}}}$ followed by $V_{s_1,s_2,\theta_0}^{-1}$, as shown in Fig.~\ref{fig:GenUncompTVS}\footnote{Although this analysis is based on the specific implementations of $V_{s_1,s_2,\theta_{\mathrm{err}}}$ and $V_{s_1,s_2,\theta_0}^{-1}$ in Fig.~\ref{fig:GenTVS}, the conclusion is applicable to other implementations as well.}. 
We denote the phase difference between the expected and unexpected outputs by $\Delta\theta \triangleq |\theta_{\mathrm{err}} - \theta_0|$.

In this sequential execution, the intermediate operations cancel, leaving an $R_z(\theta_{\mathrm{err}} - \theta_0)$ gate and two $H$ gates. According to the circuit equation $HR_z(\theta)H \equiv R_x(\theta)$ (see Eq. (\ref{equ:rxrz})), this sequence is equivalent to an $R_x(\Delta\theta)$ gate on a single qubit. Thus, a phase difference between the expected and unexpected outputs finally reduces to an $R_x$ gate.

For a single $R_x(\Delta\theta)$ gate, the state transform is given by
\begin{equation*}
R_x(\Delta\theta)\ket{0} = \cos\frac{\Delta\theta}{2}\ket{0} - i\sin\frac{\Delta\theta}{2}\ket{1}.
\end{equation*}
\noindent Hence, measuring the qubit yields a result of 0 with probability $\cos^2\frac{\Delta\theta}{2}$ and a non-zero result with probability $\sin^2\frac{\Delta\theta}{2}$. For a correct output, where $\Delta\theta = 0$, the measurement result is always 0. For an incorrect output, the measurement probabilities depend on the phase difference $\Delta\theta$ relative to the correct output. Intuitively, the smaller $\Delta\theta$ is, the closer the probability of obtaining a zero result is to 1, increasing the likelihood of misjudgment. The PRUM process must be repeated multiple times to limit the probability of misjudgment.

We set a distinguishable phase angle $\Delta\theta$ and an allowed misjudgment probability $\alpha$. Repeating the PRUM process $N_{tv}$ times, the probability that all results are 0 (i.e., the misjudgment probability) is $\left(\cos^2\frac{\Delta\theta}{2}\right)^{N_{tv}}$. To ensure the the misjudgment probability does not exceed $\alpha$:
\begin{equation*}
\left(\cos^2\frac{\Delta\theta}{2}\right)^{N_{tv}}\leq \alpha,
\end{equation*}
\noindent the following inequality must hold:
\begin{equation*}
	N_{tv} \geq \frac{\ln \alpha}{\ln (\cos^2\frac{\Delta\theta}{2})}.
\end{equation*}
\noindent Eq. (\ref{equ:Ntv}) is thus proved.

Finally, we prove the asymptotic relation $\frac{\ln \alpha}{\ln (\cos^2\frac{\Delta\theta}{2})} = O\left(\frac{\log(1/\alpha)}{(\Delta\theta)^2}\right)$. Note that this relation concerns a small but nonzero $\Delta\theta$. For convenience, we restrict $\Delta\theta$ to $\left(-\pi,\pi\right)\backslash\{0\}$. To establish this, it suffices to prove the following inequality:
\begin{equation}
\label{equ:lncos2}
\frac{1}{\ln(\cos^2\frac{\Delta\theta}{2})} \geq -\frac{4}{(\Delta\theta)^2}.
\end{equation}
\noindent Multiplying both sides of inequality (\ref{equ:lncos2}) by $\ln \alpha$ (noting that $\ln \alpha < 0$ for $0 < \alpha < 1$), we obtain:
\begin{equation*}
	\frac{\ln\alpha}{\ln(\cos^2 \frac{\Delta\theta}{2})} \leq \frac{-4\ln\alpha}{(\Delta\theta)^2} = O\!\left(\frac{\log(1/\alpha)}{(\Delta\theta)^2}\right),
\end{equation*}
\noindent which gives the desired asymptotic relation. The remaining task is to prove inequality~(\ref{equ:lncos2}).

Consider the function
\begin{equation*}
f(x) = \ln\!\left(\cos^2\frac{x}{2}\right) + \frac{x^2}{4}, \qquad x \in (-\pi,\pi).
\end{equation*}
\noindent Taking the derivative of $f(x)$ with respect to $x$ and setting it to zero yields
\begin{equation*}
f'(x) = -\tan\frac{x}{2} + \frac{x}{2} = 0.
\end{equation*}
\noindent This equation has a unique solution $x=0$ in $(-\pi,\pi)$, making $f(0)=0$ the unique extremum of $f(x)$. Next, we show that this point is a maximum. Taking the second, third, and fourth derivatives of $f(x)$ gives
\begin{align*}
& f''(x) = \frac{1}{2}\left(-\sec^2\frac{x}{2}+1\right), &&f''(0) = 0; \\
& f^{(3)}(x) =  -\frac{1}{2}\sec^2\frac{x}{2}\tan\frac{x}{2}, &&f^{(3)}(0) = 0;\\
& f^{(4)}(x) = -\frac{1}{4}\sec^4\frac{x}{2} - \frac{1}{2}\sec^2\frac{x}{2}\tan^2\frac{x}{2}, &&f^{(4)}(0) = -\frac{1}{4} < 0.
\end{align*}
\noindent Since the first nonzero derivative at $x=0$ is the fourth derivative (an even order) and $f^{(4)}(0) < 0$, we conclude that $f(0)=0$ is a local maximum of $f(x)$ in $(-\pi,\pi)$, i.e., $f(x)\leq 0$ for all $x\in(-\pi,\pi)$. Consequently, we have
\begin{equation*}
\ln\!\left(\cos^2\frac{x}{2}\right) \leq -\frac{x^2}{4}, \qquad x \in (-\pi,\pi).
\end{equation*}
\noindent Taking the reciprocal (excluding $x=0$) and substituting $x$ by $\Delta\theta$ yields
\begin{equation*}
\frac{1}{\ln(\cos^2\frac{\Delta\theta}{2})} \geq -\frac{4}{(\Delta\theta)^2}, \qquad \Delta\theta \in (-\pi,\pi)\backslash\{0\},
\end{equation*}
\noindent thereby completing the proof of inequality~(\ref{equ:lncos2}).

\section{Proof of Proposition~\ref{prop:aExp_bNExp}}
\label{appd:aExp_bNExp}

Without loss of generality, we assume that the $R_y(\theta)$ gate is applied to the first qubit of $\ket{\psi_{\mathrm{expected}}}$. Since $\ket{\psi_{\mathrm{expected}}}$ is a computational-basis state (i.e., the output obtained under a computational-basis input), it can be represented as a binary string. Formally, by considering the first qubit separately, it can be written as $\ket{0}\ket{x}$ (if the first bit of the binary string is 0) or $\ket{1}\ket{x}$ (if the first bit is 1), where $\ket{x}$ represents the remaining qubits.

Note that
\begin{align*}
&R_y(\theta)\ket{0} =
\begin{bmatrix}
    \cos\frac{\theta}{2} & -\sin\frac{\theta}{2}\\
    \sin\frac{\theta}{2} & \cos\frac{\theta}{2}
\end{bmatrix}
\begin{bmatrix}
    1\\
    0
\end{bmatrix}
=
\begin{bmatrix}
    \cos\frac{\theta}{2}\\
    \sin\frac{\theta}{2}
\end{bmatrix}
=
\cos\frac{\theta}{2}\ket{0} + \sin\frac{\theta}{2}\ket{1}, \\[2ex]
&R_y(\theta)\ket{1} =
\begin{bmatrix}
    \cos\frac{\theta}{2} & -\sin\frac{\theta}{2}\\
    \sin\frac{\theta}{2} & \cos\frac{\theta}{2}
\end{bmatrix}
\begin{bmatrix}
    0\\
    1
\end{bmatrix}
=
\begin{bmatrix}
    -\sin\frac{\theta}{2}\\
    \cos\frac{\theta}{2}
\end{bmatrix}
=
-\sin\frac{\theta}{2}\ket{0} + \cos\frac{\theta}{2}\ket{1}.
\end{align*}
\noindent For the case of $\ket{0}\ket{x}$, applying the $R_y(\theta)$ gate to the first qubit yields
\begin{equation*}
\left(\cos\frac{\theta}{2}\ket{0} + \sin\frac{\theta}{2}\ket{1}\right)\ket{x}
= \cos\frac{\theta}{2}\ket{0}\ket{x} + \sin\frac{\theta}{2}\ket{1}\ket{x}
= \cos\frac{\theta}{2}\ket{\psi_{\mathrm{expected}}} + \sin\frac{\theta}{2}\ket{\psi_{\mathrm{expected}}^{\bot}}.
\end{equation*}

\noindent When we choose $\theta = 2\arccos(a)$, the amplitude of the expected component satisfies $\cos\frac{\theta}{2} = a$. For the case of $\ket{1}\ket{x}$, applying the $R_y(\theta)$ gate to the first qubit similarly yields
\begin{equation*}
\left(-\sin\frac{\theta}{2}\ket{0} + \cos\frac{\theta}{2}\ket{1}\right)\ket{x}
= \cos\frac{\theta}{2}\ket{1}\ket{x} - \sin\frac{\theta}{2}\ket{0}\ket{x}
= \cos\frac{\theta}{2}\ket{\psi_{\mathrm{expected}}} - \sin\frac{\theta}{2}\ket{\psi_{\mathrm{expected}}^{\bot}}.
\end{equation*}

\noindent In this case as well, the amplitude of the expected component is $\cos\frac{\theta}{2} = a$.

\end{document}